\documentclass[twocolumn,astrosymb]{aastex7}

\usepackage{amsmath}
\usepackage{apjfonts}
\usepackage{natbib}
\usepackage{CJK}

\newcommand\ii{{\sc ii}}
\newcommand\iii{{\sc iii}}

\newcommand\hii{\ion{H}{2} }

\newcommand\oi{[\ion{O}{1}]}
\newcommand\oii{[\ion{O}{2}]}
\newcommand\oiii{[\ion{O}{3}]}

\newcommand\siii{[\ion{S}{3}]}
\newcommand\nii{[\ion{N}{2}]}
\newcommand\sii{[\ion{S}{2}]}
\newcommand\ariii{[\ion{Ar}{3}]}
\newcommand\ariv{[\ion{Ar}{4}]}
\newcommand\te{T$_e$}
\newcommand\den{n$_e$}

\shorttitle{Physical Conditions and Direct-Method Abundances at $z\sim2-3$}
\shortauthors{Rogers et~al.}
\accepted{to ApJL on December 8th, 2025}

\begin{document}
\begin{CJK*}{UTF8}{gbsn}

\title{CECILIA: Gas-Phase Physical Conditions and Multi-Element Chemistry at Cosmic Noon}

\author[0000-0002-0361-8223]{Noah S. J. Rogers}
\affiliation{Center for Interdisciplinary Exploration and Research in Astrophysics (CIERA), Northwestern University, 1800 Sherman Avenue, Evanston, IL, 60201, USA}
\email{noah.rogers@northwestern.edu}

\author[0000-0001-6369-1636]{Allison L. Strom}
\affiliation{Center for Interdisciplinary Exploration and Research in Astrophysics (CIERA), Northwestern University, 1800 Sherman Avenue, Evanston, IL, 60201, USA}
\affiliation{Department of Physics and Astronomy, Northwestern University, 2145 Sheridan Road, Evanston, IL, 60208, USA}
\email{allison.strom@northwestern.edu}

\author[0000-0002-8459-5413]{Gwen C. Rudie}
\affiliation{Carnegie Observatories, 813 Santa Barbara Street, Pasadena, CA 91101, USA}
\email{gwen@carnegiescience.edu}

\author[0000-0002-6967-7322]{Ryan F. Trainor}
\affiliation{Department of Physics and Astronomy, Franklin \& Marshall College, 637 College Avenue, Lancaster, PA 17603, USA}
\email{ryan.trainor@fandm.edu}

\author[0000-0002-6034-082X]{Caroline von Raesfeld}
\affiliation{Center for Interdisciplinary Exploration and Research in Astrophysics (CIERA), Northwestern University, 1800 Sherman Avenue, Evanston, IL, 60201, USA}
\affiliation{Department of Physics and Astronomy, Northwestern University, 2145 Sheridan Road, Evanston, IL, 60208, USA}
\email{CarolinevonRaesfeld2027@u.northwestern.edu}

\author[0009-0008-2226-5241]{Menelaos Raptis}
\affiliation{Department of Physics and Astronomy, Franklin \& Marshall College, 637 College Avenue, Lancaster, PA 17603, USA}
\email{mraptis@fandm.edu}

\author[0000-0003-2385-9240]{Nathalie A. Korhonen Cuestas}
\affiliation{Center for Interdisciplinary Exploration and Research in Astrophysics (CIERA), Northwestern University, 1800 Sherman Avenue, Evanston, IL, 60201, USA}
\affiliation{Department of Physics and Astronomy, Northwestern University, 2145 Sheridan Road, Evanston, IL, 60208, USA}
\email{nathaliekorhonencuestas2029@u.northwestern.edu}

\author[0000-0001-8367-6265]{Tim B. Miller}
\affiliation{Center for Interdisciplinary Exploration and Research in Astrophysics (CIERA), Northwestern University, 1800 Sherman Avenue, Evanston, IL, 60201, USA}
\email{timothy.miller@northwestern.edu}

\author[0000-0002-4834-7260]{Charles C. Steidel}
\affiliation{Cahill Center for Astronomy and Astrophysics, California Institute of Technology, MS 249-17, Pasadena, CA 91125, USA}
\email{ccs@astro.caltech.edu}

\author[0000-0003-0695-4414]{Michael V. Maseda}
\affiliation{Department of Astronomy, University of Wisconsin-Madison, 475 N. Charter Street, Madison, WI 53706, USA}
\email{maseda@astro.wisc.edu}

\author[0000-0003-4520-5395]{Yuguang Chen (陈昱光)}
\affiliation{Department of Physics, The Chinese University of Hong Kong, Shatin, N.T., Hong Kong SAR, China}
\email{yuguangchen@cuhk.edu.hk}

\author[0000-0002-9402-186X]{David R. Law}
\affiliation{Space Telescope Science Institute, 3700 San Martin Drive, Baltimore, MD 21218, USA}
\email{dlaw@stsci.edu}


\correspondingauthor{Noah S. J. Rogers}
\email{noah.rogers@northwestern.edu}

\begin{abstract}

Galaxies at Cosmic Noon (\emph{z}$\sim$2-3) are characterized by rapid star formation that will lead to significant metal enrichment in the interstellar medium (ISM). While much observational evidence suggests that these galaxies are chemically distinct from those in the local Universe, directly measuring the ISM chemistry in large samples of high-\emph{z} galaxies is only now possible with the observational capabilities of JWST.
In this first key paper of the CECILIA program, we present the direct-method physical conditions and multi-element abundances in twenty galaxies at Cosmic Noon. Using a combination of archival Keck/MOSFIRE and new $\sim$30-hr NIRSpec spectroscopy, we measure multiple electron gas densities and the temperature structure from the O$^+$ and S$^{2+}$ ions. We find that \den\oii\ and \den\sii\ are comparable but elevated with respect to \den\ in local star-forming galaxies, and the simultaneous \te\oii\ and \te\siii\ generally agree with photoionization model \te\ scaling relations.
The O abundances in the CECILIA galaxies range from 12+log(O/H)$=$7.76-8.81 (12-131\% solar O/H), representing some of the highest direct-method metallicities and lowest \te\ (\te\oii$\approx$6500 K) measured with JWST to date.
The CECILIA galaxies exhibit significantly sub-solar S/O and Ar/O a signature of predominant enrichment from core collapse supernovae. The N/O-O/H trends in the CECILIA galaxies generally agree with the abundance trends in local nebulae, but the large scatter in N/O could be sensitive to the star-formation history.
The CECILIA observations demonstrate that exceptionally deep JWST spectroscopy can unveil the multi-element ISM abundance patterns in typical high-\emph{z} galaxies.

\end{abstract}

\section{Introduction}\label{sec:intro}

The evolving chemistry of galaxies is key to understanding many outstanding questions in contemporary astrophysics. The composition of protostellar gas affects the initial mass function, multiplicity, and planetary populations of stellar systems \citep[e.g.,][]{badenes2018,moe2019,sharda2022}. The metallicity of the massive stars formed from this gas affects their evolution and ionizing photon production, which in turn impacts the timescale of cosmic reionization \citep[e.g.,][]{stanway2016,gotberg2017}
and the rates of transient and gravitational-wave events \citep[e.g.,][]{eldridge2019,neijssel2019,vanson2025}.
Likewise, determining chemical abundance patterns advances our understanding of galaxy star formation histories, gas accretion and outflows, and the circum- and intergalactic media \citep[e.g.,][]{tinsley1979,mcwilliam1997}.

For these reasons, reconstructing metallicities and chemical abundance patterns across cosmic time is a high-priority goal for the astronomical community. This is especially the case for galaxies during the epoch of Cosmic Noon, the period at \emph{z}$\sim$2-3 when the cosmic star-formation rate (SFR) density peaked \citep[e.g.,][]{madau2014}. Given the rapid star formation taking place in these galaxies, it is unsurprising that they show properties distinct from their local counterparts. For example, large rest-optical surveys using ground-based NIR spectrographs have found that the typical star-forming galaxy (SFG) at 2$\lesssim$\emph{z}$\lesssim$3 shows elevated emission from collisionally-excited lines (CELs) of metals relative to local SFGs \citep[e.g.,][]{masters2014,steidel2014,shapley2015}. The increased CEL emission has been attributed to ionization from a harder spectrum, likely due to low-metallicity (i.e., low Fe/H) stellar populations at fixed gas-phase O/H \citep{steidel2016,strom2017,shapley2019,clarke2023}. This O/Fe enhancement has been corroborated with observations of the rest-UV spectrum of SFGs at Cosmic Noon, which are best reproduced with low-metallicity stellar continua \citep{steidel2016,cullen2019,kashino2022,chartab2024}. Simultaneous observations and photoionization modeling of the rest-UV continuum and rest-optical metal CELs indicates that the relative abundance of O/Fe in these galaxies is enhanced by more than a factor of two relative to the solar abundance ratio \citep{strom2018,topping2020a,topping2020b,cullen2021,stanton2024}. O and Fe are sensitive to different enrichment sources, where O enrichment occurs on short timescales via core-collapse supernovae (CCSNe, $\sim$10 Myr) while the bulk of Fe enrichment from Type Ia supernovae occurs on significantly longer timescales \citep[$>$0.1 Gyr,][]{kobayashi2009}.
It may be expected, therefore, that abundance ratios like O/Fe evolve significantly during Cosmic Noon owing to the intense star formation and overall abbreviated star-formation histories (SFHs).

These distinct chemical abundance patterns affect the ionizing spectrum of the massive stars considerably \citep[e.g.,][]{eldridge2009}, but they also fundamentally shape the physical conditions in the Interstellar Medium (ISM), specifically the electron gas temperature (\te) and density (\den). Optical emission from forbidden metal CELs is an efficient coolant in the ISM \citep{osterbrock2006}, such that altering the overall metallicity and relative chemical abundance patterns may change the cooling structure within the ISM. Ground-based surveys have also provided evidence of higher ISM \den\ during Cosmic Noon relative to local SFGs \citep{hainline2009,bian2010,sanders2016}. High \den\ can significantly decrease CEL emission through collisional de-excitation, producing CEL ratios that mirror those produced in extremely metal-poor environments \citep[e.g.,][]{topping2024}. Therefore, understanding the chemical evolution of galaxies at Cosmic Noon requires a complementary appreciation of how the ISM \te\ and \den\ respond to these unique chemical abundance patterns.

Metal CEL emission is sensitive to \te, \den, gas-phase metallicity, and the ionization state of the gas, thus the intensity of any one CEL is degenerate with multiple parameters. However, the relative intensity ratio of two CELs from the same metal ion eliminates these degeneracies and is sensitive to \te\ or \den\ alone. With \te, \den, and the metal CELs, it is then possible to directly measure the metal abundances in the gas \citep{dinerstein1990,peimbert2017,maiolino2019}. The direct abundance technique presents an opportunity to assess the chemical enrichment of the ISM, but has one major observational hurdle: the rest-optical metal auroral lines required to measure \te\ are exceptionally faint, typically $\lesssim$1\% the intensity of the strong \ion{H}{1} recombination lines. As a result, direct method abundances have been largely inaccessible from ground-based spectroscopy of high-\emph{z} SFGs \citep[see examples in][]{stark2013,james2014,berg2018,sanders2020,clarke2023,citro2024}.

The prospects for applying the direct abundance method at Cosmic Noon have greatly improved in the years following the launch of JWST. Indeed, JWST/NIRSpec observations have produced a windfall of high-quality rest-optical and NIR spectroscopy of distant galaxies with auroral line detections, allowing \te, \den, and chemical abundance patterns to be measured in galaxies during Cosmic Noon \citep[e.g.,][]{rogers2024,welch2024,welch2025,morishita2025,cataldi2025,scholte2025} and at \emph{z}$\geq$6 \citep[e.g.,][]{arellano-cordova2022,schaerer2022,curti2023,laseter2024,topping2024,chakraborty2025,cullen2025,pollock2025}. 
Despite these gains afforded by JWST spectroscopy, the evolution of ISM physical conditions and chemical abundance patterns of multiple elements remain uncertain. For example, JWST observations have provided further evidence of ISM density increasing with redshift, both as traced by the low-ionization gas of S$^+$ and O$^+$ \citep{abdurrouf2024} and high-ionization gas of C$^{2+}$ and N$^{3+}$ \citep{curti2025,topping2024,topping2025}. 
However, due to the narrow wavelength separation of the \oii$\lambda\lambda$3727,29 doublet and the low spectral resolution of most JWST/NIRSpec data, simultaneous \den\sii\ and \den\oii\ are often unavailable in high-\emph{z} galaxies \citep[c.f.,][]{li2025,welch2025}.
While \den\sii\ and \den\oii\ show good agreement in local nebulae \citep{mendez-delgado2023}, the low-ionization potential (IP) of S$^+$ (10.6 eV) means that it exists outside the \hii\ region. A departure of \den\sii\ from \den\oii, should one exist, would point to a changing density structure in the ISM relative to local galaxies and would provide insight into the star formation taking place at Cosmic Noon.

Additionally, most direct \te\ measurements from JWST/NIRSpec observations are made using the \oiii$\lambda$4364 auroral line, a consequence of the high ionization and low metallicities typical of high-\emph{z} SFGs. In the local Universe, \te\ measured from different ions in the ISM are strongly correlated and parameterized by \te\ scaling relations \citep{kennicutt2003,esteban2009,croxall2016,rogers2021}. These \te\ scaling relations, either calibrated with empirical \te\ data or photoionization model predictions \citep{campbell1986,garnett1992,pagel1992}, allow for the inference of \te\ in different components of the ISM when a direct temperature is unavailable. However, the unique abundance patterns in high-\emph{z} galaxies could significantly affect the \te\ structure within the ionized gas and change the functional form of the \te\ scaling relations. Therefore, \te\ scaling relations represent a significant systematic uncertainty for chemical abundance surveys that cannot directly resolve the complex \te\ structure of the multi-phase ISM \citep{arellano-cordova2020,rogers2022}.

Finally, JWST observations have verified the distinct chemical abundance patterns in high-\emph{z} SFGs relative to SFGs in the local Universe, both in terms of the bulk metallicity and relative abundance patterns. Direct abundance measurements in galaxies at \emph{z}$>$2 have revealed unexpected trends such as N-rich, O-poor galaxies at \emph{z}$>$2 \citep{welch2024,welch2025,arellano-cordova2025} and super-solar N/O at \emph{z}$>$6 \citep{senchyna2024,topping2024,topping2025no,curti2025}.
Additionally, there is growing evidence to suggest that Ar/O is sub-solar in high-\emph{z} galaxies \citep{rogers2024,bhattacharya2025A,stanton2025}. Since Ar is produced in Type Ia SNe \citep{kobayashi2020SNIa}, the Ar/O ratio should be sensitive to the relative enrichment from CC and Type Ia SNe and, therefore, the SFH of a galaxy. Such a dependence may produce different Ar/O abundance patterns in high-\emph{z} galaxies compared to typical \emph{z}$=$0 SFGs with solar Ar/O \citep[e.g.,][]{berg2020,arellano-cordova2024a}.
The abundance patterns of other elements, such as S, remain largely unconstrained at high-\emph{z} \citep[c.f.,][]{welch2025,morishita2025} owing to small sample sizes and the faint emission lines required for the direct abundance method. Therefore, it remains unclear if these peculiar abundance patterns are representative of the typical SFG at Cosmic Noon, or if they are a reflection of prompt enrichment from recent phases of intense star formation \citep{senchyna2024,kobayashi2024}.

Uncovering the evolution of the ISM physical conditions, chemical abundance patterns, and the prevalence of different enrichment mechanisms, therefore, requires a large sample of high-\emph{z} SFGs with direct-method metallicities. In this paper, we present such a sample, acquired as part of the Chemical Evolution Constrained using Ionized Lines in Interstellar Aurorae \citep[CECILIA, PID2593,][]{strom2023} program. CECILIA was designed to detect the \te-sensitive auroral lines of \oii\ and \siii\ in SFGs at Cosmic Noon using ultra-deep NIRSpec observations and complementary Keck/MOSFIRE spectroscopy of the strong optical CELs. These deep data have enabled studies of the faintest emission lines observed in the rest-optical and NIR spectra of high-\emph{z} galaxies \citep{strom2023}, the chemical abundance patterns in the ISM at \emph{z}$\sim$3 \citep{rogers2024}, and the ionization conditions and star formation in low-mass, faint galaxies at Cosmic Noon \citep{raptis2025}. We now examine the full sample of \den, \te, and chemical abundances measured in CECILIA, 
including 9 galaxies with direct \te\siii\ and 17 with \te\oii, representing a $\sim$70\% and 44\% increase in the number of galaxies with direct \te\siii\ and \te\oii, respectively, at \emph{z}$>$1.

This manuscript is organized as follows: we describe the CECILIA survey, MOSFIRE and NIRSpec data reduction, emission line fitting, and reddening corrections in \S\ref{sec:data}. In \S\ref{sec:physconditions}, we outline the methods used for \den\ and \te\ calculations in the CECILIA galaxies, and we discuss the simultaneous \den\oii-\den\sii\ and \te\oii-\te\siii\ trends measured at Cosmic Noon. In \S\ref{sec:abun}, the chemical abundance trends in the full sample are presented and analyzed. We summarize our conclusions in \S\ref{sec:conclusions}, and we elaborate on specific aspects of the data reduction and analysis in Appendices \ref{app:gbkg} and \ref{app:ebv}. Throughout this paper, we assume a $\Lambda$CDM cosmology with $H_0$ = 70 km/s/Mpc, $\Omega_\Lambda$ = 0.7, and $\Omega_m$ = 0.3. Solar metallicities are adopted from \citet{asplund2021}: 12+log(O/H)$_\odot$ $=$ 8.69$\pm$0.04, log(N/O)$_\odot$ $=$ $-$0.86$\pm$0.08, log(S/O)$_\odot$ $=$ $-$1.57$\pm$0.05, and log(Ar/O)$_\odot$ $=$ $-$2.31$\pm$0.11. We refer to the position of spectral features using their vacuum wavelengths. When discussing individual galaxies in the text, we use the galaxy ID and omit the "Q2343-" field name.

\section{The CECILIA Survey}\label{sec:data}

\subsection{Galaxy Sample}\label{sec:sample}

CECILIA includes galaxies that were first identified and observed as part of the Keck Baryonic Structure Survey (KBSS; \citealt{steidel2010,rudie2012,strom2017}). KBSS is a large spectroscopic survey of $1.5\lesssim z\lesssim3.5$ galaxies in 15 fields surrounding bright quasars; most KBSS galaxies are selected based on their rest-UV continuum colors, with a subset of continuum-faint galaxies selected on the basis of their Ly$\alpha$ emission in narrow-band imaging. The new JWST/NIRSpec observations were conducted in the Q2343+125 field due to the (1) high density of sources with spectroscopic redshifts, (2) large catalog of narrow-band-selected Ly$\alpha$ emitters (LAEs) at $z\approx2.55$, and (3) existing HST/WFC3 F140W mosaic that provided precision astrometry for mask design and galaxy size measurements. The ancillary data available for the Q2343+125 field and, by extension, for CECILIA include $J$, $H$, and $K$ NIR (rest-optical) spectroscopy from Keck/MOSFIRE, deep optical (rest-UV) spectroscopy from Keck/LRIS, and imaging in $U_n$ through $K_s$, F140W, F160W, IRAC Ch~$1-4$, MIPS 24 $\mu$m, and narrow-band (rest-frame) Ly$\alpha$ filters.

Given the explicit goal of measuring electron temperatures and direct-method metallicities from the JWST/NIRSpec observations, we prioritized galaxies with $2.1\leq z\leq2.64$, where all of the required emission lines fall within the G235M, G395M, and ground-based spectral bandpasses. As described in more detail in \citet{strom2023}, galaxies were also assigned a higher priority if they have smaller than average sizes, high SFRs, and/or large observed nebular [\ion{O}{3}]$\lambda5007$ line fluxes, all of which increase the ease of detecting the faint temperature-sensitive auroral lines. The full targeted sample spans the parameter space occupied by typical SFGs in various line ratio diagrams such as the N2-BPT \citep{baldwin1981} and $O_{32}$-$R_{23}$ (or ionization-excitation diagram) in order to provide a useful metric of the ISM properties across the parent galaxy population. We also included a number of narrow-band selected Ly$\alpha$ emitters (LAEs) from \citet{trainor2015} with spectroscopic detections of Ly$\alpha$ and [\ion{O}{3}]$\lambda5007$ or H$\alpha$; the first results for these faint galaxies are presented in \citet{raptis2025,raptis2025b} and are not the main focus of the present work.

\subsection{SED Fitting}\label{sec:sed}

We determine the spectral energy distribution (SED) for each CECILIA target using the archival photometry described above. Note that the Palomar/WIRC photometry in $K_s$ is generally of worse quality than other bands, but in the case of non-detections in the IRAC channels, it is required to constrain the SED redward of the Balmer break. As a result, we only include the $K_s$ photometric point in the fit if the galaxy is undetected in IRAC Ch 1, 2, and 3.

We use BAGPIPES \citep{carnall2018} for SED fitting, using similar methods to those described by \citet{korhonen-cuestas2025}. We employ the Bayesian nested sampler Nautilus \citep{lange2023} and the BPASS v.2.2.1 \citep{stanway2018} stellar grids assuming a \cite{kroupa2001}-like IMF with upper mass slope of $-$2.35 and a maximum stellar mass of 100 $M_\odot$. Cloudy \citep{ferland2017} grids are used to model the nebular continuum. We do not include nebular emission lines in the SED model, as the photometry has been corrected for the contributions of emission lines using the measured JWST/NIRSpec equivalent widths as well as previously-measured emission line fluxes from MOSFIRE where available. We adopt uniform priors on the dust extinction, $A_V$, logarithm of the ionization parameter, stellar metallicity, and logarithm of the stellar mass, allowing each parameter to vary between $[0, 2]$, $[-3.5, -2.5]$, $[0.05, 0.6]\,Z_\odot$, and $[0, 13]\,\textrm{M}_\odot$ respectively. Gas phase metallicity is fixed to the stellar metallicity.\footnote{Note that the SED fits are insensitive to gas-phase metallicity because the emission lines are subtracted from the photometry and are not fit with the BAGPIPES modeling.} Dust extinction is modeled assuming the \cite{gordon2003} SMC dust reddening curve. The adopted priors are based on the stellar population synthesis results reported in \citet{steidel2016} for a stacked UV spectrum of KBSS galaxies; however, as discussed in \S\ref{sec:oxygen}, some of the CECILIA galaxies have gas-phase O/H greater than the upper limit permitted by the metallicity prior. When repeating the SED fitting with [0.05,1.0]$Z_\odot$, we find that stellar masses and SFRs are unchanged. While the median stellar metallicity increases by $\sim$0.15 dex, the typical uncertainty on $Z_*$ is 0.37 dex. Since we do not analyze the stellar metallicities here, we adopt the narrower prior on stellar metallicity.

The SFH is fit using a non-parametric, continuity SFH, as described in \cite{leja2019}. We define time bins of 0, 10, 100, 250, 500, 1000, 1500, 2000, 2500, 3000, and 3500 Myr before the time of observation and fit for the change in SFR between each bin. In practice, the last time bin is sometimes excluded by the redshift of the galaxy when older than the age of the Universe at that epoch. The prior for changes in SFR between adjacent bins is set to the BAGPIPES default for a continuity SFH; i.e., a Student's t distribution centered at 0 with a scale $\sigma=0.3$ and 2 degrees of freedom. Such a prior is weighted against dramatic changes in SFR between time bins, generally producing smoother SFHs.  Notably, the nominal SFR reported by BAGPIPES is measured on 100 Myr timescales by default, which produces an apparent maximum specific star formation rate (sSFR) of $10^{-8}\,\textrm{yr}^{-1}$ and artificial linear features to the distribution of objects in the SFR-M$_*$ plane. However, the underlying SFH used for the SED fitting and total mass estimation includes the full set of time bins described above. The resulting distribution of SED-fit SFRs and stellar masses for the CECILIA sample and their parent population are discussed further in Sec.~\ref{sec:bpt}.

\subsection{Keck/MOSFIRE Spectroscopy}

The Keck/MOSFIRE observations of KBSS galaxies have been described extensively in past work \citep[]{steidel2014,strom2017}. In brief, KBSS used MOSFIRE's configurable slit unit (CSU) to observe $\sim20-30$ galaxies at a time in a single atmospheric window; individual galaxies have observations spanning the $J$, $H$, and $K$ bands, which collectively cover $\lambda_{\rm rest}\approx3700-7000$~\AA\, for our highest-priority targets. Individual mask configurations were designed with $0\farcs7$-wide slits, resulting in a spectral resolution of $R\approx3300-3700$. Observations were acquired using a two-position dither sequence along the slit, using the standard dither spacing of $3\farcs0$. Individual integration times were 180~s in $K$ band and 120~s in the $J$ and $H$ bands; total integration times range from $\simeq1.5-10$~hr/band, with a typical total integration time of $\sim3$~hr/band. In general, once a spectroscopic redshift was determined, galaxies continued to be observed until the strongest emission lines (including [\ion{O}{2}]$\lambda\lambda3727,29$, H$\beta$, [\ion{O}{3}]$\lambda5008$, H$\alpha$, and [\ion{N}{2}]$\lambda6585$) were detected at $\gtrsim3\sigma$, so some of the fainter galaxies in the sample have substantially longer exposure times.

The raw MOSFIRE data were reduced using the publicly-available data reduction pipeline,\footnote{\url{https://keck-datareductionpipelines.github.io/MosfireDRP/}} which produces rectified and wavelength-calibrated 2D spectra for individual galaxies. Relative flux calibration and correction for telluric absorption were accomplished using wide- and narrow-slit observations of a standard star, and the spectra were shifted to account for the heliocentric velocity at the time of observation. For galaxies that were observed more than once per spectral band, the individual 2D spectra were co-added to create a final, inverse variance-weighted 2D spectrum for each object.

1D spectra are extracted using MOSPEC,\footnote{\url{https://github.com/allisonstrom/mospec}} an interactive IDL-based analysis tool developed for faint emission line spectroscopy with MOSFIRE. We adopt a Gaussian weight profile and optimally extract the spectra using the approach from \citet{horne1986}. Examples of 1D MOSFIRE data of KBSS galaxies can be found in \citet{steidel2014}. To account for slit losses, which can differ from band to band depending on observing conditions, we use the Markov Chain Monte
Carlo (MCMC) method from \citet{strom2017}. First, we compare observations of stars that are observed on each MOSFIRE mask with their NIR magnitudes. Then, using these correction factors as priors, we determine the optimal correction factor for each MOSFIRE mask, making the reasonable assumption that independent measurements of the strongest emission lines in individual galaxies' spectra should be the same; i.e., the adopted mask correction factors should minimize the scatter between the line flux measurements for the same object on different masks. For galaxies observed on a single mask, these mask corrections are used to account for slit losses, but if a galaxy has been observed on more than one mask, the line fluxes from the weighted average spectrum are corrected to match the weighted average of the corrected line fluxes for individual observations. Overall, the median correction factor is $\sim2$ with a median uncertainty of 4\%, and the uncertainties on the slit-loss corrections are later included in the final line flux uncertainties for the MOSFIRE data.

\subsection{JWST/NIRSpec Spectroscopy}\label{sec:jwstdata}

The JWST/NIRSpec observations including mask design and strategy are described in \citet{strom2023}. The data were taken using a single configuration of the micro-shutter assembly (MSA), at an aperture position angle of 19.3~deg. The MSA shutters have an open area of approximately $0\farcs20\times0\farcs46$ and are arranged in a fixed grid, so we opted to use the ``midpoint'' centering criterion to jointly maximize the sample size and signal-to-noise (S/N) for individual galaxies. Given these constraints, the optimal MSA design included 34 sources. Each object's slitlet was composed of at least 3 shutters, with additional shutters included to elongate slits in cases where there were not other open shutters in the same MSA row. Additional shutters were opened to sample the background in areas with no contamination from other sources based on deep \textit{HST} images.

Two NIRSpec disperser-filter combinations were used to obtain spectral coverage spanning the rest-optical and rest-NIR, both utilizing medium-resolution ($R\sim1000$) gratings. Ultra-deep 29.5-hr spectra in G235M/F170LP ($\lambda_{\rm obs}=1.66-3.07~\mu$m) cover the observed wavelengths of multiple auroral features, including [\ion{S}{3}]$\lambda6314$ and \oii$\lambda\lambda7322,32$, for galaxies $1.63\lesssim z\lesssim3.19$, including 30/34 CECILIA sources; a shorter 1.1-hr exposure sequence in G395M/F290LP ($\lambda_{\rm rest}=2.87-5.10~\mu$m) is used to measure the corresponding nebular [\ion{S}{3}]$\lambda\lambda9071,9533$ doublet for the 14 CECILIA galaxies at $z\gtrsim2.4$. In both cases, the observations were taken using the default three-point nod pattern and the recommended NRS IRS2 detector readout mode. 

As CECILIA observations were taken early in Cycle 1, the performance of the MSA shutters had not yet been quantified. The data were unfortunately impacted by both failed shut and failed open shutters. Failed shut shutters impacted 5 high-priority targets (BX348, BX418, BX274, BX350, BX391). Among them, BX391 is excluded from this work as $>80\%$ of the exposures were lost. The remaining three galaxies are included in this analysis with loss of 30-60\% of their exposures. Notably all three have detections of auroral  \oii\ but none have detections of the fainter \siii\ auroral line.  Failed open shutters contribute extra background light to unassociated shutters in the same or adjacent MSA rows. These shutters were flagged and affected portions of each exposure were masked before combining as described below. All these sources were recoverable and they are included in the remaining analysis, with only BX216 experiencing contamination of the science object spectrum by the additional background light. For this object, 4 of the 12 exposures were masked at $\lambda_\textrm{obs}\approx2.7-3.2\,\mu$m ($\lambda_\textrm{rest}\approx0.87-1\,\mu$m).  

Initial reductions of the CECILIA NIRSpec data have been described in \citet{strom2023} and \citet{rogers2024}, but we take this opportunity to discuss the general procedures applied to the full sample and to highlight improvements from these initial reductions. The raw, uncalibrated data files are processed through Level 1 of \textsc{calwebb} v1.15.1 with \textsc{crds\textunderscore context} 1251.pmap \citep{calwebb_v1.15.1} using \textsc{grizli} v1.11.11 \citep{grizli_v1.9.11}. This step performs initial data quality flagging, corrections for linearity and persistence, and cosmic ray (or "snowball") masking to eliminate jumps in the science groups. Once accounted for, a ramp is fit to the science groups to obtain the calibrated \textsc{rate} files. While recent versions of \textsc{calwebb} now implement a 1/\emph{f} noise correction, we have elected to apply \textsc{NSClean} \citep{rauscher2024} directly to the \textsc{rate} files to remove this noise from the NIRSpec data. We first mask areas of the \textsc{rate} files illuminated by science or background light and pixels where residual artifacts could affect the fit to the low-frequency 1/\emph{f} noise. The default \textsc{NSClean} parameters were found to introduce high-frequency noise to the data, so we update the critical frequency, kill width, and Gaussian smoothing standard deviation to fc$=$1/2048, kw$=$fc/4, and buffer\textunderscore sigma$=$3, respectively, which reduces the RMS noise in the unmasked \textsc{rate} files for both NRS1 and NRS2.

The cleaned files are further processed through portions of the Level 2 \textsc{calwebb} pipeline using \textsc{msaexp} v0.8.5 \citep{msaexp}. For each group of exposures: a bias term is removed from the data, slits are identified using the slit meta data file, WCS is assigned to the \textsc{rate} file and individual slits, and the photometric calibration step is run. We adopt the \textsc{calwebb} pathloss solution for a uniformly illuminated slit. This procedure corrects for the flux loss at higher wavelengths due to the increasing full width at half maximum (FWHM) of the point spread function (PSF). The uniform solution may not be appropriate for the most compact targets or objects that are observed away from slit center. However, some of these issues are mitigated by performing a spectrophotometric correction to the 1D spectra, as described later in this section.

\begin{figure*}[t]
   \centering
   \includegraphics[width=0.99\textwidth]{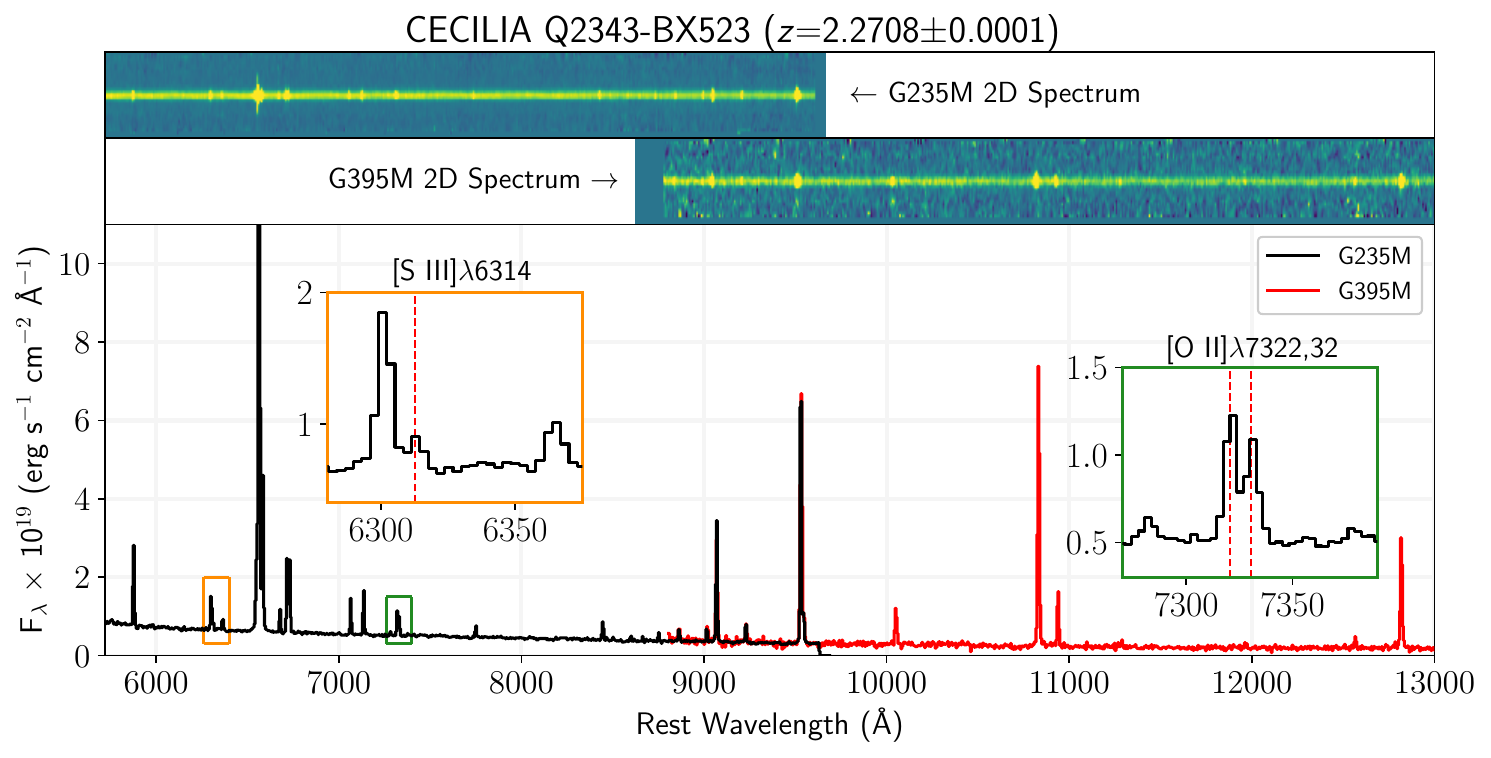}
   \caption{NIRSpec spectrum of BX523, a star-forming galaxy at \emph{z} $=$ 2.2708$\pm$0.0001 in the CECILIA sample. The top two panels plot the 2D spectra from each grating, where the global background and bar shadow removal procedures have resulted in a featureless background with no self subtraction in the strong emission lines. The optimally-extracted 1D spectra, shifted to the rest-frame of BX523, are plotted in the main panel. The G235M spectrum (black) has been resampled to match the G395M data (red) for plotting purposes only. We highlight the high-S/N detections of the \siii$\lambda$6314 and \oii$\lambda\lambda$7322,32 \te-sensitive auroral lines in the inset panels (plotted at the native sampling of G235M). The depth of the CECILIA observations reveal many other faint emission lines in both the 2D and 1D spectra.}
   \label{fig:bx523_spec}
\end{figure*}

The slit cutouts are saved in individual files to be combined and processed later. These \textsc{phot} files have bar shadow artifacts introduced by the bar structure between the each micro-shutter in MSA. Bar shadows can be corrected using \textsc{calwebb} or \textsc{msaexp}, although in practice we find that these solutions did not work well for the deep CECILIA data and left residual noise that complicated the measurement of a global background as well as extended source emission. Instead, we follow an iterative procedure in which we create a fiducial 1D background model similar to the approach described by \citet{strom2023}, but we also create a separate model for each detector and grating combination. We then compare the spatial variations in the background emission relative to this 1D background model among all the individual object spectra to construct a custom 2D bar shadow model. Next, we evaluate the bar shadow model on the spatial-wavelength grid for each object and divide the \textsc{phot} file by this model. Finally, we estimate the off-source background from each bar shadow-corrected \textsc{phot} file and apply an additive low-order polynomial correction to match the object-specific background to the fiducial global background solution. During this stage, we also identify and mask portions of individual exposures that are affected by open or closed shutters in the MSA. A comparison of this global background approach and the commonly-adopted local/nod background method is provided in Appendix \ref{app:gbkg}.

The depth of the CECILIA data have revealed a number of pixels with data quality issues that are unflagged in the bad pixel masks available in \textsc{crds\textunderscore context} 1251.pmap, and the default hot pixel algorithm tends to flag and remove intense emission lines when combining the science exposures. Instead, for each science observation we use all dither positions to create a median image, effectively removing the continuum and line emission from the target. This results in an image where bad pixels are easy to flag and remove from individual 2D spectra with a simple outlier detection algorithm. When applied to the CECILIA observations, this approach successfully removes unflagged or residual bad pixels without removing object emission. With the background, bar shadows, and bad pixels removed from the \textsc{phot} files, we combine the science observations and extract 1D spectra using the \textsc{slit\textunderscore combine} package in \textsc{msaexp}. This program determines the object trace and profile, performs a rectification of the 2D data, and extracts the 1D science spectrum in a manner similar to \citet[][]{horne1986}. The width of the modeled extraction profile increases with wavelength to account for the increasing size of the PSF.

Finally, given the potential deviations from a uniform pathloss solution and general uncertainties with the flux calibration of the NIRSpec MSA, we apply a flux correction to the 1D spectrum of each target. To do so, we take the ratio of observed continuum in NIRSpec and the SED continuum predicted from the BAGPIPES fitting for each galaxy, using a rolling median that excludes emission lines and residual data artifacts. This ratio is fit with a quadratic polynomial as a function of wavelength, then multiplied to the 1D NIRSpec data to correct the observed continuum to the model continuum. Similar approaches have been taken for other NIRSpec MSA programs, such as CEERS \citep{arrabal2023}, Blue Jay \citep{belli2024}, and EXCELS \citep{stanton2025}, and we apply this technique to all sources with significant continuum detections (for an analysis of the continuum-faint CECILIA galaxies, see \citet{raptis2025}.

Figure \ref{fig:bx523_spec} shows the final reduced 1D spectrum of BX523, a galaxy in the CECILIA sample with a spectroscopic redshift of 2.2708$\pm$0.0001. The deep G235M data reveal strong CEL emission from numerous ions and the \te-sensitive auroral lines \siii$\lambda$6314 and \oii$\lambda\lambda$7322,32 (also plotted in the inset panels). Despite the relatively shallow G395M observations, the strong NIR lines of \siii\ and \ion{He}{1}, as well as numerous Paschen recombination lines, are detected at high S/N. The combination of these data with ground-based rest-optical spectra from MOSFIRE enable the direct measurement of gas-phase physical conditions and chemical abundances of numerous elements.

\subsection{Line Measurements}\label{sec:linefits}

Emission lines in the NIRSpec data are modeled using Gaussian profiles and the SED continua discussed in \S\ref{sec:sed}. We fit the emission lines in spectral windows (up to 13 in G235M and 6 in G395M) within which we assume all emission lines have the same Gaussian FWHM and redshift. While the observed continuum generally matches the SED continuum due to the application of the spectrophotometric correction, there are instances where faint emission lines are over- or under-fit if the SED continuum is used instead of a local continuum. To capture the shape of the local continuum, we also fit a residual linear function to the difference between the observed and SED continuum and subtract this from the spectrum. The total model parameters within each window, therefore, are the local continuum slope and intercept, Gaussian FWHM, redshift, and the amplitude of each Gaussian. We subtract the SED continuum from the spectrum and use the \textsc{astropy} \citep{astropy2013,astropy2018,astropy2022} modeling package to perform a least squares fit and determine the model parameters.

This general approach is further simplified using the expected ratios of certain metal CELs. For example, the \nii$\lambda\lambda$6550,85 emission lines originate from the $^1$D$_2$ level, such that the \nii$\lambda$6585/\nii$\lambda$6550 is fixed at 2.94 from the atomic data of \citet{froesefischer2004} and \citet{tayal2011}. We fix the \oiii\ and \nii\ line flux ratios to the theoretical ratios, decreasing the number of free parameters. This matches the approach required for the MOSFIRE observations of the CECILIA galaxies, as the strong line detections in the ground-based spectra are generally lower S/N and require additional model constraints to provide robust fits to the rest-optical CELs. The \sii$\lambda\lambda$6718,33 and \oii$\lambda\lambda$3727,29 CELs are density sensitive, but the ratio of these emission lines vary within an expected range based on the atomic data. For NIRSpec and MOSFIRE we further constrain the model parameters by restricting the range of \sii$\lambda$6718/\sii$\lambda$6733 and \oii$\lambda\lambda$3729/\oii$\lambda\lambda$3727, respectively, to the ratios predicted for an electron gas at 1$<$\den$<$10$^5$ cm$^{-3}$.

We calculate the line flux uncertainty via the errors on the emission line amplitude and FWHM returned from the \textsc{astropy} modeling covariance matrix.
We consider individual emission lines detected if their S/N ratio is $\geq$ 3, while line doublets (e.g., \oii$\lambda$7322 and \oii$\lambda$7332) are detected if the sum of the emission lines has S/N $>$ 3.
This procedure is applied for the G235M and G395M spectra individually, but we include an additional uncertainty term when taking the ratio of emission lines measured in different gratings. As discussed in Appendix \ref{app:ebv}, we find that the average ratio of emission line fluxes simultaneously measured in the G235M and G395M spectra is consistent with unity but with a scatter of 9\%. Since this could represent a lingering flux calibration issue between the two gratings, we add an extra 9\% uncertainty in quadrature to all cross-grating line flux ratios.

The MOSFIRE data are fit using the same \textsc{astropy} method with a few notable exceptions. The MOSFIRE observations show an increased level of noise, especially at wavelengths that coincide with atmospheric OH emission lines. Therefore, we do not fit local deviations from the stellar continuum and take a different approach for the error estimation. Using the reliable error spectrum measured from the MOSFIRE data, we bootstrap resample the MOSFIRE spectra 500 times and refit the emission lines using the same \textsc{astropy} least squares approach. The standard deviation of the resulting flux distribution is taken as the uncertainty on the MOSFIRE line fluxes. We compare these line fluxes and uncertainties from the \textsc{astropy} modeling to the same quantities measured from direct integration of the isolated emission lines, and we find that the percent difference between the resulting line fluxes and uncertainties from the two methods are 1\% and 5\%, respectively. We also adopt a more constrained fit to the \oii$\lambda\lambda$3727,29 lines in $J$ band, as this portion of the NIR spectrum exhibits the most significant sky emission and the \oii\ lines can be unresolved depending on the spatial extent of the galaxy. To reliably fit the \oii\ doublet, we leave the redshift of the line doublet as a free parameter and use the FWHM measured from the \oiii\ lines in $K$ band to infer the FWHM of the \oii\ lines in $J$ band, accounting for the change in instrumental resolution between the two bands. We perform a visual inspection of all \oii$\lambda\lambda$3727,29 doublets, and in instances where both lines are resolved and uncontaminated by sky emission (fBM40, MD43, MD41, and C31) we leave the FWHM of the \oii\ lines as a free parameter.

\subsection{Combining MOSFIRE and NIRSpec Data}\label{sec:combine}

\begin{figure}[t]
   \centering
   \includegraphics[width=0.47\textwidth]{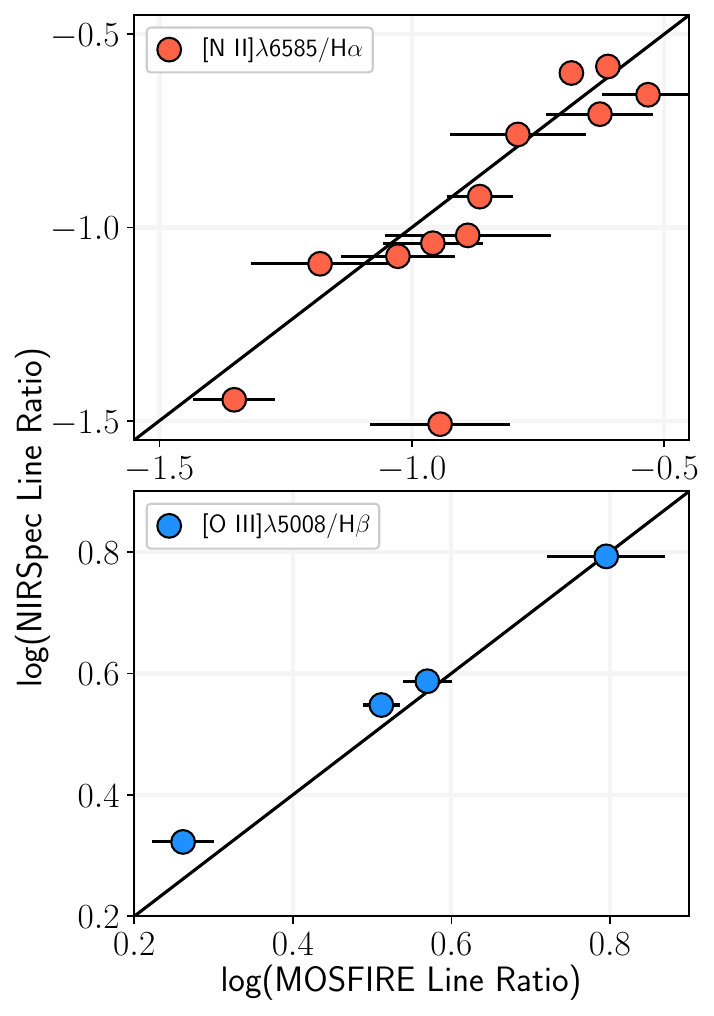}
   \caption{Comparison of NIRSpec and MOSFIRE line ratios: log(\nii$\lambda$6585/H$\alpha$) and log(\oiii$\lambda$5008/H$\beta$) in the top and bottom panels, respectively. There is good consistency between the independent measurements of the line ratios in the CECILIA sample, where only two MOSFIRE line ratios are inconsistent with the same ratio in NIRSpec at $>$2$\sigma$. The similar flux ratios observed in the two detectors allows for a combination of the ground-based strong lines (e.g., from \oii\ and \oiii) with the faint emission lines detected from JWST/NIRSpec.}
   \label{fig:nirmos_comp}
\end{figure}

The CECILIA survey was designed to use the deep JWST/NIRSpec observations of the \oii\ and \siii\ auroral lines in concert with optical strong line detections from Keck/MOSFIRE. Given the wavelength coverage of the G235M grating and the redshift of the targets, the nebular \oii$\lambda\lambda$3727,29 line doublet is not observed in any of the NIRSpec observations.
To determine \te\oii\ requires a combination of the \oii\ nebular and auroral line fluxes measured from MOSFIRE and NIRSpec, respectively.

To make this comparison, we normalize the \oii\ emission lines by the intensity of H$\alpha$ or H$\beta$ observed in MOSFIRE and NIRSpec. This approach assumes that all line ratios (e.g., \oii$\lambda\lambda$3727,29/H$\beta$) remain constant between the NIRSpec and MOSFIRE observations of the same galaxy. However, such an assumption may be invalid if the different observations trace distinct components of the ionized gas. The archival MOSFIRE observations of the CECILIA galaxies are acquired at a variety of PA, with slits of width 0\farcs7, and median extraction length 1\farcs8 \citep{strom2017}. The MOSFIRE slits are significantly larger than the MSA slits, which have width 0\farcs20 and length ranging from 3-8 shutters (corresponding to 1\farcs38-3\farcs68) for the CECILIA galaxies. The MSA slits were selected based on the assigned PA range at time of observations \citep[see discussion in][]{strom2023}, and small differences in the positioning of the MSA slit relative to the MOSFIRE observations could produce different line ratios in the NIRSpec and MOSFIRE spectra. 

To assess these effects, Figure \ref{fig:nirmos_comp} compares two line flux ratios measured independently in NIRSpec and MOSFIRE. We compare the log(\nii$\lambda$6585/H$\alpha$) ratios in the top panel, where the \nii\ lines represent some of the faintest CELs that can be robustly measured in the MOSFIRE spectra. Eleven galaxies have the requisite S/N in \nii$\lambda$6585 to make this comparison, and nine of these galaxies show MOSFIRE \nii$\lambda$6585/H$\alpha$ ratios that are consistent with those measured in NIRSpec to within 2$\sigma$, suggesting that the spectra from NIRSpec and MOSFIRE trace similar line-emitting regions. The significant outlier, BX216, shows very low S/N in \nii\ and a possible companion source in the MOSFIRE 2D spectra that is unseen in the NIRSpec data. This combination could alter the \nii/H$\alpha$ ratio, as the companion galaxy may not have the same ionization or chemical conditions as the main source. The bottom panel compares the log(\oiii$\lambda$5008/H$\beta$) ratios measured in NIRSpec and MOSFIRE; although the sample is limited to moderately high-\emph{z} galaxies where H$\beta$ is detected in NIRSpec, all galaxies show consistent \oiii/H$\beta$ ratios to within 2$\sigma$.

The stability of the CEL-to-\ion{H}{1} line ratios is reassuring, although we note that a similar comparison cannot be made for the \oii\ nebular lines. Normalizing the \oii\ lines to H$\beta$ or H$\alpha$ in MOSFIRE is sensitive to the relative flux calibration of different NIR observations and can be impacted by slit losses \citep[see discussion and approach adopted for KBSS in][]{strom2017}. While we acknowledge that slit losses and observational effects may alter the comparison of emission line ratios, the agreement observed in Figure \ref{fig:nirmos_comp} motivates us to use the relative emission line ratios in the NIRSpec and MOSFIRE data to carry out the following \te\ and abundance analysis.

\subsection{Reddening Corrections}\label{sec:ebvtrends}

We correct for dust attenuation using the H$\alpha$ and H$\beta$ recombination lines in the NIRSpec and MOSFIRE data. For the galaxies studied in this work, a measurement of the Balmer decrement at S/N $>$ 5, the recommended S/N for measuring $E(B-V)$ in \citet{strom2017}, is available in the NIRSpec and MOSFIRE data for 7 and 16 galaxies, respectively. We compare the observed flux ratio to the theoretical Case B H$\alpha$/H$\beta$ ratio at \te\ $=$ 1.25$\times$10$^4$ K and \den\ $=$ 300 cm$^{-3}$, which is 2.82 using the atomic data of \citet{storey1995}. While the Balmer lines are sensitive to underlying stellar absorption, this is generally a small correction for H$\alpha$ and H$\beta$ \citep[6\% to the H$\beta$ flux in KBSS,][]{strom2017}, and the absorption component is accounted for in the SED continuum. The required $E(B-V)$ to produce the observed Balmer decrement is computed assuming a \citet{reddy2020} attenuation curve, which is calibrated on the emission line trends of galaxies at Cosmic Noon from the MOSFIRE Deep Evolution Field survey \citep[MOSDEF,][]{kriek2015}. The functional form of this attenuation curve is similar to the Galactic attenuation curve of \citet{cardelli1989}, although recent JWST observations suggest that the shape of the attenuation curve may evolve at \emph{z}$>$2 \citep{markov2024,sanders2024,reddy2025}. When H$\gamma$ is available in the G235M data for galaxies at \emph{z}$>$2.82, we derive $E(B-V)$ as a weighted average of the reddening determined from the H$\alpha$/H$\beta$ and H$\gamma$/H$\beta$ ratios.

While the Balmer decrement is unavailable in NIRSpec for most CECILIA targets, we can use the significant Paschen line detections in the deep G235M data along with H$\alpha$ to measure $E(B-V)$. This is discussed further in Appendix \ref{app:ebv}, although we note that these calculations rely on the shape of the \citet{reddy2020} attenuation curve in an area where it is not well calibrated. As such, when reporting $E(B-V)$ for the CECILIA galaxies we prioritize $E(B-V)$ measured from the NIRSpec Balmer decrement, followed by the MOSFIRE Balmer decrement, and finally an average $E(B-V)$ from the available Paschen lines in the G235M or G395M data relative to H$\alpha$. For six galaxies with $E(B-V)$ measurements from the Balmer decrement and multiple Paschen lines, we find that $E(B-V)$ from the Balmer lines is, on average, larger by 0.09 mag. This is opposite to the average $E(B-V)$ trends measured in galaxies at \emph{z}$>$1.5 by \citet{reddy2025}, and we note that a lack of high-order Balmer lines that could better constrain $E(B-V)$ in the rest-optical may contribute to this offset in the CECILIA galaxies. For galaxies RK120 and BX523, we use $E(B-V)$ from the Paschen lines owing to uncertain MOSFIRE slit-loss corrections.

The adopted $E(B-V)$ in each galaxy is reported in Table \ref{t:abun}. Emission line ratios most sensitive to $E(B-V)$ include the auroral-to-nebular line ratios of \oii\ and \siii\ in addition to \oii$\lambda\lambda$3727,29/H$\beta$ required to measure O$^+$/H$^+$.
Systematic uncertainties in the attenuation curve for high-\emph{z} galaxies could significantly bias the physical conditions and chemical abundances reported in the CECILIA galaxies. Therefore, we adopt an approach that uses neighboring \ion{H}{1} lines for reddening-insensitive \te\ and abundance calculations, which is discussed in \S\ref{sec:tecalcs}.

\subsection{CECILIA Properties and Emission Line Trends}
\label{sec:bpt}

\begin{figure*}[!t]
   \centering
   \includegraphics[width=0.80\textwidth]{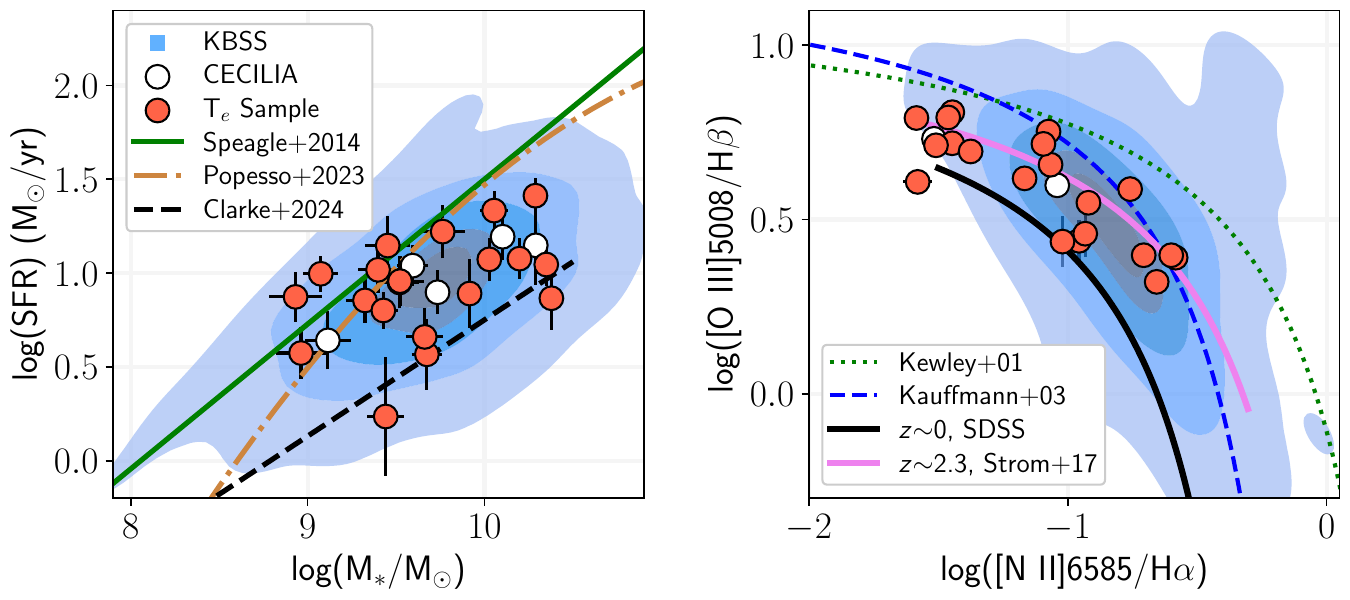}
   \caption{General properties of CECILIA and KBSS, the CECILIA galaxies' parent sample. \textit{Left:} The star-forming main sequence, log(SFR) vs.\ log(M$_*$). The SFMS observed in KBSS is represented in blue contours (at 5\%, 25\%, 50\%, 75\%, and 95\% densities),
   the CECILIA galaxies are denoted with white circles, where red circles represent the subsample of galaxies with at least one direct \te. We also provide fits to the SFMS from \citet[][at \emph{z}$=$2.5]{speagle2014}, \citet[][at \emph{z}$=$2.5]{popesso2023}, and \citet[][at 1.4$<$\emph{z}$\leq$2.7]{clarke2024} as solid green, dot-dashed brown, and dashed black lines, respectively.
   The \te\ sample presented here is both representative of the average properties of KBSS and typical galaxies at Cosmic Noon.
   \textit{Right:} The N2-BPT, log(\oiii/H$\beta$) vs.\ log(\nii/H$\alpha$). The ratios plotted on the horizontal axis are measured from NIRSpec. \oiii/H$\beta$ is measured from NIRSpec when available, otherwise they are measured from the archival MOSFIRE data. The green dotted and blue dashed lines represent the \citet{kewley2001} and \citet{kauffmann2003} demarcations for the line ratios predicted from an extreme starburst and AGN ionization, respectively. The line ratios measured in KBSS are represented as the blue contours, and the average trend derived by \citet{strom2017} is plotted in solid pink. The average N2-BPT line ratio trend for \emph{z}$\sim$0 SFGs in SDSS is plotted as a black line. The \te\ sample spans almost 1 dex in \nii/H$\alpha$ and follow the ionization trends predicted from the full KBSS sample.}
   \label{fig:bpt}
\end{figure*} 

In Figure \ref{fig:bpt}, we investigate whether the star-formation and ionization conditions of the CECILIA sample is representative of the parent sample, KBSS.
In both panels, we denote the CECILIA galaxies as white circles, where the red circles represent the galaxies with at least one direct \te\ measurement. We plot the star-forming main sequence (SFMS), log(SFR) vs.\ log(M$_*$/M$_\odot$), for KBSS (blue contours at the 5\%, 25\%, 50\%, 75\%, and 95\% densities)\footnote{Note that the full KBSS distribution displays an artificial linear feature corresponding to an apparent maximum sSFR of $10^{-8}\,\textrm{yr}^{-1}$. As noted in \S\ref{sec:sed}, this feature is due to the SED-fit SFRs being averaged over a $10^8\,$yr timescale that is longer than the $10^7\,$yr bursts allowed by our continuity SFH model.} and CECILIA. Here, the SFR and M$_*$ for all galaxies are determined through BAGPIPES SED fitting to the available KBSS photometric data using non-parametric SFHs \citep[see \S\ref{sec:sed} and][]{korhonen-cuestas2025}.
The CECILIA galaxies are consistent with the SFMS obtained from the full KBSS sample, where the \te\ sample spans log(SFR) $=$ [0.22,1.69] and log(M$_*$/M$_\odot$) $=$ [8.32,10.55], and is not significantly elevated in SFR relative to the parent sample.

We also provide fits to the SFMS at Cosmic Noon from \citet{speagle2014}, \citet{popesso2023}, and \citet{clarke2024}. We have selected the SFMS most appropriate for comparison to the CECILIA sample: for \citet{speagle2014} and \citet{popesso2023}, the plotted relations are determined at \emph{z}$=$2.5, while the \citet{clarke2024} relation is derived from SFGs at 1.4$<$\emph{z}$\leq$2.7 with SED-based SFRs. In general, the CECILIA galaxies with \te\ measurements scatter between the various measurements of the SFMS at Cosmic Noon. We note that using the H$\alpha$ luminosities measured from the G235M spectra produce SFRs that are, on average, 0.26 dex larger than measured from the SED fits. This reflects recent ($\lesssim$ 10 Myr) star formation within the CECILIA galaxies that the SED fitting is insensitive to on account of the line-corrected photometry.

Plotted in the right panel of Figure \ref{fig:bpt} is the N2-BPT diagram \citep[see][]{baldwin1981,veilleux1987}, which uses the emission line ratios of \oiii/H$\beta$ and \nii/H$\alpha$. The evolution of these line ratios in SFGs is sensitive to the metallicity of the ISM, the electron density, and the ionizing spectrum incident upon the gas \citep[e.g.,][]{kewley2013}. All CECILIA galaxies plotted in this diagram have \nii/H$\alpha$ measured from NIRSpec, but NIRSpec \oiii/H$\beta$ ratios are only available in seven galaxies. For the remainder, we use the line ratios measured from the MOSFIRE spectra.

The demarcation curves separating line ratios observed in local AGN \citep[][dotted green line]{kauffmann2003} and predicted for extreme starbursts \citep[][blue dashed line]{kewley2001} are plotted, as are the average line ratios measured in local SFGs from the Sloan Digital Sky Survey \citep[SDSS,][solid black]{york2000} and KBSS sample by \citet[][solid pink]{strom2017}. The CECILIA line ratios agree with the KBSS line ratios in the N2-BPT diagram, indicating that CECILIA is a representative subsample of KBSS in terms of average ionization and metallicity. Furthermore, the line ratios are consistent with the general emission line trends at high-\emph{z}, namely an offset to more intense CEL emission relative to \emph{z}$\sim$0 SFGs \citep[see][]{steidel2014,shapley2015,kashino2017}. We note that galaxies with low \oiii/H$\beta$ in KBSS were not prioritized for the CECILIA sample (see \S\ref{sec:sample}), as auroral line emission is prohibitively faint in metal rich galaxies with low-ionization conditions.

In summary, the stellar and ionizing properties of the CECILIA galaxies are representative of the KBSS sample and they are in good agreement with the general trends observed in galaxies at Cosmic Noon. Crucially, we observe that direct \te\ measurements are possible in a diverse population of galaxies, including those on the SFMS at low and high stellar masses.
The 20 galaxies with at least one \te\ measurement span the full range in log(M$_*$) and log(SFR) targeted in the CECILIA galaxies and have ionization conditions of log($O_{32}$) $=$ [$-$0.44, 0.87], and log($R_{23}$) $=$ [0.76, 1.14].
Most prior \te\ measurements at high-\emph{z} have been made in galaxies significantly above from the SFMS \citep{scholte2025}, which is a consequence of the ISM conditions necessary for auroral line excitation. As Figure \ref{fig:bpt} demonstrates, it is possible to significantly detect faint auroral line emission in a variety of SFGs with deep NIRSpec observations, allowing us to characterize the physical conditions and metal production in the typical SFG at Cosmic Noon.

\section{Physical Conditions and Chemical Abundances}\label{sec:physconditions}

\begin{deluxetable}{lccc}  
\tablecaption{Atomic Data for \te, \den, and Abundances \label{t:atomic}}
\tablewidth{\columnwidth}
\tabletypesize{\footnotesize}
\tablehead{ 
  \colhead{Ion}	&
  \colhead{Transition Probabilities}	&
  \colhead{Collision Strengths}
  }
\startdata
N$^+$ 	& \citet{froesefischer2004}  &  \citet{tayal2011}  \\ 
O$^+$ 	& \citet{froesefischer2004}  &  \citet{kisielius2009} \\
O$^{2+}$   & \citet{froesefischer2004}  &   \citet{storey2014} \\
S$^+$   & \citet{irimia2005}  & \citet{tayal2010} \\
S$^{2+}$   & \citet{froesefischer2006}  &   \citet{hudson2012}   \\
Ar$^{2+}$  &   \citet{mendoza1983}    &   \citet{munozburgos2009}  \\
\enddata
\end{deluxetable}

With the combination of the NIRSpec and MOSFIRE data, we can now determine the gas-phase physical conditions and chemical composition of the ISM in the CECILIA galaxies. We employ \textsc{PyNeb} v1.1.18 \citep{luridiana2015,morisset2020} for these calculations, and we summarize the assumed atomic data in Table \ref{t:atomic}. We consider the ISM to be composed of three zones, separated based on the IPs of the metals within the gas. A two-zone model \citep[e.g.,][]{esteban2009} is often invoked when measuring the ionic abundances of O, where the conditions of the \oii\ and \oiii\ emitting gas define the low- and high-ionization zones, respectively. These zones contain other metal ions with similar IPs: N$^+$, O$^+$, and S$^+$ reside in the low-ionization zone (10.1 eV $<$ IP $<$ 23.4 eV), while the high-ionization zone contains O$^{2+}$ (IP $=$ 35.1 eV). However, S$^{2+}$ and Ar$^{2+}$ have IPs that do not align with either O$^+$ or O$^{2+}$, which has motivated the use of a third, intermediate-ionization zone \citep[e.g.,][]{garnett1992} for ions with 23.4 eV $<$ IP $<$ 35.1 eV. Indeed, spectroscopic observations of local star-forming nebulae indicate that the \te\ trends in the intermediate-ionization zone are distinct from the \te\ in the simple two-zone model \citep{kennicutt2003,croxall2016,rogers2021,mendez-delgado2023}, requiring either a directly measured or inferred \te\ to accurately measure the abundances of S$^{2+}$ and Ar$^{2+}$. Given our numerous detections of the \siii\ auroral line, we can constrain the \te\siii\ and ionic abundances of S$^{2+}$ and Ar$^{2+}$ in this intermediate-ionization zone within the CECILIA galaxies.

\subsection{Electron Densities}\label{sec:densities}

There are two \den-sensitive doublets routinely detected in the CECILIA spectroscopic data. \den\sii\ is measured from \sii$\lambda\lambda$6718,33 in the NIRSpec data, and \den\oii\ is constrained using \oii$\lambda\lambda$3727,29 from the higher resolution MOSFIRE rest-optical spectra (R $\sim$ 3300 in $J$ band).
This allows us to compare the \den\ inferred from different low-ionization species, which would not be possible from the NIRSpec medium-resolution gratings (R $\sim$ 1000) as the \oii\ strong lines require R $\gtrsim$ 1350 to resolve.
This is true even for compact targets where the effective resolution of the medium gratings is $>$ 1000 \citep{degraaf2024b}, requiring the use of high-resolution gratings to reliably separate the \oii\ lines and constrain \den\oii\ \citep[e.g.,][]{li2025,welch2025}. Owing to the observational configuration of KBSS, the \oii\ strong lines are generally observed in $J$ band, which has many bright OH lines. Since \den\ measurements require an accurate fit to the shape of the emission line doublet, we use the FWHM of \oiii$\lambda$5008 to constrain the FWHM of the \oii\ lines (see \S\ref{sec:linefits}) and inspect all \oii\ emission lines in the MOSFIRE spectra to ensure that both lines are reliably fit and uncontaminated from sky line residuals. In instances where the doublet cannot be reliably fit (five galaxies total), we do not attempt a \den\oii\ measurement.

Other optical density diagnostics from scarce ions such as Cl$^{2+}$ and Ar$^{3+}$ are either not covered in the CECILIA observations or are too faint to detect even in the ultra-deep G235M observations. The density-sensitive [\ion{Cl}{3}] lines have yet to be detected in JWST spectroscopy of any high-\emph{z} SFG; while the [\ion{Ar}{4}] doublet has been reported in a handful of high-\emph{z} nebulae \citep{rogers2024,morishita2025,welch2025}, the neighboring \ion{He}{1} $\lambda$4715 emission line must be properly modeled to accurately measure \den[\ion{Ar}{4}]. The CECILIA observations do not cover the near-UV [\ion{C}{3}]$\lambda$1907,09 doublet, which has provided a tracer of the electron density in the high-ionization ISM of high-\emph{z} SFGs \citep[e.g.,][]{topping2024,topping2025}.
While this line doublet is generally too faint to detect in the Keck/LRIS observations of individual KBSS galaxies, a \den[\ion{C}{3}] of 1470$\pm$660 cm$^{-3}$ was reported in a stack of 30 KBSS galaxies \citep{steidel2016}.

For \den\ calculations, we use the \textsc{PyNeb} \textsc{getTemDen} function, which takes the \den-sensitive line intensity ratio and an assumed gas temperature and returns the density required to produce the observed line ratio. While the intensity of an individual CEL in the doublet is sensitive to both \te\ and \den, the line ratio mitigates much of the \te\ dependence because the CELs have similar excitation energies. To calculate \den, we assume a \te\ of 1.25$\times$10$^4$ K based on the initial temperature measurements in the CECILIA galaxies \citep{strom2023,rogers2024}. To estimate uncertainties on \den, we resample the line intensity ratio 1500 times from a normal distribution with standard deviation equal to the uncertainty on the line ratio, then recalculate \den\ for the distribution excluding any ratios outside the range permitted by the atomic data. We calculate the asymmetric errors on \den\ as the 16th and 84th percentiles of the distribution.

\begin{figure}[t]
   \centering
   \includegraphics[width=0.47\textwidth]{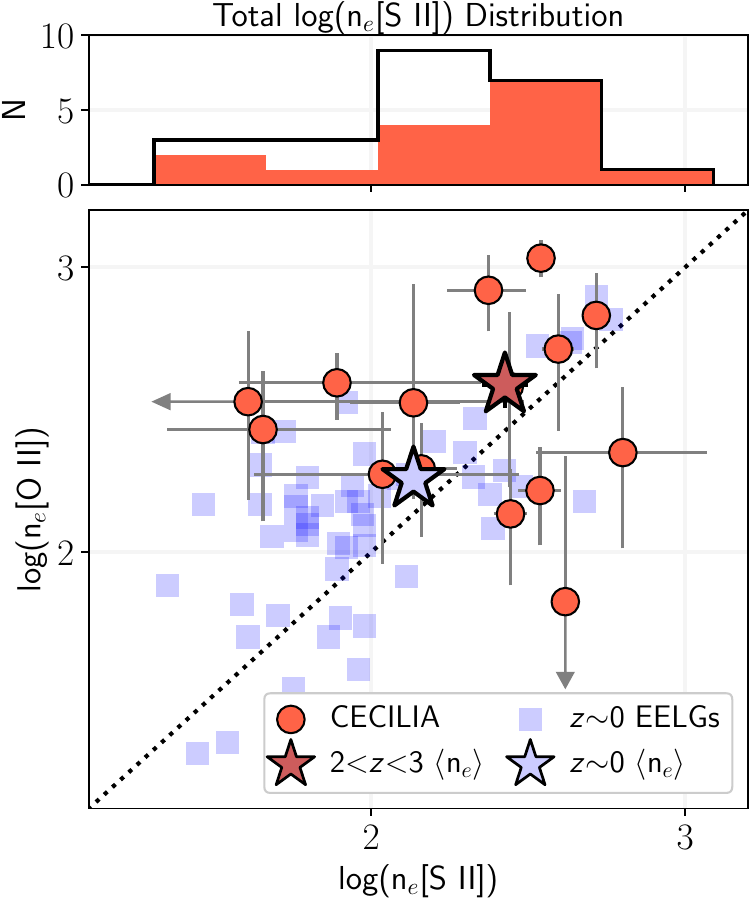}
   \caption{Density trends in the CECILIA galaxies (red circles) and local EELGs (light blue squares). Each point represents the simultaneous measurement of \den\oii\ and \den\sii, and the dotted line reflects equivalent densities. The dark red and light blue stars are the average \den\ measured in CECILIA and at \emph{z}$=$0, respectively. The top panel plots the histogram distribution of 23 \den\sii\ measurements in the CECILIA sample, where the red bins represent the galaxies with simultaneous \den\oii. In agreement with previous findings, the direct measurements of \den\sii\ and \den\oii\ from the CECILIA galaxies indicate that the ISM density at Cosmic Noon is elevated with respect to the \den\ measured in the local Universe. Furthermore, \den\sii\ and \den\oii\ show good agreement, as may be expected for two ions primarily originating in the same ionization zone. }
   \label{fig:do2ds2}
\end{figure}

The lower panel of Figure \ref{fig:do2ds2} plots the 15 simultaneous \den\oii\ and \den\sii\ measured from the MOSFIRE and NIRSpec observations, respectively, as red circles. The average densities measured in the CECILIA galaxies with both \den\sii\ and \den\oii\ (plotted as a dark red star) are: $\langle$\den\sii$\rangle$$=$267$\pm$44 cm$^{-3}$ and $\langle$\den\oii$\rangle$$=$386$\pm$70 cm$^{-3}$, where the uncertainties are estimated via bootstrap resampling the average statistic. The average \oii\ density is slightly elevated compared to the \sii\ density, although the individual \den\ measurements scatter around the one-to-one line (dotted line). This trend may be expected for two ions originating in the same volume of the ISM, although the IP of S$^+$ is such that this ion may also reside outside the \hii\ region in the photodissociation region (PDR) and diffuse ionized gas (DIG).

The typical electron density of the DIG is on the order of 0.01-0.1 cm$^{-3}$ \citep{reynolds1991,berkhuijsen2008}, which is much lower than the \den\ within the \hii\ regions. A substantial DIG contribution to the \sii\ intensity could bias \den\sii\ low relative to \den\oii, which is measured within the volume of the \hii\ region. However, \citet{oey2007} found a decreasing DIG fraction in local galaxies with high H$\alpha$ surface brightness, such that galaxies with high SFR surface densities are expected to have minimal DIG contributions to the integrated galaxy spectrum \citep[see also][]{zhang2017}. \citet{shapley2019} used the SFR surface densities in the MOSDEF sample to predict the DIG fraction as a function of \emph{z} \citep[their Equation 2, based on fit by][]{sanders2017}, finding that the DIG contribution is negligible for galaxies at \emph{z}$\sim$2.3. 

The slightly elevated \den\oii\ compared to \den\sii\ in the CECILIA galaxies, therefore, may be related to a marginal contribution of the DIG to the \sii\ line intensities, where the density measured from these CELs is more reflective of the \den\ in the \hii\ regions.
Alternatively, the DIG at \emph{z}$>$2 may exhibit elevated electron density relative to the DIG in local SFGs. If the DIG and \hii\ regions are at comparable densities, then \den\sii\ is unbiased by the inclusion of the DIG in the integrated spectrum. The agreement between \den\sii\ and \den\oii\ could also be related to the size and position of the MSA slits: if the small slits target the brightest \hii\ regions and do not incorporate a significant DIG component in the integrated spectrum, then \den\sii\ should better reflect the \hii\ region density, similar to \den\oii.
In summary, if low-density DIG is present in the integrated spectra of the CECILIA galaxies, then its contribution to the \sii\ line doublet only slightly decreases the average \den\sii\ relative to \den\oii, or the DIG is described by marginally higher \den\ that does not produce a substantial bias to \den\sii.

Overall, the density trends measured in the CECILIA galaxies are in good agreement with prior estimates of ISM \den\ at Cosmic Noon.
Large ground-based spectroscopic surveys of galaxies at Cosmic Noon have consistently found that the average \den\oii\ and \den\sii\ range from $\sim$150-300 cm$^{-3}$ \citep{sanders2016,strom2017,davies2021}
These findings have been corroborated with JWST observations \citep[e.g.,][]{isobe2023,li2025,stanton2025}. Recently, \citet{topping2025} reported a gradual increase in \den\sii\ with \emph{z} from 51 galaxies at \emph{z}$>$1.38 observed as part of the AURORA survey \citep{shapley2024}. The average \den\sii$=$268$_{-49}^{+45}$ cm$^{-3}$ measured from the 37 AURORA galaxies at \emph{z}$\sim$2 is in excellent agreement with the average reported in Figure \ref{fig:do2ds2}.

The density trends observed at Cosmic Noon are relatively uncommon in local SFGs, where \den\sii\ is typically $\lesssim$100 cm$^{-3}$ \citep{kaasinen2017}. As a comparison, we plot the \den\oii\ and \den\sii\ trends in a sample of 55 metal-poor, Extreme Emission Line Galaxies (EELGs) at \emph{z}$=$0 (Rogers et al., in preparation) as blue squares in Figure \ref{fig:do2ds2}. The average \den\oii\ and \den\sii\ in this sample are 180$\pm$21 cm$^{-3}$ and 136$\pm$18 cm$^{-3}$, respectively, indicating that these two densities are closely related \citep[see also][]{mendez-delgado2023}. The median \emph{z}$\sim$0 densities are in good agreement with the \den\oii\ and \den\sii\ observed in samples of other \emph{z}$\sim$0 EELGs \citep{senchyna2017,mingozzi2022}, some of which are selected to have high SFRs typical of those in SFGs at Cosmic Noon. However, the average densities measured in the \emph{z}$=$0 EELGs are approximately a factor of two less than the average ISM density observed in the CECILIA galaxies. The offset is observed when considering all 23 \den\sii\ measurements in the CECILIA sample (i.e., including eight galaxies without \den\oii), where the average is 212$\pm$34 cm$^{-3}$. This is highlighted in the top panel of Figure \ref{fig:do2ds2}, which plots the histogram distribution of log(\den\sii) for the full sample. A portion of the CECILIA galaxies are described by \den\sii$\leq$100 cm$^{-3}$ similar to the local EELGs, but the peak of the distribution is offset to higher \den\sii.

The higher \den\ at Cosmic Noon is potentially linked to the higher SFRs measured in these galaxies \citep[][]{reddy2023mosdefne}. SFR surface density is proportional to the gas surface density by the Kennicutt-Schmidt law \citep{kennicutt1998,kennicutt2012}, where giant molecular clouds with larger H$_2$ densities could fuel the larger SFR. The massive stars, in turn, ionize the dense molecular cloud and establish a higher initial electron gas density by \den$\approx$2$\times$n$_{H_2}$. Prior studies have found tentative evidence of a correlation between \den\ and SFR or SFR surface density, although there are other scenarios that could promote high \den\ in distant galaxies \citep[see discussion in][]{davies2021,topping2025}. In this respect, CECILIA may provide insight into the \den\ evolution of typical SFGs at Cosmic Noon (i.e., extending to low stellar mass and SFRs characteristic of the SFMS), although such an analysis is beyond the scope of this paper.

\subsubsection{Density Prioritization}\label{sec:denprior}

It is important to consider the higher ISM densities in the CECILIA galaxies when calculating \te\ and ionic abundances.
The emissivities of all relevant optical CELs are not strongly dependent on density when \den\ $\lesssim$100 cm$^{-3}$. Therefore, \te\ and ionic abundances can be measured at fixed \den\ if the ISM is in this low-density limit, which is often invoked in local SFGs based on the direct \den\sii\ trends in these nebulae. At the higher \den\sii\ and \den\oii\ at Cosmic Noon, the low-density limit no longer applies: the critical densities of the \oii\ $^2$D state are on the order of 10$^3$ cm$^{-3}$, such that the \oii$\lambda\lambda$3727,29 emission and the auroral-to-nebular line ratio is sensitive to \den.
Assuming a single \den\ value for these calculations can introduce systematic biases in the resulting \te\oii. For example, if the low-density limit is assumed for the average SFG at Cosmic Noon, where there is significant collisional de-excitation of the \oii$\lambda\lambda$3727,29 lines, then the auroral-to-nebular \oii\ line ratio and \te\oii\ will be systematically biased high, resulting in erroneously low O/H. Alternatively, some galaxies at \emph{z}$>$2 are consistent with the low-density limit (top panel of Figure \ref{fig:do2ds2}), such that adopting a larger characteristic density (e.g., 300 cm$^{-3}$) will significantly bias \te\oii\ low and O$^+$ abundances high.

To avoid these systematic uncertainties, we use the multiple \den\ measurements available to determine an appropriate \den\ for \te\ and abundance calculations. We adopt an average of \den\sii\ and \den\oii\ when both are directly measured, otherwise we simply use \den\sii\ when \den\oii\ is unavailable.
This is similar to the approach proposed by \citet{mendez-delgado2023}, although we favor the average of \den\sii\ and \den\oii\ even when \den\sii\ $<$ 100 cm$^{-3}$ because there is evidence of higher density from the \oii\ line ratios.
The measured and adopted \den\ in the \te\ sample are reported in Table \ref{t:abun}. Note that a direct \den\sii\ is available in all galaxies in the \te\ sample.

\subsection{Electron Temperatures}\label{sec:temperatures}

\subsubsection{Temperature Calculations in the CECILIA Galaxies}\label{sec:tecalcs}

With the observational setup of CECILIA, it is possible to measure \te\ from four ions: \oii, \nii, \siii, and \oiii. The temperature-sensitive emission line ratios \nii$\lambda$5756/\nii$\lambda$6585, \siii$\lambda$6314/\siii$\lambda$9071,9533, and \oiii$\lambda$4364/\oiii$\lambda$4960,5008 are measured purely from the NIRSpec data. Given the nominal wavelength coverage of the NIRSpec G235M grating, it is only possible to measure \oiii$\lambda$4364 in the two CECILIA galaxies at \emph{z}$>$2.8: C31 and D40. Unfortunately, the \oiii\ auroral line is not significantly detected in C31, such that D40 is the only galaxy in CECILIA with a measurement of \te\oiii\ \citep[see][]{rogers2024}. Combining the NIRSpec \oii$\lambda\lambda$7322,32 auroral line detections with the MOSFIRE \oii$\lambda\lambda$3727,29 nebular emission allows for a constraint on the \te\ structure of the low-ionization zone. Cutouts of the \siii\ and \oii\ auroral lines detected in the CECILIA galaxies are provided in Appendix \ref{app:auroral} (Figures \ref{fig:auroral_1} and \ref{fig:auroral_2}). We note that some galaxies with auroral line detections do not have direct \te\ owing to a lack of coverage for the corresponding nebular lines or H$\alpha$ in NIRSpec.

Unlike the \den-sensitive lines, the auroral and nebular CELs required to measure \te\ have large wavelength separations, necessitating accurate correction for dust attenuation and flux calibration between the different instruments used to acquire the spectra. As discussed in \citet{rogers2024}, the reddening-corrected \ion{H}{1} line intensity ratios in D40 deviate from those predicted by Case B recombination, indicating a persistent, wavelength-dependent flux calibration error in the G235M/G395M data or a difference in the shape of the attenuation curve. For reliable \te\ measurements, a more careful approach is required for measuring the auroral-to-nebular line ratios.

Consistent with the procedures in \citet{rogers2024} and similar to the method recommended by \citet{stasinska2023} when the shape of the attenuation curve is unknown, we analyze all metal CEL fluxes relative to neighboring \ion{H}{1} recombination lines. For CELs and \ion{H}{1} lines of similar wavelength, the CEL-to-\ion{H}{1} flux ratio is reddening insensitive and eliminates potential systematics owing to wavelength-dependent flux calibration errors. When we require the relative intensities of two metal CELs, we first divide the two CEL-to-\ion{H}{1} ratios, then correct the ratio using the emissivities of the two \ion{H}{1} lines assuming Case B recombination at \te$=$1.25$\times$10$^4$ K and the adopted \den\ of the galaxy. For example, the ratio of \siii$\lambda$6314/\siii$\lambda$9533 is measured by first calculating the flux ratios of \siii$\lambda$6314/H$\alpha$ and \siii$\lambda$9533/P8$\lambda$9549. Although these ratios are nearly reddening independent owing to the small wavelength separations between the lines, we apply a reddening correction based on the $E(B-V)$ outlined in \S\ref{sec:ebvtrends} using the \citet{reddy2020} attenuation law. Next, we determine (\siii$\lambda$6314/H$\alpha$)/(\siii$\lambda$9533/P8) and multiply this ratio by the theoretical H$\alpha$/P8 ratio, ending with \siii$\lambda$6314/\siii$\lambda$9533 that is pseudo-reddening corrected. The depth of the CECILIA NIRSpec observations enable this approach, as the faint Paschen lines in the NIR are required to normalize the flux of the \siii\ nebular lines.

When applying this method to measure the \siii\ auroral-to-nebular line ratio, the \siii$\lambda$6314 auroral line is always normalized to H$\alpha$ in the G235M data. We use the sum of the \siii\ nebular lines and prioritize P8$\lambda$9549 as the normalizing \ion{H}{1} line when the \siii\ nebular lines are measured in G235M. We use the flux of P8$\lambda$9549 because it has the smallest wavelength separation relative to \siii$\lambda$9533, and it is generally the most intense, resolved Paschen line in the G235M data. When measuring \siii\ lines in G395M, we prioritize P7$\lambda$10053 for \ion{H}{1} normalization because \siii$\lambda$9533 and P8 are partially blended in the G395M data. When \siii$\lambda$9533 is undetected, we use the ratio of \siii$\lambda$9071/P9$\lambda$9232 to infer the \siii\ auroral-to-nebular ratio. Since \siii$\lambda$6314 is only detected in the G235M observations, we also consider an additional 9\% uncertainty on the resulting auroral-to-nebular line ratio when using \siii\ strong lines in G395M to account for potential flux calibration errors between configurations/gratings (see Appendix \ref{app:ebv}). The corresponding normalization method is straightforward for the \oiii\ lines, as \oiii$\lambda$4364 and \oiii$\lambda$4960,5008 neighbor the \ion{H}{1} lines of H$\gamma$ and H$\beta$, respectively. When measuring \te\nii, we use the reddening-corrected \nii$\lambda$5756/\nii$\lambda$6585 ratio directly, as H$\alpha$ would be used for the normalization of each \nii\ line.

This normalization approach is challenging to apply when measuring \te\oii, as no high-order Balmer lines are significantly detected in the ground-based MOSFIRE observations of the CECILIA galaxies. Therefore, the \oii$\lambda\lambda$3727,29 emission lines must be normalized to H$\beta$, H$\alpha$, or \oiii$\lambda$5008.
To measure \te\oii, we determine the reddening-corrected \oii$\lambda\lambda$3727,29/H$\beta$ or \oii$\lambda\lambda$3727,29/H$\alpha$ intensity ratio in MOSFIRE (depending on which \ion{H}{1} line is available). We then measure \oii$\lambda\lambda$7322,32/H$\alpha$ in NIRSpec, and finally account for the theoretical \ion{H}{1} line ratio to obtain the \oii\ auroral-to-nebular line ratio; note that if \oii$\lambda\lambda$3727,29/H$\alpha$ is used, then no correction to theory is required.
There are two galaxies, C31 and BX350, where H$\beta$ and H$\alpha$ are unavailable in the MOSFIRE data. Fortunately, \oiii$\lambda$5008 is measured in both MOSFIRE and NIRSpec in these galaxies, so we determine the \oii\ auroral-to-nebular line ratio by way of the reddening-corrected \oii$\lambda\lambda$3727,29/\oiii$\lambda$5008 in MOSFIRE and \oii$\lambda\lambda$7322,32/\oiii$\lambda$5008 in NIRSpec.
We note here that the \oii\ temperatures are most sensitive to reddening uncertainties and cross-band slit loss corrections, systematic uncertainties present regardless of normalization to a \ion{H}{1} line or the \oiii\ CELs.

For \te\ calculations, we use the \textsc{PyNeb} \textsc{getTemDen} function with the auroral-to-nebular line ratio and the adopted density from \S\ref{sec:denprior}. \te\ uncertainties are calculated using a distribution of 500 auroral-to-nebular line intensity ratios sampled from a normal distribution, where we recalculate \te\ for each line ratio and take the standard deviation of the ensemble as the uncertainty on \te. The \textsc{PyNeb} \textsc{getCrossTemDen} function can be used to simultaneously solve for \den\ and \te\ using line ratios from ions in the same ionization zone. For galaxies where \den\oii\ and \te\oii\ can be simultaneously constrained, the \oii\ lines produce the same \te\ and \den\ determined from the above procedures except for two galaxies. The galaxies BX461 and BX336 have \den\oii\ less than the adopted \den\ from \S\ref{sec:denprior}, which produces the difference between the \te\oii\ measured from the adopted and \textsc{getCrossTemDen} approaches. We choose to maintain a consistent approach and adopt the average density discussed in \S\ref{sec:denprior} for \te\ calculations, which leverages the deeper, higher S/N NIRSpec observations to constrain the ISM density when \den\oii\ is unavailable.

\subsubsection{Simultaneous Temperature Trends}\label{sec:tetrends}

Before the launch of JWST, direct \te\ measurements in high-\emph{z} SFGs were limited in number and variety. Despite significant investment in ground-based spectroscopy, \oiii$\lambda$4364 was only observed in a handful of unlensed galaxies at \emph{z}$>$1.6 \citep[][]{sanders2020,clarke2023}, and most other auroral lines from low-ionization species were undetected \citep[c.f.,][]{sanders2023o2}. The UV \ion{O}{3}]$\lambda$1660,66 auroral lines were detected in individual lensed galaxies \citep[e.g.,][]{stark2013,james2014,berg2018,citro2024}, while stacking techniques probed faint auroral line emission in the UV and optical \citep{steidel2016,trainor2016}.
The high-quality NIR spectroscopy from JWST has greatly improved our ability to constrain the gas temperature in distant galaxies, although the auroral lines of \oiii\ remain the most commonly detected in the era of JWST. Even so, the number of high-\emph{z} galaxies with multiple \te\ from different ions \citep{rogers2024,welch2024,welch2025,chakraborty2025,stanton2025,morishita2025,sanders2025,cataldi2025} is still small relative to the growing archive of \te\oiii\ measurements.

\begin{figure}[t]
   \centering
   \includegraphics[width=0.47\textwidth]{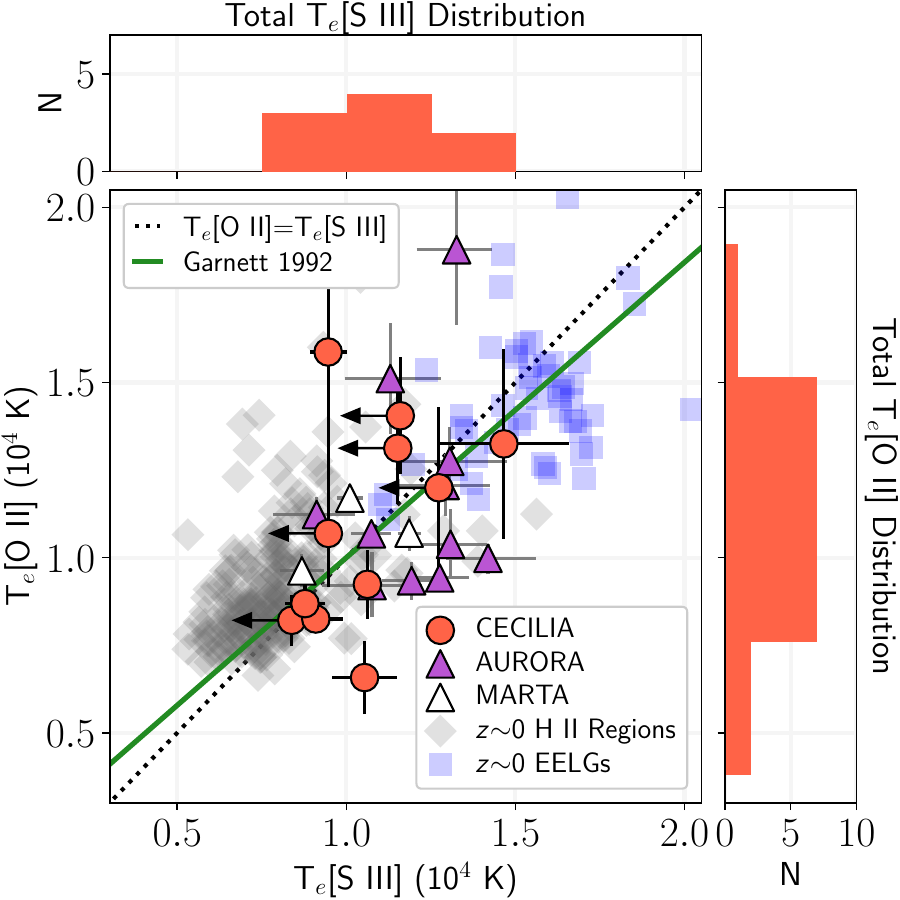}
   \caption{Simultaneous \te\oii\ and \te\siii\ measured in the CECILIA sample (red circles, 3$\sigma$ \te\siii\ upper limits provided with black arrows), local EELGs (blue squares), and the metal-rich CHAOS \hii\ regions (gray diamonds). We also include direct \te\ from the AURORA \citep{sanders2025} and MARTA surveys \citep{cataldi2025,curti2025} as purple and white triangles, respectively. The theoretical \te\ scaling relation from \citet{garnett1992} is plotted in green, while the black dotted line indicates equivalent \te\oii\ and \te\siii. The histogram distribution of the 17 \te\oii\ and 9 \te\siii\ measurements in CECILIA are provided in the right and top panels, respectively. The high-\emph{z} \te\ are generally consistent with those measured in the local Universe and from photoionization model predictions, but there is currently insufficient data to assess the scatter in \te\oii-\te\siii\ at high-\emph{z}.}
   \label{fig:to2ts3}
\end{figure}

We present the simultaneous \te\oii\ and \te\siii\ trends measured in the CECILIA galaxies as red circles in Figure \ref{fig:to2ts3}. Owing to the depth and wavelength coverage of the CECILIA observations, this sample increases both the number of \te\ measurements from ions other than O$^{2+}$ and the number of galaxies with simultaneous \te\ from multiple ions. In total, there are 17 and 9 galaxies in CECILIA with \te\oii\ and \te\siii\ measurements, respectively, where 6 galaxies have simultaneous \te\oii\ and \te\siii. We also provide upper limits on \te\siii\ as black arrows when direct \te\oii\ are available. The total \te\oii\ and \te\siii\ distributions are included in the top and right panels of Figure \ref{fig:to2ts3}. This represents one of the largest homogeneous samples of direct \te\oii\ and \te\siii\ in SFGs at \emph{z}$>$2. We note that \te\siii\ can be measured using the nebular \siii\ doublet in both G235M or G395M for galaxies at 2.15$<$\emph{z}$<$2.3; we find that \te\siii\ is unchanged depending on the choice of \siii\ in G235M or G395M, so we favor \te\siii\ measured from the G235M strong lines, if available, owing to the higher S/N.

In addition to the \te\ trends of the CECILIA galaxies, we plot the direct \te\oii\ and \te\siii\ measurements from the same \emph{z}$\sim$0 EELG sample from Figure \ref{fig:do2ds2}. We also include the \te\ measurements in metal-rich \hii\ regions observed as part of the CHemical Abundances of Spirals \citep[CHAOS,][]{berg2015} project. We include the \hii\ regions presented in \citet{berg2020}, \citet{rogers2021}, and \citet{rogers2022}, where all \te\ and abundances of the CHAOS regions have been recomputed using the methods described in \citet{rogers2022}. We also plot the \citet{garnett1992} photoionization model \te\ scaling relation as a solid green line.
Finally, we include other high-\emph{z} SFGs with simultaneous \te\siii\ and \te\oii\ available in the literature, including three SFGs from the MARTA survey reported in \citet[][white triangles]{cataldi2025} and another eleven galaxies in the AURORA survey from \citet[][purple triangles]{sanders2025}. The literature galaxies span a redshift of 1.91 $\leq$ \emph{z} $\leq$ 4.41, complementary to the CECILIA galaxies.

The CECILIA and literature SFGs in Figure \ref{fig:to2ts3} span over 10000 K in \te\oii\ and provide insight into the \te\ structure of high-\emph{z} galaxies.
Figure \ref{fig:to2ts3} reveals
that the \te\oii-\te\siii\ trends at Cosmic Noon are comparable to the \te\ observed in local star-forming nebulae. Indeed, five of the six CECILIA galaxies with direct \te\oii\ and \te\siii\ are within 2$\sigma$ of the trends predicted by the photoionization model \te\ scaling relation. Similar agreement to local scaling relations was found when comparing the \te\oiii-\te\siii\ trends in D40\footnote{\citet{rogers2024} discuss D40's \te\ and abundance trends in detail, although our new data reduction indicates \te\oiii\ is 14300$\pm$700 K owing to a slightly higher \oiii$\lambda$4363 flux and $E(B-V)$. This \te\ is consistent with the original \te\oiii\ measurement, \te\oiii\ $=$ 13300$\pm$500 K, and the simultaneous \te\oiii\ and \te\siii\ are in agreement with the \citet{garnett1992} scaling relation.}, the only CECILIA galaxy with a direct \te\oiii\ measurement \citep{rogers2024}. Such agreement may suggest that the gas temperatures in the different ionization zones are well described by current photoionization modeling, despite different chemical abundance patterns and ionizing sources in the distant galaxies. As \te\ is primarily set by the available metals in the ISM, the presence of the abundant O \citep[roughly 37 times more abundant than S in the Sun,][]{asplund2021} may play a more principal role in shaping the temperature structure in the ISM. 

The temperature trends observed in the CECILIA galaxies, local \hii\ regions and EELGs, and those predicted by the photoionization model \te\ scaling relation suggest that \te\oii\ and \te\siii\ are comparable from $\sim$5000 K up to 15000 K. This behavior is related to the presence of O$^+$ in both the low- and intermediate-ionization zones. Owing to the high excitation energies of the optical CELs, auroral lines are preferentially emitted in the high-\te\ ISM. This results in a known bias, where the optical auroral lines indicate a higher \te\ than the volume average temperature \citep[e.g.,][]{peimbert1967}. Since the volume of O$^+$ gas also contains higher ionization ions like S$^{2+}$, the measurement of \te\oii\ may be biased to the higher \te, higher-ionization gas, resulting in \te\oii-\te\siii\ that scatter around the one-to-one line.
The \te\ stratification in the ISM can be significant when comparing direct \te\ from ions that do not overlap in IP \citep[e.g., see relations in][]{rogers2021,mendez-delgado2023}.

When discussing these simultaneous temperatures, it is important to note the dispersion in the direct \te.
In local nebulae, \te\ scaling relations involving \te\oii\ generally display large dispersion relative to other common \te\ scaling relations \citep[e.g.,][]{kennicutt2003,berg2020},
which is evident in the local nebulae and high-\emph{z} SFGs plotted in Figure \ref{fig:to2ts3}. Large differences in \te\ are even observed when comparing direct \te\oii\ to other low-ionization zone \te, such as \te\nii\ and \te\sii\ \citep{rogers2021}. Such discrepancies are also observed in the two CECILIA galaxies with direct \te\nii\ (BX523 and BX474): in both cases, \te\nii\ is determined to be in excess of 1.3$\times$10$^4$ K \citep[see also][]{strom2023} while \te\oii$<$10$^4$ K and \te\siii$\approx$1.1$\times$10$^4$ K. These two galaxies do not follow the simple assumption that \te\oii$\approx$\te\nii\ or the \te\nii-\te\siii\ scaling relations available in the literature \citep[e.g.,][]{rogers2021},
either suggesting a more complex \te\ structure in high-\emph{z} SFGs or that the large variation in \te\oii\ observed in local SFGs is also present at Cosmic Noon. At this time, there are too few high-\emph{z} SFGs with direct \te\nii\ to quantitatively evaluate the variations in different low-ionization zone temperatures. 

Numerous studies have attempted to assess the physical mechanisms influencing the scatter in \te\oii, but no physical condition is clearly correlated with the offset in \te\oii\ from the best-fit scaling relation \citep{yates2020,cataldi2025}.
This dispersion, therefore, represents a large systematic uncertainty when \te\oii\ is unconstrained or required to infer a missing \te, as the functional form of the \te\ scaling relation cannot capture the empirical scatter in \te\oii\ at fixed \te\siii.
Currently, there is insufficient data to derive new empirical \te\ scaling relations or assess the degree of scatter in \te\oii\ at Cosmic Noon.
Constraining the shape of these \te\ scaling relations remains a necessary component for reliable chemical abundance surveys at high-\emph{z} with future JWST observations.

\subsection{Direct Ionic and Total Abundances}\label{sec:abuncalcs}

Gas-phase ionic abundances are calculated using the intensity of a metal CEL and emissivity of the given transition, the latter of which is a function of \te\ and \den. The \textsc{PyNeb} \textsc{getIonAbundance} function performs ionic abundance calculations relative to H$^+$, taking the line intensity ratio normalized to H$\beta$ as input. Similar to the method adopted for \te, we normalize CELs relative to a neighboring \ion{H}{1} line, then correct this ratio using the emissivities of the \ion{H}{1} line and H$\beta$ assuming Case B recombination.
The line flux ratio and \te\ uncertainties are propagated through to estimate the uncertainty on the ionic abundance. This assumes the uncertainty on \den\ has a negligible impact on the ionic abundance measurements, which is valid at high critical densities. Even in instances where uncertainty in the density may affect ionic abundance calculations (e.g., O$^{+}$ and S$^+$), the uncertainty on \te\ has a larger fractional change to the emissivities than the uncertainty on \den, such that the \te\ error drives the uncertainty in ionic abundance calculations.

In general, it is possible to measure the ionic abundance of N$^+$, S$^+$, S$^{2+}$, and Ar$^{2+}$ using only the NIRSpec observations of the CECILIA galaxies, where measurements of O$^{2+}$ are possible when \emph{z}$>$2.32. We measure the abundance of O$^+$ and O$^{2+}$ (the latter at \emph{z}$<$2.32) using the MOSFIRE \oii$\lambda\lambda$3727,29 and \oiii$\lambda$5008 emission line fluxes. A measurement of O$^+$ is possible using the auroral \oii\ lines, although these lines are relatively faint, are sensitive to variations in \te, and can be produced by dielectronic recombination \citep{rubin1986,liu2000}. This combination leads to potentially large systematic biases in the O$^+$ abundances \citep{rodriguez2020,mendez-delgado2023}, so we prioritize the MOSFIRE measurements of the nebular \oii\ line intensities relative to H$\beta$. The exception is when calculating the relative N$^+$/O$^+$ with \te\oii, as the \nii$\lambda$6585/\oii$\lambda\lambda$7322,32 ratio is measured entirely in NIRSpec, is insensitive to $E(B-V)$, and is typically measured at higher S/N than \nii$\lambda$6585/\oii$\lambda\lambda$3727,29.

Ionic abundance calculations are performed using the prioritized density discussed in \S\ref{sec:denprior} and the relevant ionization zone \te. If both \te\nii\ and \te\oii\ are available, we follow the recommendations of previous chemical abundance surveys and adopt \te\nii\ as the low-ionization zone \te\ \citep{berg2020,rogers2021,mendez-delgado2023}. When a direct \te\ is missing, we use a \te\ scaling relation to infer the missing ionization zone temperature. For the low- and intermediate- ionization zones, we adopt the empirical scaling relations reported in \citet{rogers2021}, accounting for the intrinsic scatter in \te\ about the relation when reporting the inferred temperature uncertainty.
For the high-ionization zone, we prioritize the \citet{garnett1992} scaling relation with \te\siii\ when available. This choice is motivated by the agreement between \te\oiii-\te\siii\ patterns in local EELGs and the photoionization model relation, with a small intrinsic scatter of $\sim$400 K (Rogers et al., in preparation). When \te\siii\ is unavailable, we apply the \te\oiii-\te\nii\ relation from \citet{rogers2021} assuming that \te\oii\ $\approx$ \te\nii\ to infer the high-ionization zone \te.

Total O abundances are determined assuming O/H $\approx$ O$^+$/H$^+$ + O$^{2+}$/H$^+$, ignoring the potential contribution of O$^{3+}$ to the total O abundance. In the few instances where the gas-phase O$^{3+}$/H$^+$ has been constrained in local EELGs, the fractional contribution to the global O/H abundance has been found to be $\lesssim$ 2\% \citep{berg2021,rickardsvaught2025}. While \oi\ emission is detected in numerous CECILIA galaxies (see Figures \ref{fig:auroral_1} and \ref{fig:auroral_2}), we assume this emission comes from outside the \hii\ regions (either in the PDR or DIG). Finally, we do not account for dust grain depletion of O. This correction can be as large as 0.1 dex \citep{peimbert2010,pena2012} and is most significant in metal-rich nebulae. This correction may be an important consideration for the most metal-rich CECILIA galaxies, but is likely too large for the metal-poor, highly-ionized ISM observed in the majority of the targets. Therefore, we choose to adopt a consistent approach for O/H calculation and forgo correction for dust depletion.

O is the only element where the most abundant ionization states are directly observable in the rest-optical/NIR. For N, S, and Ar abundances, we require Ionization Correction Factors (ICFs) to correct for the abundance of unobserved ionization states. Empirically-derived ICFs, while available for some elements \citep[e.g.,][]{dors2013,esteban2015}, are relatively uncommon owing to the observational challenges associated with detecting all relevant metal ionization states. A simple approach to ICFs is to use ions of similar IP, where it is assumed each ion comprises a similar ionic fraction in the ISM. A notable example that is adopted in the present work is that of N$^+$ and O$^+$, where the relative N/O abundance is often assumed to be N/O $\approx$ N$^+$/O$^+$ and ICF(N) $=$ O/O$^+$ \citep{peimbert1969}. This ICF has been widely adopted for N abundance calculations from the optical \nii\ lines in local \citep{vanZee2006,croxall2015,berg2020} and high-\emph{z} SFGs \citep{arellano-cordova2025,morishita2025,welch2025} and is generally valid to within 10\% of the true N/O abundance \citep{nava2006,amayo2021}.

For total S/H and Ar/H, we require ICFs derived with photoionization model predictions. As discussed in \citet{rogers2024} and in many other chemical abundance studies \citep[see recent works by][]{arellano-cordova2024a,esteban2025}, the choice of ICF represents a large source of systematic uncertainty when deriving total abundances. This problem is exacerbated at high-\emph{z}, where typical SFGs are characterized by harder ionizing spectra and distinct chemical abundance patterns \citep{steidel2014,shapley2015,shapley2024,strom2017,strom2018,cullen2019}. The harder ionizing spectra will produce emission line ratios that deviate from those observed in local SFGs; as such, ICFs calibrated on photoionization models selected to reproduce the \emph{z} $=$ 0 BPT locus \citep[e.g.,][]{amayo2021} are inappropriate to apply at high-\emph{z}. The derivation of new ICFs from a grid photoionization models tailored for the chemical and ionization trends of high-\emph{z} SFGs is necessary, but beyond the scope of the present work. Instead, we adopt the S and Ar ICFs of \citet{izotov2006}, which have been utilized for S and Ar abundances in other high-\emph{z} SFGs \citep[e.g.,][]{stanton2025}.

ICF(S) accounts for missing S$^{3+}$ in the highly-ionized ISM, and we choose a ICF(Ar) that corrects for unobserved Ar$^+$ and Ar$^{3+}$. This is required because the typical CECILIA galaxies do not have coverage of the optical \ariv\ emission lines \citep[for tentative detection in D40, see][]{rogers2024}. The \citet{izotov2006} ICFs are polynomial functions of O$^+$/O, with different calibrations for metal-poor and metal-rich nebulae. None of the CECILIA galaxies are described by 12+log(O/H) $<$ 7.2, and we adopt the high-Z calibrations for galaxies with 12+log(O/H) $\geq$ 8.2. For galaxies with 7.6 $\leq$ 12+log(O/H) $<$ 8.2, we follow the recommendations of \citet{izotov2006} and linearly interpolate between the intermediate- and high-metallicity ICFs.
Similar to the methods of \citet{rogers2024}, we assume a 10\% uncertainty on the ICF to partially account for the systematic uncertainty related to the choice of ICF.
We report the ionic and total abundances for the CECILIA galaxies in Table \ref{t:abun}.

\section{Chemical Abundance Trends at Cosmic Noon}\label{sec:abun}

\subsection{Total O Abundance}\label{sec:oxygen}

\begin{figure*}[!t]
   \centering
   \includegraphics[width=0.85\textwidth]{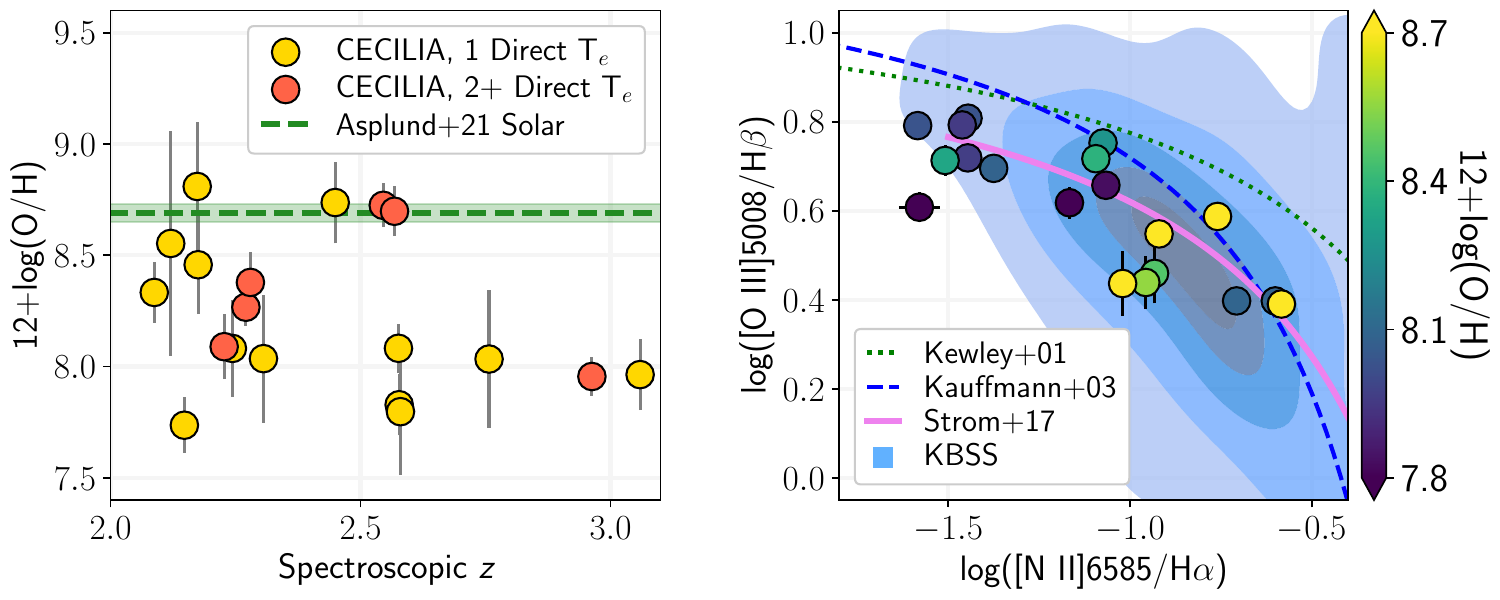}
   \caption{\textit{Left:} Direct gas-phase O/H vs.\ \emph{z} measured in 19 CECILIA galaxies. The color of the circles represents the number of direct \te\ measured in each galaxy: red circles have two or more direct \te, yellow circles have a single \te. The solar O/H ratio is plotted as a dashed green line and shaded area.
   While the highest \emph{z} galaxies tend to show low O/H, several CECILIA galaxies show gas-phase O/H consistent with the solar ratio at \emph{z}$\gtrsim$2.
   \textit{Right:} The N2-BPT is reproduced with the \te\ sample, color-coded by 12+log(O/H). The CECILIA sample follows the expected trends in this relation: galaxies with low \nii/H$\alpha$ and high \oiii/H$\beta$ exhibit relatively low O/H, and O/H increases with \nii/H$\alpha$ as \oiii/H$\beta$ decreases.}
   \label{fig:oh_n2bpt}
\end{figure*} 

We now discuss the direct abundance trends in the CECILIA galaxies, providing a look into the chemical makeup of SFGs during the epoch of Cosmic Noon. We start with the gas-phase O/H, which traces the enrichment from massive stars and is sensitive to feedback mechanisms such as pristine gas inflows and metal-rich outflows.
In the left panel of Figure \ref{fig:oh_n2bpt}, we show the redshift evolution of O/H in the CECILIA galaxies. In this figure, the color of the point represents the number of direct \te\ measured in the galaxy: galaxies with multiple direct \te\ are provided as red circles, galaxies with a single \te\ are plotted as yellow circles. The \te\ sample ranges from \emph{z}$=$2.08 to 3.06, although an evolution in 12+log(O/H) with \emph{z} is not clear from the sparse sampling.
We observe that the three CECILIA galaxies with direct \te\ at \emph{z}$>$2.7 have 12+log(O/H)$\sim$8.0 dex (or 20\% solar abundance), and there are galaxies at \emph{z}$\sim$2.1 with similarly low O/H.

Figure \ref{fig:oh_n2bpt} also reveals the presence of relatively metal-rich galaxies in the CECILIA sample. The left panel plots the \citet{asplund2021} solar O/H ratio and its uncertainty as a green dashed line and shaded area, respectively. Of the 19 galaxies plotted, 6 are consistent with the solar O/H\footnote{This excludes RK120, a galaxy with direct 12+log(O/H) $>$ 9.0 and other peculiar abundance patterns. The MOSFIRE cross-band slit-loss corrections for RK120 are poorly constrained, which affects the inferred reddening and strong-line ratios that get propagated into the \te\ and abundances. We do not discuss RK120 in the remainder of our analysis.}. Such high metallicities are uncommon in high-\emph{z} galaxies. For example, the highest gas-phase metallicity measured in the EXCELS SFGs is 12+log(O/H)$=$8.38$\pm$0.30 at \emph{z}$=$4.234 \citep{stanton2025}, which is consistent with the upper metallicity measured in the CEERS \citep{finkelstein2025} auroral line sample in \citet{sanders2024sl}.
Despite the growing sample of high-\emph{z} galaxies with direct \te\ and abundances, including 16 and 41 SFGs at \emph{z}$>$1.3 reported in \citet{cataldi2025} and \citet{sanders2025}, respectively, there are only three galaxies in the literature with solar O/H abundance in the ISM. The number is more than doubled with CECILIA, and the three CECILIA galaxies with 12+log(O/H)$>$8.70 dex constitute the largest direct-method metallicities measured with JWST to date.

In general, it is challenging to apply the direct abundance method at solar O/H, as metal CEL emission is the dominant cooling mechanism in the ISM. As metallicity increases, collisional excitation from free electrons and spontaneous emission from the abundant metals acts to remove energy from the gas and significantly lower the overall temperature of the ISM.
At these low \te, emission from metal fine-structure lines in the infrared becomes more significant as the intensity of the optical CELs decreases \citep{dopita2003}.
The decrease in intensity is especially notable for the optical \te-sensitive auroral lines, which originate from energy levels with high excitation energies ranging from 3.36 eV for \siii$\lambda$6314 to 5.35 eV for \oiii$\lambda$4364. Even from large samples of metal-rich nebulae in the local Universe, there are relatively few \hii\ regions with direct-method 12+log(O/H)$\geq$8.69 dex \citep[see compilation and discussion in][]{mendez-delgado2024}. Given these trends, it is notable that roughly a quarter of the CECILIA galaxies with direct O/H, including two with multiple \te\ measurements, show O/H consistent with the solar abundance ratio.

To assess if the direct O/H measurements are consistent with the physical conditions and emission line properties of the CECILIA galaxies, we first consider the \te\ distributions plotted in Figure \ref{fig:to2ts3}. The median \te\oii\ measured in the CECILIA galaxies is 10700 K, similar to the \te\oii\ measured in the metal-rich CHAOS \hii\ regions and less than the \te\oii\ measured in the metal-poor EELGs at \emph{z}$=$0. We then compare the O/H abundances from the CECILIA and CHAOS samples to determine if CECILIA is offset to higher metallicities than anticipated from the \te\ distribution. Considering the CHAOS \hii\ regions with \te\oii\ $=$ 10700$\pm$500 K, the average 12+log(O/H) in the 24 regions is 8.41 dex with a standard deviation of 0.14 dex. This is comparable with the average 12+log(O/H) in the CECILIA galaxies, 8.29 dex, indicating that the combination of optical CELs and direct \te\ produces O/H that is consistent with nebulae of similar \te\ conditions. 

Next, we reproduce the N2-BPT in the right panel of Figure \ref{fig:oh_n2bpt}, where we have focused on the \te\ sample and color-coded each point by the direct O/H. The observed trend matches the expected behavior of galaxies in this line ratio diagram, which is sensitive to ionization, total O/H, and the relative abundance of N/O. At low-O/H, the highly-ionized gas cools via \oiii$\lambda$5008 emission and results in large \oiii/H$\beta$, while low N abundance and weak low-ionization emission lines result in low \nii/H$\alpha$. As metallicity increases, so too do the available coolants in the form of O and secondary N, while the hardness of the ionizing spectrum decreases. This combination leads to lower \oiii/H$\beta$ as \nii/H$\alpha$ increases. In CECILIA, the galaxies with log(\nii/H$\alpha$) $<$$-$1.4 and log(\nii/H$\alpha$) $>$$-$1.0 have average 12+log(O/H) of 8.02 and 8.48 dex, respectively, in qualitative agreement with the predicted evolution in the BPT diagram. Finally, an evolution in O/H is expected from the mass-metallicity relation \citep[MZR,][]{lequeux1979,tremonti2004}. The average stellar masses of the CECILIA galaxies with 12+log(O/H) $>$ 8.5 dex and $<$ 7.5 dex are log(M$_*$/M$_\odot$) $=$ 9.9 and 9.3, respectively. This is in qualitative agreement with the trends of other high-\emph{z} galaxies \citep[e.g.,][]{erb2006metal,sanders2021,strom2022,curti2024}, although we defer a detailed study of the MZR at Cosmic Noon for a future work \citep[see][]{raptis2025b}.

\begin{figure*}[!t]
   \centering
   \includegraphics[width=0.90\textwidth]{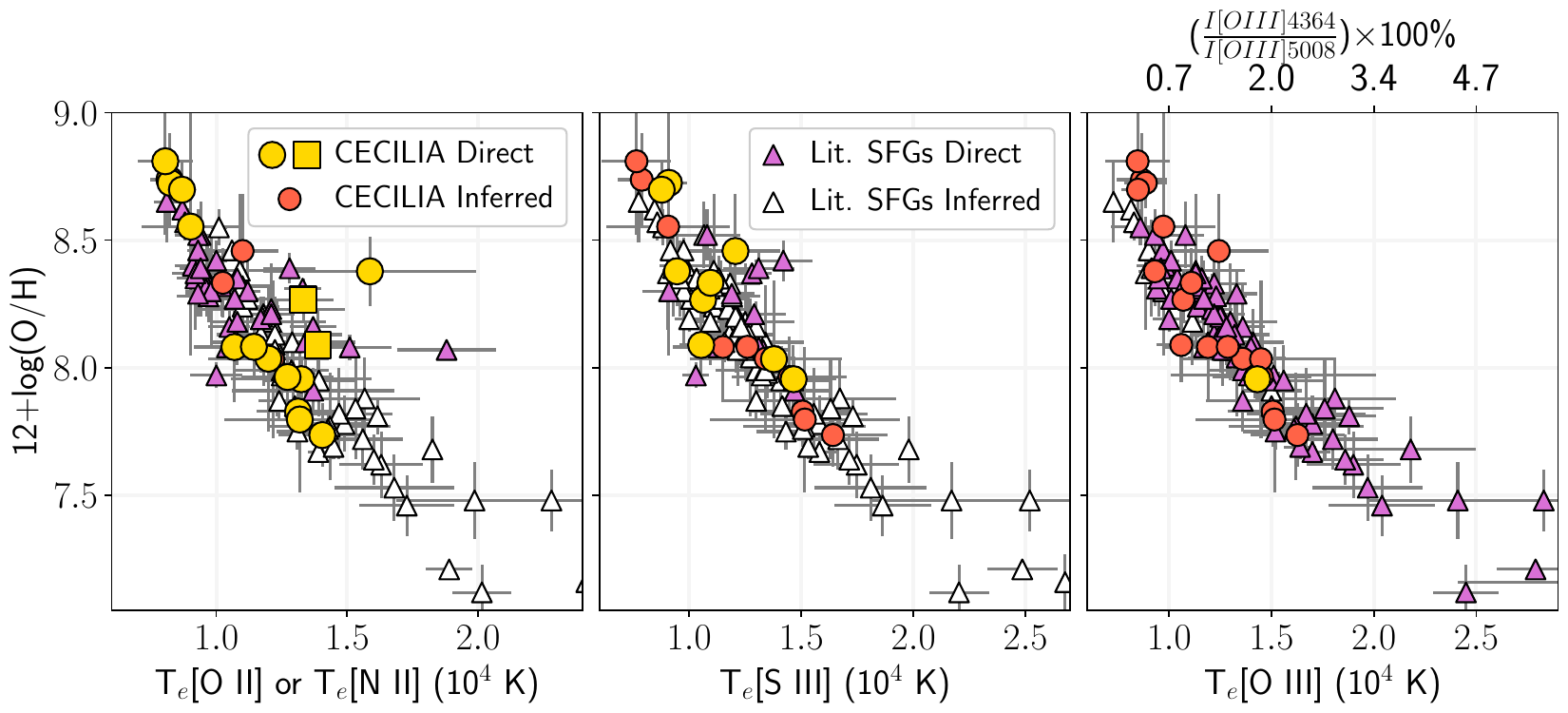}
   \caption{Direct 12+log(O/H) measured in galaxies at \emph{z}$>$1.3 are plotted against \te\oii\ or \te\nii\ when available (\textit{Left}), \te\siii\ (\textit{Middle}), and \te\oiii\ (\textit{Right}). The shapes represent whether the \te\ in each panel are directly measured or inferred from a \te\ scaling relation: yellow circles are direct \te\ measurements from CECILIA (yellow squares represent direct \te\nii); red circles are CECILIA galaxies with inferred \te; purple and white triangles are literature SFGs where the specific \te\ has been directly measured and inferred, respectively. The expected anti-correlation between O/H and \te\ is observed, and the CECILIA galaxies with low direct \te\oii\ and \te\siii\ extend the trend to the highest O/H. Above the right panel, we provide the auroral-to-nebular \oiii\ line ratio required to produce the direct or inferred \te. Few high-\emph{z} galaxies have \oiii\ auroral-to-nebular line ratios $\leq$0.007 predicted in the most metal-rich CECILIA galaxies.  
   }
   \label{fig:oh_te}
\end{figure*} 

Systematic uncertainties could bias the reported O abundances in CECILIA, the most plausible being the use of the \oii\ nebular lines from the Keck/MOSFIRE spectrum to derive \te\oii\ (see discussion in \S\ref{sec:combine}).
One method to assess this is to use the \te\siii\ measurement and a \te\ scaling relation to infer the low-ionization zone temperature, then apply that temperature with the \oii\ auroral lines to measure O$^+$ from the NIRSpec data alone. This method was adopted for the chemical abundance study of D40 \citep{rogers2024}, and is only possible for the galaxies with simultaneous \siii$\lambda$6314 and \oii$\lambda\lambda$7322,32 detections (i.e., those plotted in Figure \ref{fig:to2ts3}). Of the four galaxies (i.e., excluding two with direct \te\nii\ where \te\oii\ is not adopted), the use of \te\siii\ alone produces O/H consistent with the reported values within 1$\sigma$, with an average difference of $-$0.03 dex.
While we acknowledge that some galaxies could be affected by a bias in \te\oii, this exercise also relies on the shape of the \te\oii-\te\siii\ scaling relation that is, presently, not well constrained on empirical \te\ data at high-\emph{z}. The \te\ scaling relations are another source of systematic uncertainty, particularly for the high-ionization zone \te\ that must be inferred for the majority of CECILIA galaxies.
Nonetheless, these consistency checks indicate that the O/H in the CECILIA galaxies is not uncharacteristic of galaxies with similar \te\ or CEL emission, suggesting that SFGs at Cosmic Noon could reach solar O/H in the gas phase.

We now compare the \te\ and O/H measurements in the CECILIA galaxies to other high-\emph{z} SFGs with published direct-method \te\ and metallicities acquired with JWST. In Figure \ref{fig:oh_te}, we plot 12+log(O/H) against \te\oii\ (or \te\nii, when available), \te\siii, and \te\oiii\ for the CECILIA sample and 85 literature SFGs at \emph{z}$>$1.3 \citep[from][]{welch2024,welch2025,morishita2025,stanton2025,arellano-cordova2025,sanders2024sl,sanders2025,cataldi2025}. In each panel, we denote whether each \te\ is directly measured or inferred from a \te\ scaling relation. For CECILIA, the direct \te\ are plotted as yellow circles (or yellow squares for \te\nii) while the inferred \te\ are represented by red circles. The purple and white triangles indicate the literature SFGs with direct and inferred \te, respectively. When an inferred \te\oii\ or \te\siii\ is not reported, we apply the \citet{garnett1992} scaling relations along with the reported \te\oiii\ to include the galaxy in Figure \ref{fig:oh_te}. Since there are so few \te\oii\ and \te\siii\ samples at \emph{z}$>$1, most of the literature \te\ in the left and middle panels are inferred from the reported \te\oiii.

As expected, there is a strong anti-correlation between gas-phase metallicity and \te\ \citep[see also][]{flury2020}, although Figure \ref{fig:oh_te} provides additional insight into the state of direct \te\ and abundances in the era of JWST.
\te\oiii\ is the most common temperature measured at high-\emph{z} and ranges from $\sim$0.85-2.5$\times$10$^4$ K. The resulting metallicities, 
7.0$\lesssim$12+log(O/H)$\lesssim$8.5, are limited by the inherent challenge of applying the direct abundance method with \te\oiii\ alone. When the ISM is highly ionized and at low-metallicity, O$^{2+}$ is the dominant ionization state of O and the ISM is sufficiently hot to permit auroral \oiii\ excitation. However, in the same environments the \oii\ and \siii\ auroral lines are relatively faint owing to the scarcity of O$^+$ and S in the ISM. This physical scenario manifests as the dearth of direct \te\oii\ at 12+log(O/H)$\lesssim$8.0 in the literature sample, and the lack of literature \te\siii\ likely originates from a combination of exceptionally faint auroral emission in tandem with insufficient wavelength coverage for the nebular \siii\ lines.

In contrast, low- and intermediate-ionization state auroral lines are easier to detect in metal-rich or low-ionization environments where \oiii$\lambda$4364 becomes prohibitively faint \citep[e.g.,][]{rogers2021}. Accordingly, most direct \te\oii\ in the literature SFGs occur between 12+log(O/H) of 8.0 and 8.5. Measuring \te\oii\ in the cooler, metal-rich ISM becomes challenging owing to the diminishing intensity of the \oii\ auroral lines with decreasing temperature. Detecting the relatively faint \oii\ and \siii\ auroral lines over a wider range of O/H is possible from the ultra-deep CECILIA observations, and these data extend the O/H-\te\ anti-correlation to \te$\approx$8000 K, consistent with the lowest \te\ and highest direct-method metallicities measured with JWST.

Focusing on the CECILIA galaxies with 12+log(O/H)$>$8.5, we investigate the implied auroral-to-nebular \oiii\ ratio produced by the inferred high-ionization zone \te. In the right panel of Figure \ref{fig:oh_te}, we provide the \oiii$\lambda$4364/\oiii$\lambda$5008 ratio corresponding to the \te\oiii\ on the bottom axis (assuming a typical density of 300 cm$^{-3}$). Very few high-\emph{z} SFGs have direct measurements of \oiii$\lambda$4364 at $<$1\% the intensity of \oiii$\lambda$5008.
This exercise, although reliant on the shape of the \te\ scaling relations, reveals that the discovery space at high-metallicity requires exceptionally deep spectroscopy and broad wavelength coverage, a combination that is feasible in SFGs at Cosmic Noon with JWST.
Only by pushing to these extremes with future JWST observations can we reveal the pathways that have resulted in significant O/H enrichment, or whether the simultaneous \te\ trends indicate that the ISM is described by a different \te\ structure that favors a lower metallicity in the high-ionization gas.

\subsection{S/O and Ar/O}

In local nebulae, the abundance ratios of S/O and Ar/O in the ISM tend to be constant and very close to the solar abundance ratios \citep{izotov2006,berg2020,amayo2021,rogers2022,esteban2025}. This trend suggests a similar production mechanism for all three elements, that being the $\alpha$ process in the interiors of massive stars. However, the Type Ia SNe yield models of \citet{kobayashi2020SNIa} indicate that S and Ar can be produced in excess of O in Type Ia SNe. If the abundance of S and Ar are sensitive to enrichment from Type Ia SNe, then the ratios of S/O and Ar/O should evolve based on the relative enrichment of these two sources and the SFH of a galaxy. Variations in the Ar/O abundance have been observed in local nebulae, such as the planetary nebulae of M31 \citep{arnaboldi2022,kobayashi2023}. Recently, \citet{bhattacharya2025} reanalyzed SFGs in SDSS and uncovered a subtle evolution in the O/Ar ratio with Ar/H and stellar mass, where high-mass galaxies show larger O/Ar at fixed Ar/H relative to low-mass galaxies. This trend reflects the lower star-formation efficiency (SFE) and gradual Type Ia enrichment in low-mass galaxies that contributes to the decrease of O with respect to Ar or other Type Ia nucleosynthetic products.

\begin{figure}[t]
   \centering
   \includegraphics[width=0.47\textwidth]{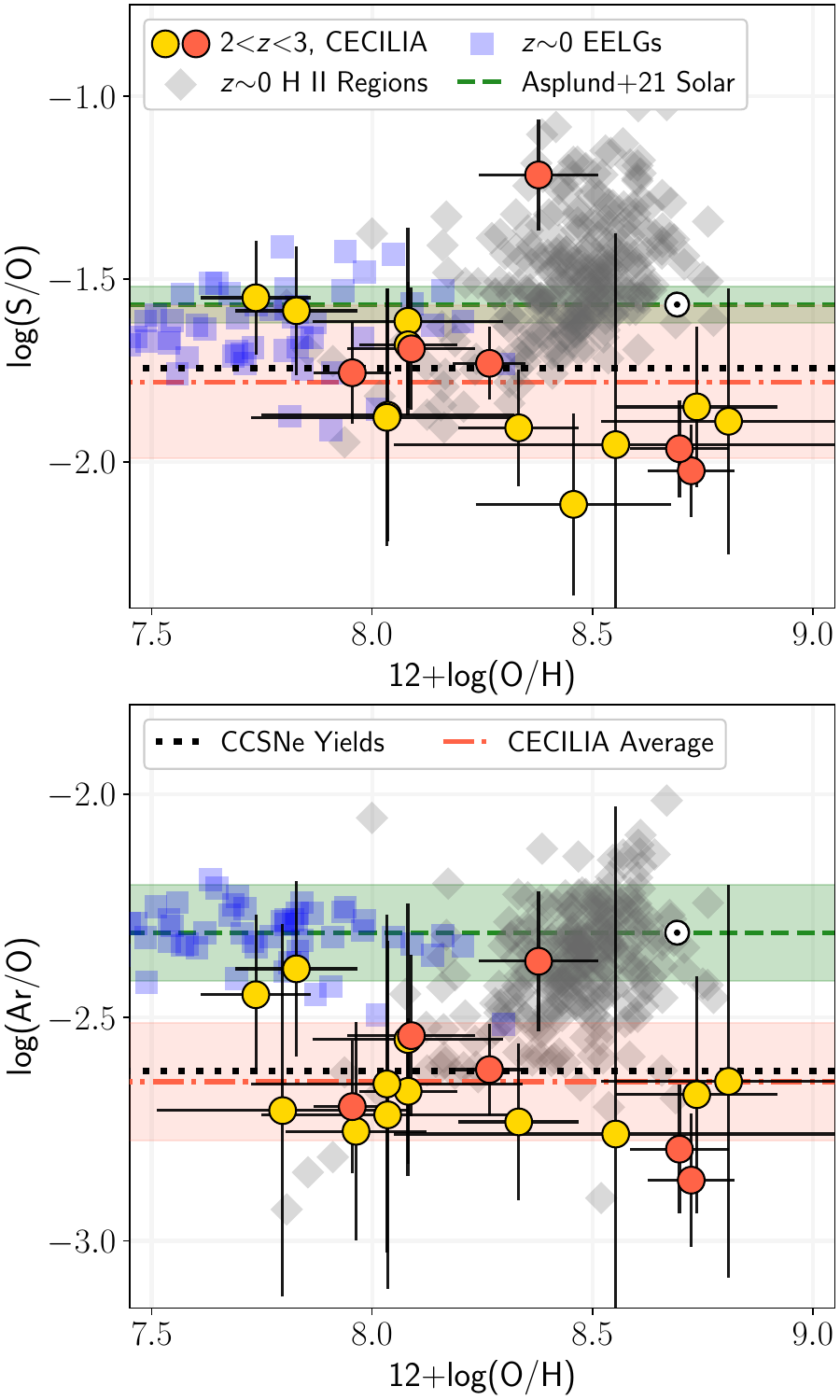}
   \caption{Direct abundance trends of S/O (\textit{Top}) and Ar/O (\textit{Bottom}) vs.\ 12+log(O/H) for the CECILIA galaxies. Shape and coloring of the data is similar to Figures \ref{fig:to2ts3} and \ref{fig:oh_n2bpt}.
   The solar abundance ratio from \citet{asplund2021} is represented as the circle dot and dashed green line with shaded area. The red dash-dotted line and shaded area represent the average S/O and Ar/O abundance ratios and their uncertainties, respectively, from the full CECILIA sample. The dotted black line is the abundance ratio predicted from metal-poor CCSNe yields of \citet{kobayashi2006}. Consistent with the abundance trends in D40 \citep{rogers2024} and other high-\emph{z} galaxies, S/O and Ar/O are significantly sub-solar in the typical SFG at Cosmic Noon. This abundance pattern is expected for a galaxy with significant CCSNe enrichment, with little Type Ia contributions owing to the abbreviated SFH at \emph{z}$>$2.}
   \label{fig:lsoaro}
\end{figure}

The chemical abundance patterns in galaxies at Cosmic Noon, which are undergoing recent and rapid star formation, may be sensitive to the relative enrichment from CC and Type Ia SNe.
In Figure \ref{fig:lsoaro}, we plot the direct S/O and Ar/O abundance ratios measured from the full CECILIA sample against the corresponding gas-phase O/H measured in each galaxy. For comparison, the S/O and Ar/O abundance patterns in the CHAOS \hii\ regions and low-metallicity EELGs at \emph{z}$=$0 are plotted as gray diamonds and blue squares, respectively. We also provide the solar abundance ratios as the circle dot and dashed green line. The CECILIA galaxies plotted in Figure \ref{fig:lsoaro} represent the largest sample of S and Ar abundances in galaxies at \emph{z}$>$2 compiled from JWST observations, which is enabled because of the design of CECILIA. Reliable S abundances require the detection of the \siii\ nebular lines in the rest-NIR \citep[see discussion in][]{amayo2021}, while the \ariii$\lambda$7138 emission line is faint ($\sim$2\% the flux of H$\alpha$ in the CECILIA galaxies) owing to the scarcity of Ar in the ISM. As such, deep spectroscopy required to measure \ariii$\lambda$7138 and the \siii\ auroral line must be matched with broad wavelength coverage to detect the nebular \siii\ lines and derive a reliable \te\siii. 

These abundance trends reveal that the typical galaxy in the CECILIA sample has lower S/O and Ar/O than the solar abundance ratio. We measure an average log(S/O) $=$ $-$1.78$\pm$0.21 dex and log(Ar/O) $=$ $-$2.64$\pm$0.13 dex (plotted as dot dashed red lines in Figure \ref{fig:lsoaro}), which are 0.21 and 0.33 dex lower than the respective solar ratios. Consistent with local chemical abundance surveys, there is no apparent evolution in S/O or Ar/O with total metallicity. From CECILIA, there is growing evidence for non-solar abundance patterns in galaxies during the epoch of Cosmic Noon. Given the contribution of Type Ia SNe to the gas-phase S and Ar abundances, a natural explanation for the enhanced O relative to these elements is the preferential enrichment from CCSNe. We provide the S/O and Ar/O abundance ratios predicted from the IMF-averaged CCSNe yields for metal-poor (5\% $Z_\odot$) stars from \citet{kobayashi2006} as black dotted lines in Figure \ref{fig:lsoaro}, which are log(S/O) $=$ $-$1.74 dex and log(Ar/O) $=$ $-$2.62 dex. These values are consistent with the minimum Ar/O predicted from chemical evolution models for the Milky Way \citep[e.g.,][]{kobayashi2020} and M31 \citep{arnaboldi2022,kobayashi2023}. We observe that the average S/O and Ar/O measured in CECILIA is in good agreement with the yield predictions and minimum ratios expected from chemical evolution models, further supporting a scenario of significant CCSNe enrichment with minimal contributions from Type Ia SNe.

The average ratios measured in the full sample are in good agreement with the abundance trends in D40, the first galaxy with direct \te\siii, S and Ar abundance measurements at \emph{z}$>$2 \citep{rogers2024}. As discussed in that work, certain systematic uncertainties could produce an offset to low S/O and Ar/O. First, D40 was previously missing a direct measurement of the low-ionization zone \te, such that the earlier O$^+$ and S$^+$ abundances could have been biased. From the combined MOSFIRE and NIRSpec data reported here, the \te\oii\ in D40 agrees with the \te\ inferred from the other direct temperatures. While the lack of \te\ information did not appear to significantly bias the abundance ratios in D40, galaxies with a single \te\ are reliant on the functional form of the adopted \te\ scaling relations. Second, the selected ICFs could fail to account for all unobserved ionization states of S and Ar in the ISM. However, the trend of sub-solar S/O and Ar/O persists when the analysis is repeated with the \citet{amayo2021} ICFs (average S/O and Ar/O are $-$1.83$\pm$0.15 dex and $-$2.65$\pm$0.12 dex, respectively), indicating that a substantial change to the shape of the ICFs is required to bring the ratios into agreement with the solar abundances \citep[see also][]{stanton2025}.

The use of the \oii\ nebular lines from the MOSFIRE spectra could bias to larger O$^+$ abundance (see discussion in previous subsection). However, when we repeat the analysis with only \te\siii\ and use the \oii\ auroral lines for the O$^+$ abundance (similar to the original analysis for D40), we find that the S/O abundances in five of the six galaxies remain consistent within uncertainties.
For BX628, using \te\siii\ to infer the low-ionization zone \te\ produces log(S/O) $=$ $-$1.68$\pm$0.15 dex; this is in good agreement with the average S/O in CECILIA and may indicate that \te\oii\ is biased high by the low-S/N MOSFIRE \oii\ lines in this galaxy. Finally, we note that accounting for dust depletion would increase O in the most metal-rich galaxies (see \S\ref{sec:abuncalcs}). While the depletion of S onto dust grains is uncertain \citep{jenkins2009}, Ar is not depleted onto grains and the Ar/O abundance ratios in some of the CECILIA galaxies may be interpreted as a maximum if dust depletion is significant. In summary,
while we cannot rule out the potentially distinct S or Ar ICFs,
the average CECILIA S/O and Ar/O trends should be representative of the chemical abundance patterns in typical SFGs at Cosmic Noon.

In principle, S should be more accessible than Ar owing to the relatively bright \sii\ and \siii\ CELs in the optical and NIR: on average, the fluxes of \siii$\lambda$9533 and \sii$\lambda$6718 are a factor of 6.8 and 3.1 times larger than the flux of \ariii$\lambda$7138 in the same CECILIA galaxies. In practice, there are relatively few high-\emph{z} galaxies with direct S/O abundances owing to the observational challenges associated with detecting the NIR \siii\ lines. Instead, we can compare these abundance trends to local SFGs analogous to high-\emph{z} galaxies in their SFRs. Recently, \citet{arellano-cordova2024a} reported an average log(S/O) $=$ $-$1.73$\pm$0.10 dex in the highly star-forming galaxies of the CLASSY survey \citep{berg2022,james2022}. This is consistent with the average reported in the CECILIA galaxies (log(S/O)$=$$-$1.78$\pm$0.21 dex, or (S/O)$=$0.62$\pm$0.30$\times$(S/O)$_\odot$) and further supports a scenario where the gas-phase S abundance is sensitive to enrichment from CC and Type Ia SNe \citep[see also][]{esteban2025}.
However, the scatter in S/O observed in the CLASSY and CECILIA galaxies is large, and additional S abundances at high-\emph{z} are necessary to assess the evolution of S/O over cosmic time.

\begin{figure}[t]
   \centering
   \includegraphics[width=0.47\textwidth]{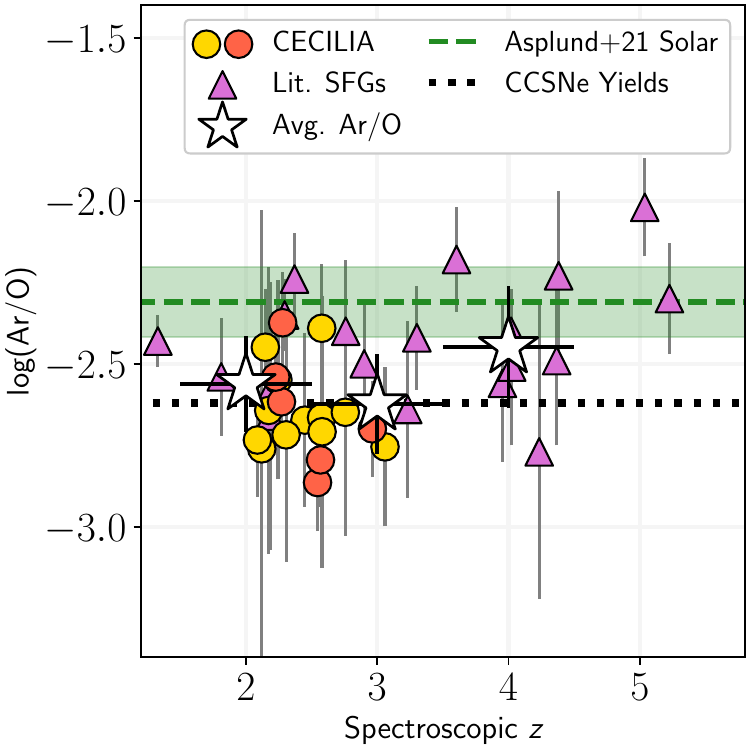}
   \caption{The redshift evolution of the relative Ar/O abundance in SFGs observed with JWST. The chemical abundance patterns in the literature galaxies (purple triangles) are taken directly from \citet{welch2024,welch2025,morishita2025,stanton2025,bhattacharya2025A}. The color of the CECILIA galaxies represents the number of direct \te\ measurements, see Figure \ref{fig:oh_n2bpt}. The average Ar/O ratios are measured in three redshift bins (1.5$\leq$\emph{z}$<$2.5, 2.5$\leq$\emph{z}$<$3.5, and 3.5$\leq$\emph{z}$<$4.5) and plotted as white stars. The SFGs at 2$<$\emph{z}$<$3.1 indicate that the gas-phase Ar/O abundance reaches (Ar/O)$=$0.56$\pm$0.05$\times$(Ar/O)$_\odot$, similar to the CCSNe yield predictions plotted in dotted black.}
   \label{fig:lar_lit}
\end{figure}

To compare with existing abundance measurements in Figure \ref{fig:lar_lit},
we focus on Ar/O measured in other high-\emph{z} galaxy surveys. We include Ar/O from the observations of SGAS 1723+34 from the TEMPLATES survey \citep{welch2024}, the Sunburst Arc \citep{welch2025}, MACSJ1149-WR1 \citep{morishita2025}, the EXCELS survey from \citet{stanton2025}, and the compilation of Ar/O from \citet{bhattacharya2025A} excluding D40 and SGAS 1723+34. We use the published Ar/O in each source except for MACSJ1149-WR1, which is reported to have log(Ar/O) $=$ $-$2.40$\pm$0.02 dex. The reported uncertainty on Ar/O is likely underestimated, considering that the uncertainty in the total O abundance is $\sim$0.1 dex and the low-ionization zone temperature has uncertainty in excess of 3000 K. We recompute the Ar/O abundance in MACSJ1149-WR1 using the reported line intensities and ionization zone temperatures in \citet{morishita2025} and find log(Ar/O) $=$ $-$2.39$\pm$0.22 dex, which we use when comparing to the CECILIA sample.

Figure \ref{fig:lar_lit} shows log(Ar/O) vs.\ redshift for 36 galaxies at 1.2$<$\emph{z}$<$5.8, where the literature abundances are plotted as purple triangles. The Ar/O abundances in the literature galaxies are generally at or below the solar abundance ratio. To assess the average trends of these high-\emph{z} galaxies, we measure the average Ar/O ratio in three redshift bins: 1.5$\leq$\emph{z}$<$2.5, 2.5$\leq$\emph{z}$<$3.5, and 3.5$\leq$\emph{z}$<$4.5. Each bin has more than five galaxies with direct abundances from which we measure the average Ar/O, which are plotted as white stars in Figure \ref{fig:lar_lit}. While the highest redshift bin has the largest uncertainty, the galaxies at 1.5$\leq$\emph{z}$<$2.5 and 2.5$\leq$\emph{z}$<$3.5 show similar log(Ar/O) and are in agreement with the CCSNe yield predictions. Excluding the CECILIA sample, the number of galaxies in the 1.5$\leq$\emph{z}$<$2.5 and 2.5$\leq$\emph{z}$<$3.5 bins decrease from 15 to 5 and 12 to 4, respectively. Without the CECILIA galaxies, the average log(Ar/O) in these bins increases to log(Ar/O) $\sim$ $-$2.48 dex, consistent with the binned averages in Figure \ref{fig:lar_lit} and over 0.15 dex below the solar Ar/O ratio.

These direct abundance patterns complement previous evidence for predominant enrichment from CCSNe during Cosmic Noon. For example, jointly modeling the emission line properties and shape of the UV continuum in SFGs at Cosmic Noon has led to the conclusion that these galaxies have a combination of Fe-poor stars and modest gas-phase O enrichment \citep{steidel2016,strom2018,cullen2019,cullen2021,stanton2024}. This O/Fe enhancement is another signature of significant CCSNe enrichment and a dearth of Type Ia SNe products, which results in O/Fe in excess of the solar ratio \citep{nomoto2006}. However, measuring this O/Fe enhancement in the gas-phase is exceptionally challenging owing to the faint lines required, significant dust depletion of Fe, and fluorescent effects of the optical Fe CELs \citep{rodriguez1999,rodriguez2005,mendez-delgado2024}. 
The magnitude of O/Fe enhancement is often inferred through the use of photoionization modeling \citep[e.g.,][]{strom2018,sanders2020}, simultaneous rest-UV and optical observations \citep{cullen2021,stanton2024}, or direct measurement from optical CELs \citep{curti2025b}, and these approaches indicate a O/Fe enhancement of $\gtrsim$2.2$\times$(O/Fe)$_\odot$ in galaxies at \emph{z}$>$2 \citep{cullen2019,strom2022}. The inverse variance-weighted average O/Ar observed in the combined CECILIA and literature sample at 2$\leq$\emph{z}$\leq$3.1 (23 galaxies total) is 1.78$\pm$0.16$\times$(O/Ar)$_\odot$, similar to the findings of \citet{stanton2025}.

Should Type Ia SNe represent a significant component of Ar enrichment, then Ar/O will vary in individual galaxies based on their SFHs. While most galaxies in CECILIA show Ar/O below those observed in the local Universe, other high-\emph{z} galaxies show abundance ratios that agree with the solar values. Only with larger samples of direct abundances, particularly using a direct measurement of \te\siii, can we assess the evolution of S/O and Ar/O in the average galaxy as a function of redshift.
Unfortunately, there is insufficient abundance data to evaluate the redshift evolution of Ar from Cosmic Noon to \emph{z}$=$0 where the typical SFG shows solar Ar/O. While sub-solar Ar/O is typical of the average galaxy at \emph{z}$\sim$2-3, further spectroscopic surveys at 0$<$\emph{z}$<$2, either from ground-based facilities or dedicated JWST observations, stand to reveal the evolution of Ar/O in the last 10 Gyr of cosmic time. Additionally, the evidence of solar Ar/O observed at \emph{z}$>$4 is intriguing, but requires larger samples to verify. While the \siii\ strong lines required for \te\siii\ can only be observed at \emph{z}$\leq$4.53, \ariii$\lambda$7138 is detectable with NIRSpec up to \emph{z}$=$6.39. With deep NIRSpec observations and appropriately calibrated \te\ scaling relations and ICFs, it will be possible to assess the evolution of Ar in the ISM and whether the relative Ar/O abundance acts as a tracer of Type Ia and CCSNe enrichment.

\subsection{N/O}

\begin{figure}[t]
   \centering
   \includegraphics[width=0.47\textwidth]{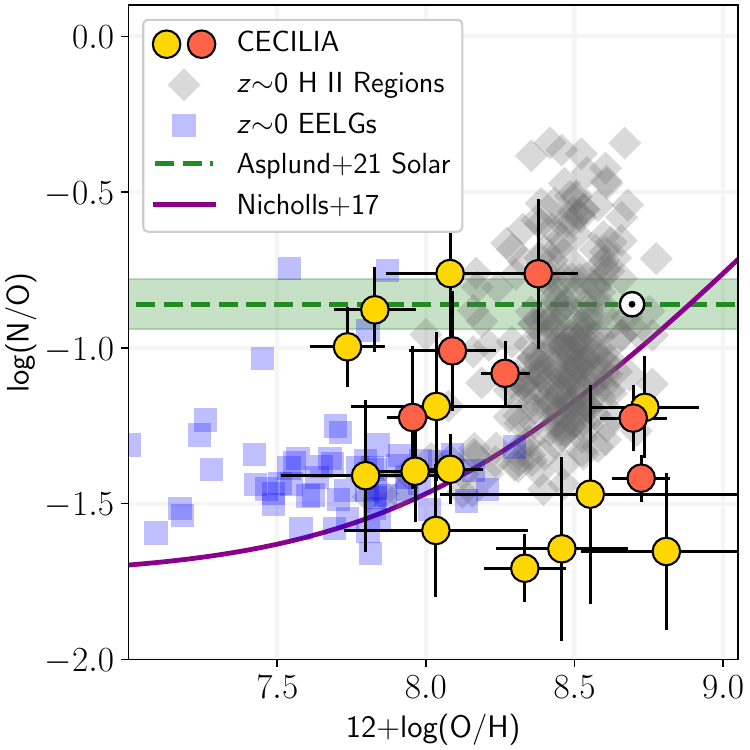}
   \caption{The N/O-O/H abundance trends in the CECILIA galaxies, local star-forming nebulae (gray diamonds and blue squares), Galactic stars \citep[solid purple line, from][]{nicholls2017}, and the Sun \citep[green dashed line and circle dot, from][]{asplund2021}. The color of the CECILIA galaxies represents the number of direct \te\ measurements, see Figure \ref{fig:oh_n2bpt}.
   The CECILIA galaxies generally scatter around the primary plateau at log(N/O) $\sim$ $-$1.45 dex and the secondary N/O locus observed in metal-rich \hii\ regions. The low N/O at 12+log(O/H) $>$ 8.5 dex in some CECILIA galaxies could be a signature of high SFE.
   High N/O at 12+log(O/H) $<$ 8.0 has been reported in other high-\emph{z} galaxies and is observed in a portion of the CECILIA sample, but is infrequently observed at \emph{z}$\sim$0.
   }
   \label{fig:nooh}
\end{figure}

The final abundance ratio we discuss in the CECILIA galaxies is N/O, which we plot against the total O/H abundance in Figure \ref{fig:nooh}. We focus our comparison relative to the local star-forming nebulae and EELGs, and we have also included the N/O-O/H evolution observed in Galactic stars \citep{nicholls2017}. 
Among the abundance ratios discussed in this work, N/O is the most sensitive to the relative flux calibration between the MOSFIRE and NIRSpec data. This is because N/O is measured using only the low-ionization zone \te; as such, any systematic uncertainty that affects \te\oii\ (e.g., reddening, MOSFIRE slit-loss corrections, comparison of \oii-to-\ion{H}{1} intensities) will directly influence N$^+$, O$^+$, and the N/O relative abundance.

In general, the CECILIA galaxies are consistent with the N/O-O/H trends of local SFGs, albeit with large uncertainty and significant scatter. This is similar to the N/O-O/H trends in the full KBSS sample measured via photoionization modeling \citep[see Figure 16 in][]{strom2018}.
At \emph{z}$=$0, there are two distinct sequences in the N/O-O/H abundances: at 12+log(O/H)$\lesssim$8.0, the relative abundance of N/O remains constant at the "primary plateau" of $-$1.45$\pm$0.10 dex \citep{garnett1990,vanZee2006}, which has been attributed to metallicity-insensitive N and O enrichment from massive stars.
As O/H increases, metallicity-sensitive (secondary) N production from the asymptotic giant branch phase in low- and intermediate-mass stars (LIMS) overtakes the N enrichment from massive stars, resulting in an increase in N/O after a sufficient time delay \citep{henry2000,vincenzo2016}.

The large scatter in the CECILIA data may suggest that some high-\emph{z} galaxies occupy two intriguing areas of the abundance ratio diagram.
First, the most metal-rich galaxies have log(N/O) only slightly above the \emph{z}$=$0 primary plateau. Second, some lower-metallicity CECILIA galaxies show log(N/O) $>$ $-$1.1 dex, more than 0.3 dex above the primary plateau. These abundance patterns are unexpected if a galaxy has had sufficient time for significant secondary enrichment from the LIMS.
However, SFE plays a significant role in the position of a galaxy in the N/O-O/H diagram. 
For example, galaxies with high SFE can rapidly enrich with O before the N/O ratio responds to the N enrichment from LIMS.
In this way, log(N/O) ratios near the primary plateau can be achieved at O/H in excess of the solar abundance ratio \citep[][]{henry2000,vincenzo2016}. Similarly, galaxies with low SFE maintain low O/H while the N abundance gradually increases through LIMS enrichment, producing high N/O at low O/H. Other mechanisms can also produce elevated N/O at low metallicity, such as local pollution from Wolf-Rayet stars \citep[e.g.,][]{james2009}, significant pristine gas infall into a moderately N-enriched ISM, or from bursty star formation and strong stellar feedback \citep[e.g.,][]{mcclymont2025,berg2025}.

Recent JWST observations suggest that elevated N/O at low O/H may be common in high-\emph{z} SFGs. Observations of galaxies at \emph{z}$>$6 have revealed the surprising detection of UV \ion{N}{3} and \ion{N}{4} emission lines, which are rarely observed in local nebulae \citep[e.g.,][]{mingozzi2022,martinez2025}. The N/O abundances implied from the UV emission lines in these high-\emph{z} EELGs are in excess of $-$0.75 dex at 12+log(O/H)$<$8.0 dex \citep{senchyna2024,marques-chaves2024,topping2024,topping2025no,curti2025}. While UV N/O may be affected by systematic uncertainties, such as the ICF and density effects, the N/O ratios measured from the optical \nii\ and \oii\ emission lines in other high-\emph{z} galaxies also indicate an enhancement in N with respect to O at moderate metallicities \citep{clarke2023,sanders2023,welch2024,welch2025,morishita2025}. Particularly, \citet{arellano-cordova2025} measured the N/O-O/H abundance patterns in two SFGs in the EXCELS survey and compiled all N/O abundances in galaxies at \emph{z}$>$1.8, including those measured from rest-optical and UV emission lines. All galaxies were found to have 12+log(O/H)$\lesssim$8.2 dex and log(N/O)$>$$-$1.10 dex, except D40 which has a N/O-O/H abundance pattern consistent with the primary plateau.

The diversity of N/O-O/H abundances measured in the CECILIA galaxies may provide additional evidence for variations in this abundance ratio beyond the expectations of local SFGs.
If the large scatter in N/O-O/H diagram is observed in other high-\emph{z} galaxies, then the differences in the chemical abundance ratios from the local nebulae have significant implications for photoionization modeling: many model grids assume a range of N/O that mirrors the N/O-O/H relation derived from local nebulae. Given the sensitivity of N/O on the SFH and SFEs, it may be necessary to fully decouple N/O and O/H when modeling the spectra of high-\emph{z} galaxies \citep[e.g.,][]{valeasari2016,strom2018}.
Chemical evolution models have been able to match the elevated N/O observed in some high-\emph{z} EELGs by invoking bursty SFHs with varying SFE \citep[e.g.,][]{kobayashi2024,mcclymont2025}. Indeed, differences in SFH and SFE can manifest as different locations of the N/O plateau, a trend that has been measured from resolved \hii\ regions within local galaxies \citep[see][]{belfiore2017,berg2020}.
Further constraints on the gas-phase N/O during Cosmic Noon will reveal the prevalence of the N-enhanced ISM and whether these trends require bursty SFHs. Additionally, comparisons between N/O abundances simultaneously measured from joint UV and optical spectroscopy (in high-\emph{z} and local SFGs) are required to understand the mechanisms of UV N emission and whether these independent measurements are representative of the average N/O abundance in the ISM of high-\emph{z} SFGs.

\section{Conclusions}\label{sec:conclusions}

In this paper, we have presented the gas-phase physical conditions and chemical abundances in CECILIA, a Cycle 1 JWST program targeting SFGs at Cosmic Noon to unveil the buildup of heavy elements during the peak of cosmic star formation. CECILIA combines ultra-deep JWST/NIRSpec observations (29.5hr in G235M, 1.1hr in G395M) with photometric and spectroscopic data of each target galaxy acquired from KBSS, enabling a detailed examination of the conditions in the multi-phase ISM. Specifically, the faint \siii\ and \oii\ auroral lines are detected, allowing for the direct determination of \te\ and the chemical abundance patterns from the strong rest-optical and NIR metal CELs in 20 SFGs at Cosmic Noon. 
The combination of these new NIRSpec data with the archival MOSFIRE observations results in one of the largest samples of direct-method abundances at \emph{z}$>$2 to date. The main aspects of our investigation are:
\begin{enumerate}
    \item For the NIRSpec data reduction, we implement a careful treatment of the MSA bar shadows and adopt a global background subtraction for the NIRSpec data (\S\ref{sec:jwstdata}). A 1D global background model is constructed from all available observations. This 1D model is used to create a custom 2D bar shadow model based on the spatial profiles at each wavelength in the 2D data, and the 2D model is used to correct the bar shadow artifacts. Once the global background and bar shadows are removed, local variations from the global background model are determined within the slit and corrected, thereby producing a featureless background. Figure \ref{fig:bx523_spec} provides an example 2D and 1D spectrum of a galaxy in the CECILIA survey, where the extended emission in the strong emission lines is preserved. The choice of the global background approach is discussed further in Appendix \ref{app:gbkg} (see also Figure \ref{fig:profs_appendix1}).
    
    \item We measure the electron density in the ISM via the \sii$\lambda\lambda$6718,33 doublet in the NIRSpec data and the \oii$\lambda\lambda$3727,29 lines in the higher-resolution MOSFIRE spectra (\S\ref{sec:densities}). A simultaneous measurement of \den\sii\ and \den\oii\ is possible in 15 CECILIA galaxies (Figure \ref{fig:do2ds2}); the average densities measured in these 15 galaxies are $\langle$\den\sii$\rangle$$=$267$\pm$44 and $\langle$\den\oii$\rangle$$=$386$\pm$70 cm$^{-3}$, approximately a factor of two larger than the typical \den\ measured in local SFGs. While the average \den\oii\ is slightly larger than \den\sii,
    the similar \den\ in the S$^+$ and O$^+$ emitting gas may be consistent with
    minimal DIG contamination to the \sii\ line ratios or higher \den\ in the DIG of galaxies at Cosmic Noon. There are galaxies in the CECILIA sample with \den\sii$<$100 cm$^{-3}$ (top panel of Figure \ref{fig:do2ds2}), where assuming a characteristic \den\ $>$250 cm$^{-3}$ is not appropriate for ionic abundance calculations.

    \item We measure \te\siii\ in 9 galaxies, while the combination of the NIRSpec and MOSFIRE data allow for the determination of \te\oii\ in 17 galaxies (\S\ref{sec:temperatures}). This represents one of the largest samples of \te\siii\ and \te\oii\ measurements at \emph{z}$>$1 to date, and we assess the \te\ trends in six CECILIA galaxies with simultaneous \te\oii\ and \te\siii\ (Figure \ref{fig:to2ts3}). The direct temperatures in these galaxies are consistent with the empirical \te\ trends in metal-rich \hii\ regions and metal-poor EELGs at \emph{z}$=$0, and the correlation between \te\oii\ and \te\siii\ at Cosmic Noon generally agrees with photoionization model \te\ scaling relations. 
    There remains a lack of direct \te\oii\ and \te\siii\ at \emph{z}$>$2, such that it is not possible to constrain the shape of this scaling relation at Cosmic Noon nor assess the scatter in simultaneous direct \te\ measurements. However, the CECILIA galaxies suggest that the low-\emph{z} \te\ scaling relations may be applicable for chemical abundance measurements in high-\emph{z} galaxies.

    \item Using the \den\ and \te\ measured in the CECILIA galaxies, we calculate the N, O, S, and Ar abundances in the ISM. The O/H abundances measured in the CECILIA galaxies range from 12+log(O/H) $=$ 7.76 to 8.81 dex (left panel of Figure \ref{fig:oh_n2bpt}). The CECILIA sample contains the highest direct-method O/H abundances measured with JWST to date, indicating that galaxies can reach solar O/H by Cosmic Noon. The O/H abundances are in qualitative agreement with the strong-line trends observed in the same galaxies (right panel of Figure \ref{fig:oh_n2bpt}). We discuss potential sources of systematic uncertainties, noting that the \te\ measured in the CECILIA galaxies are more akin to metal-rich \hii\ regions and that the same abundance results are recovered if \te\siii\ (which is measured entirely from NIRSpec) is used to measure O/H.

    \item The S/O and Ar/O abundances measured in the CECILIA galaxies show no evolution with O/H, but they are significantly sub-solar (Figure \ref{fig:lsoaro}). This trend is inconsistent with the chemical abundance patterns of many local SFGs, which mostly exhibit solar S/O and Ar/O. However, both S and Ar have a non-negligible Type Ia enrichment component \citep{kobayashi2020SNIa} such that S/O and Ar/O are sensitive to the delay between CCSNe and Type Ia enrichment. The low S/O and Ar/O, therefore, suggests that the typical galaxy at Cosmic Noon has been predominantly enriched with the nucleosynthetic products of CCSNe. This trend is consistent with prior inferences of the Fe/O abundance in similar galaxy populations. We examine the redshift evolution of the Ar/O ratio using CECILIA and a compilation of direct Ar/O measurements from the literature, finding that the average Ar/O at \emph{z}$\sim$2-2.5 is in good agreement with IMF-averaged CCSNe yield predictions for metal-poor stars (Figure \ref{fig:lar_lit}).

    \item The CECILIA galaxies exhibit N/O that are broadly consistent with the trends of local SFGs, although there is large scatter in N/O-O/H (Figure \ref{fig:nooh}). Variations in the N/O-O/H diagram could be related to differences in SFH and SFE, which can shift these abundance ratios away from those measured at \emph{z}$\sim$0. Larger samples are required to assess the role SFH has in shaping N/O-O/H abundance patterns observed in SFGs at Cosmic Noon and beyond.
\end{enumerate}

The direct abundance patterns in the CECILIA galaxies provide insight into galaxy enrichment during the period of Cosmic Noon. It is clear that galaxies during this time are chemically distinct from galaxies at \emph{z}$=$0, being offset to lower S and Ar relative to O than observed in local nebulae and the Sun. When the transition from sub-solar to solar Ar/O occurs remains unclear, as does the variation in the N/O abundances observed in some CECILIA galaxies and the literature at large. Future JWST observations and ground-based surveys stand to reveal the average abundance patterns in galaxies at 0$<$\emph{z}$<$2, although we emphasize that this requires the gas-phase physical conditions and, preferably, numerous \te\ and \den\ measurements to resolve the multi-phase ISM. Very deep spectroscopy capable of detecting the \te-sensitive auroral lines of \oiii, \siii, and \oii\ will unveil the redshift evolution of \te\ scaling relations, mitigating a source of significant systematic uncertainty in all high-\emph{z} chemical abundance studies. The direct metallicities discussed in this manuscript will enable future studies, including the calibration new strong line diagnostics for large-scale galaxy surveys with JWST, measuring the shape of the mass-metallicity relation $\sim$2-3 Gyr after the Big Bang, and exploring the diversity of chemical and ionization conditions in the ISM at Cosmic Noon.

\begin{acknowledgements}
We thank the anonymous referee for their thoughtful review and useful feedback, which helped elucidate further detail in the analysis. We'd also like to thank Bernie Rauscher for useful discussions concerning \textsc{NSClean}, as well as Gabe Brammer for the assistance with the \textsc{msaexp} reduction of the CECILIA NIRSpec data.

NSJR is supported by JWST-GO-02593.004-A, provided by NASA through a grant from the Space Telescope Science Institute, which is operated by the Association of Universities for Research in Astronomy, Inc., under NASA contract NAS 5-03127. ALS, GCR and RFT acknowledge partial support from the JWST-GO-02593.008-A, JWST-GO-02593.004-A, and JWST-GO-02593.006-A grants, respectively. ALS is also supported by the David and Lucile Packard Foundation (Packard Fellowship, grant ID 2024-77399) and performed aspects of the work contained in this paper at the Aspen Center for Physics, which is supported by National Science Foundation grant PHY-2210452. RFT also acknowledges support from the Pittsburgh Foundation (grant ID UN2021-121482) and the Research Corporation for Scientific Advancement (Cottrell Scholar Award, grant ID 28289). TBM was supported by a CIERA Postdoctoral Fellowship.

This work is primarily based on observations made with NASA/ESA/CSA JWST, associated with PID 2593, which can be accessed via doi:\dataset[10.17909/x66z-p144]{https://doi.org/10.17909/x66z-p144}. The data were obtained from the Mikulski Archive for Space Telescopes (MAST) at the Space Telescope Science Institute, which is operated by the Association of Universities for Research in Astronomy, Inc., under NASA contract NAS 5-03127 for JWST. The ground-based spectroscopy included in the analysis were obtained at W.M. Keck Observatory, which is operated as a scientific partnership between the California Institute of Technology, the University of California, and NASA. Keck access was provided by NASA, the California Institute of Technology, as well as Northwestern University and the Center for Interdisciplinary Exploration and Research in Astrophysics (CIERA). The Observatory was made possible by the generous financial support of the W. M. Keck Foundation, and the authors wish to recognize and acknowledge the significant cultural role and reverence that the summit of Maunakea has within the indigenous Hawaiian community.
\end{acknowledgements}

\facilities{Keck:I (MOSFIRE), JWST (NIRSpec)}
\software{BPASSv2 \citep{stanway2016,eldridge2017}, Cloudy \citep{ferland2013}, \texttt{GalDNA} \citep{strom2018}, JWST Calibration Pipeline \citep{calwebb_v1.10.0}, \texttt{grizli} \citep{grizli}, \texttt{msaexp} \citep{msaexp}, \texttt{PyNeb} \citep{luridiana2015}}

\appendix
\section{Global Background Subtraction}\label{app:gbkg}

As discussed in \S\ref{sec:jwstdata}, we adopt a global background subtraction approach for the CECILIA galaxies as opposed to a local background obtained via nod differencing within the slit. We adopt this approach to maximize the overall S/N in the data and to mitigate self-subtraction in the strong emission lines: some CECILIA galaxies are more extended than the nod offset of the MSA, such that extended emission in the position of the dither pattern could be included in the local background estimate. The nod background approach is the default for most NIRSpec observations and produces a characteristic off-source subtraction pattern in the resulting 2D data when the emission is extended. To prevent self-subtraction in the strong lines and enable future analyses of the extended ionized gas observed in the ultra-deep G235M CECILIA data, we prioritize a global background subtraction method. This method is outlined in \S\ref{sec:jwstdata}, and in this Appendix we provide examples of the spatial profiles and line fluxes recovered from the global background method.

To examine the differences between the global and nod background methods, we run \textsc{msaexp} using different sets of input 2D spectra. We start with the global background-subtracted, bar shadow, and artifact-corrected 2D spectra discussed in \S\ref{sec:jwstdata} from individual dither positions/observations (i.e., the \textsc{phot} files). These data are combined using the pseudo-drizzling approach in \textsc{msaexp} without nod differencing,
and we apply the default optimal extraction algorithm to obtain the 1D spectrum for each galaxy. For the nod background subtraction approach, we correct for slit artifacts like stuck open shutters and bar shadows, but we do not subtract the global background. We re-run \textsc{msaexp} to combine the data using nod differencing for a local background subtraction, then reapply the optimal extraction algorithm to obtain the 1D spectrum. We do not fix the optimal extraction parameters to match those determined from the global background approach, as our goal is to examine the difference in spatial extents (from the 2D data) and fluxes of the emission lines (from the 1D spectra) observed when applying the different background subtraction techniques.

\begin{figure}[t]
   \centering
   \includegraphics[width=0.75\textwidth]{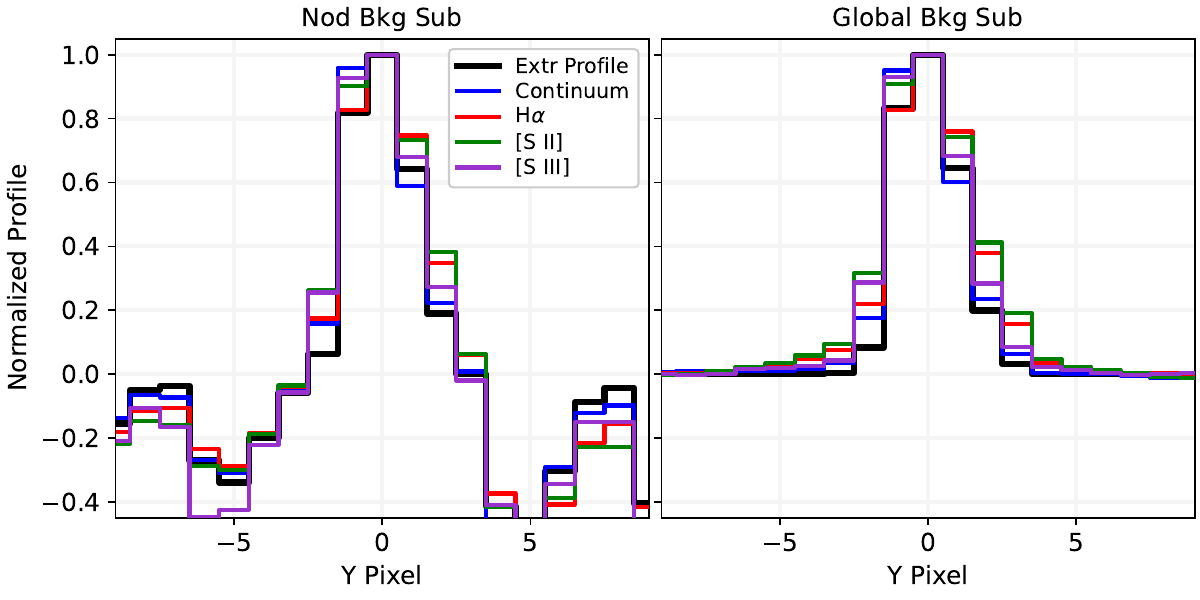}
   \caption{Emission in the spatial direction of the 2D NIRSpec data of the galaxy BX474. The profiles are normalized by the peak emission measured in the center row. The \textit{Left} and \textit{Right} panels compare the profiles obtained from the nod and global background subtraction approach, respectively. The spatial profiles plotted are H$\alpha$ (red), \sii$\lambda$6718 (green), \siii$\lambda$9533 (purple), the stellar continuum (blue), and the optimal 1D extraction profile determined from \textsc{msaexp} (black). The nod background subtracted data shows self-subtraction residuals in both the continuum and emission lines, while the global background method described in \S\ref{sec:jwstdata} conserves the extended emission.}
   \label{fig:profs_appendix1}
\end{figure}

To illustrate the impact of the global background subtraction approach, Figure \ref{fig:profs_appendix1} compares the spatial profiles of the strong emission lines observed in the galaxy BX474. This galaxy is relatively compact, with an optimal extraction profile FWHM $\sim$2.5 pixels in the spatial direction (after accounting for the FWHM of the PSF). The profile for a given emission line is acquired from the final rectified 2D spectrum; the profile is determined as the sum of the flux in an area $\pm$2 pixels
around the predicted line center in the observed frame, normalized to the maximum flux in the spatial profile. The various colored lines represent the spatial extent of the strong H$\alpha$, \sii$\lambda$6718, and \siii$\lambda$9533 emission lines, as well as the stellar continuum at 7500 \AA\ (plotted in blue). The solid black line represents the optimal profile at H$\alpha$ line center used to extract the 1D spectrum of BX474.

The left panel plots the spatial profiles measured using the nod background subtraction method, which shows the self-subtraction residuals at $\pm$5 and $\pm$10 pixels from the row of peak line intensity. The self-subtraction is observed in both the strong lines and the stellar continuum, and the magnitude of the optimal extraction profile becomes negative to match these observed trends. There is little spatial information that can be gleaned from the the profile in the left panel, as any marginally extended emission is washed out by the self-subtraction residuals. The right panel plots the spatial emission line and continuum profiles obtained from the global background subtraction method. While the spatial profile within $\pm$3 pixels from the center of the object appears relatively consistent with the nod background approach, the global background approach preserves the extended emission in the strong lines and produces a featureless off-source background. The extended emission is of particular interest: in the case of BX474, significant H$\alpha$ emission is more intense than the continuum emission at a separation larger than 5 pixels from the object center. We also observe different spatial profiles for emission lines related to the ionization state of the gas, such as \sii\ and \siii. The combination of the ultra-deep observations and global background approach will enable future studies concerning the radial change in ionization, gas density, and chemical composition in the CECILIA galaxies

\begin{figure}[t]
   \centering
   \includegraphics[width=0.50\textwidth]{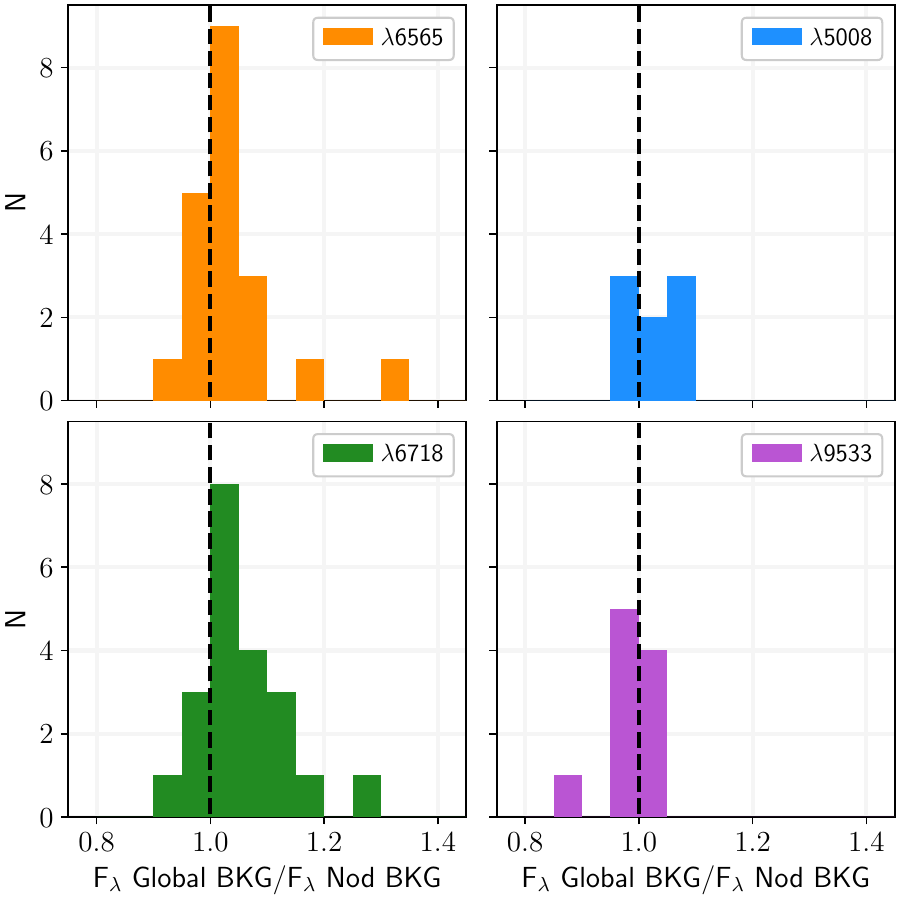}
   \caption{Comparison of the line fluxes measured from the global and nod background subtraction approaches when using an optimal extraction algorithm. The distribution of flux ratios for a given emission line is plotted in each panel, clockwise from top left: H$\alpha$, \oiii$\lambda$5008, \siii$\lambda$9533, and \sii$\lambda$6718. The black dashed lines represents equivalent line fluxes from the two techniques. While many CECILIA galaxies show consistent line fluxes when using a global or nod background subtraction method, there is large scatter and some emission lines (e.g., \sii) show slightly larger fluxes when using the global background method.}
   \label{fig:hist_appendix1}
\end{figure}

For the purposes of the present study, we are primarily interested in the change in line flux introduced when adopting either a global or local background method, as this can potentially alter the inferred physical conditions and chemical composition of the ISM. In measuring the line fluxes from the 1D spectra, we apply the same methodology as outlined in \S\ref{sec:jwstdata} and \S\ref{sec:linefits}: we apply a spectrophotometric correction to produce agreement between the measured stellar continuum and the predicted SED continuum, subtract the SED continuum from the 1D spectrum, then fit the emission lines with Gaussian profiles. Figure \ref{fig:hist_appendix1} plots the distribution of line flux ratios measured from the CECILIA galaxies, where unity represents consistent line fluxes between the two approaches.

It is reassuring that both methods produce relatively consistent strong line fluxes; indeed, other studies have found consistent line fluxes when using \textsc{msaexp}'s recently-implemented global background strategy and the default nod background approach \citep[e.g.,][]{degraaf2024,degraaf2025}. This trend is related to the optimal extraction profiles, which provide the most weight to the central 1-2 pixels in the spatial direction when producing the 1D spectra. We do note that there is relatively large scatter in the observed line flux ratios, where the global background approach produces generally stronger \sii$\lambda$6718. The change in emission line flux is related to the chosen background approach and the resulting optimal 1D extraction, which can weight different portions of the 2D spectra based on the assumed profile.
In summary, the global background approach preserves the extended strong line emission while producing line fluxes similar to the nod background method when using an optimal extraction algorithm. For future spatial analyses of the CECILIA galaxies and to capture the total emission from lines such as \sii\ and H$\alpha$, we adopt the global background subtraction method for the determination of \te\ and chemical abundances.

\section{Reddening Corrections and Cross Grating Uncertainties}\label{app:ebv}

Dust attenuation must be accounted for before the measurement of physical conditions and chemical abundances. As discussed in \S\ref{sec:ebvtrends}, $E(B-V)$ is constrained using the observed \ion{H}{1} recombination line ratios and the theoretical emissivities under the assumption of Case B recombination. H$\alpha$/H$\beta$ has been the most accessible line ratio for ground-based surveys of high-\emph{z} galaxies, but it is only possible to measure H$\alpha$/H$\beta$ from the NIRSpec observations of seven CECILIA galaxies. H$\alpha$/H$\beta$ can be measured from the archival MOSFIRE observations, but it is also possible to measure faint, high-order Paschen lines in the NIRSpec data, allowing for an additional check on $E(B-V)$. In this Appendix, we compare the $E(B-V)$ inferred from the ratio of Paschen lines to H$\alpha$ and discuss lingering flux uncertainties between the G235M and G395M data.

The Paschen lines are relatively faint, comparable in intensity to the \te-sensitive auroral lines. Additionally, the \ion{H}{1} recombination lines with the longest wavelength separation are most sensitive to dust attenuation. Therefore, we take the following approach to estimate $E(B-V)$ from the Paschen lines: 1. We utilize Paschen lines up to P9 $\lambda$9232; 2. We require at least two Paschen lines to be significantly detected; 3. $E(B-V)$ is determined as a weighted average from the individual reddening estimates from each Paschen line relative to H$\alpha$. This calculation is performed for both the G235M and the G395M Paschen line ratios, but we note that the lines utilized change depending on the grating. In the G235M data we use at most three Paschen lines (P9, P8, P7), while there are four potential Paschen lines (P8 through P5) detectable in the G395M spectra of the CECILIA galaxies.

\begin{figure}[t]
   \centering
   \includegraphics[width=0.37\textwidth]{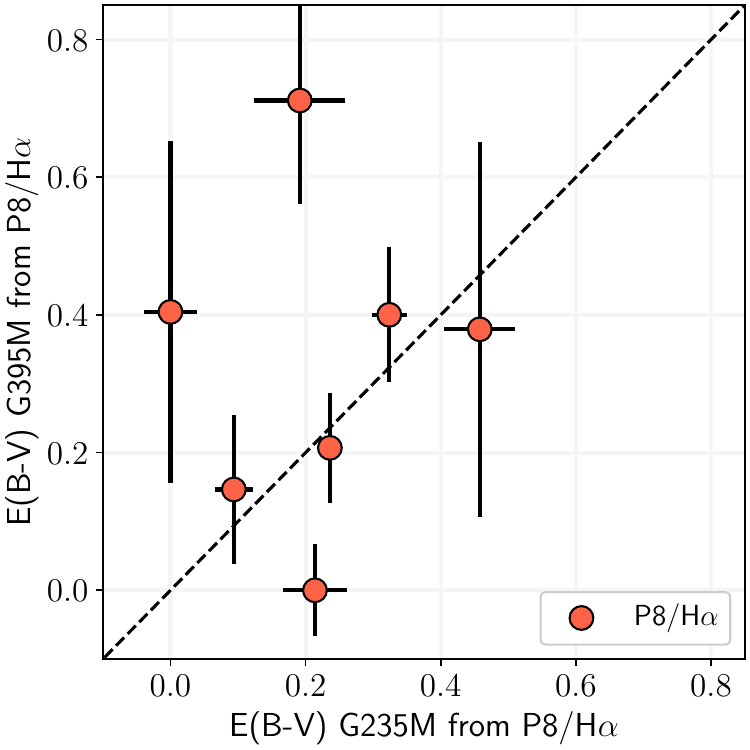}
   \hskip 6ex
   \includegraphics[width=0.37\textwidth]{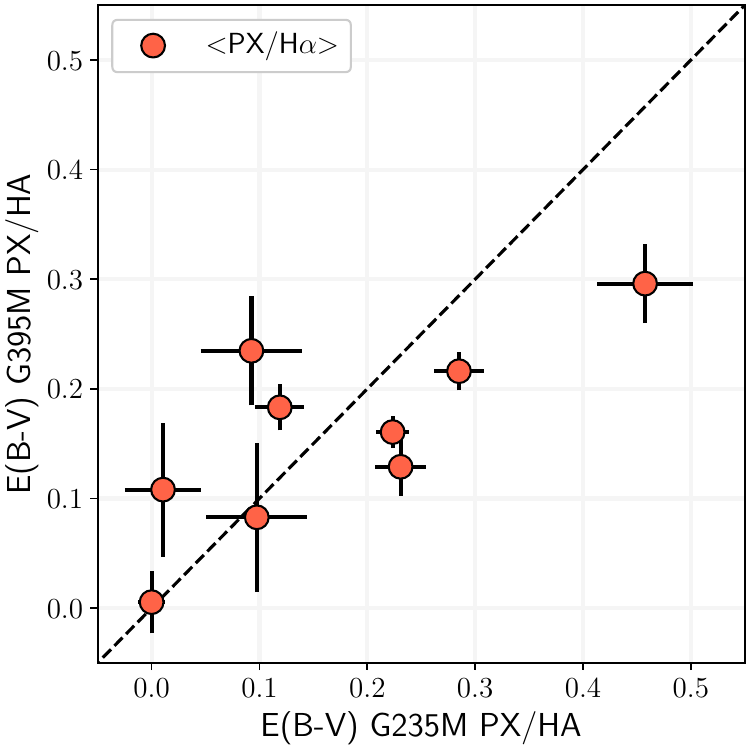}
   \caption{$E(B-V)$ determined from the Paschen line ratios measured in the CECILIA galaxies. \textit{Left:} $E(B-V)$ from individual P8/H$\alpha$ ratios. P8 is measured in G235M on the horizontal axis and G395M on the vertical axis. H$\alpha$ is measured from G235M in all galaxies. The individual $E(B-V)$ measurements scatter around the one-to-one line (black dashed line). Differences in $E(B-V)$ could be caused by the blend of \siii$\lambda$9533 and P8 in the G395M data, whereas the lines are resolved in most G235M spectra. \textit{Right:} The average $E(B-V)$ measured from the available Paschen-to-H$\alpha$ line ratio in G235M (horizontal) and G395M (vertical). Offsets in $E(B-V)$ measured in individual galaxies could be related to differences in the attenuation curve or lingering flux calibration issues between the two gratings.
   }
   \label{fig:ebv_appendix1}
\end{figure}

Figure \ref{fig:ebv_appendix1} plots the $E(B-V)$ trends measured from the Paschen-to-H$\alpha$ line ratios in the CECILIA sample. The left panel focuses on the $E(B-V)$ inferred from the P8/H$\alpha$ ratio. This Paschen line is most frequently detected in the overlap portion of the G235M and G395M spectra, permitting a comparison between the $E(B-V)$ predicted from the different detectors. While the uncertainties on the G395M-derived $E(B-V)$ are large on account of the faint P8 line and noisy continuum, four of the seven galaxies show agreement between the independent $E(B-V)$ measurements in G235M and G395M. Observational challenges associated with robustly measuring P8 may lead to discrepant $E(B-V)$. For example, the P8 line is in close proximity to \siii$\lambda$9533: while both lines are resolved in the average G235M spectrum, the two lines are blended in most G395M data.

Differences to the line fits resulting from the coarser wavelength sampling and noisier spectra can contribute to the resulting scatter about the one-to-one line in the left panel, so in the right panel of Figure \ref{fig:ebv_appendix1} we plot $E(B-V)$ measured from G235M and G395M using the average $E(B-V)$ from multiple Pachen-to-H$\alpha$ line ratios. The $E(B-V)$ from G235M and G395M scatter around the one-to-one line, although $E(B-V)$ measured in individual galaxies can differ depending on the available Paschen lines. This comparison assumes the shape of the \citet{reddy2020} is valid at wavelengths longer than 8000 \AA, an area where it is not well calibrated.
Differences in $E(B-V)$ from the different Paschen lines, therefore, could be the result of a different attenuation law shape at NIR wavelengths. Indeed, early evidence from JWST observations of high-\emph{z} galaxies indicate that the attenuation curve may deviate from both the \citet{cardelli1989} and \citet{reddy2020} calibrations. For example, \citet{sanders2024} measure the attenuation curve from eleven Balmer and Paschen lines observed in a galaxy at \emph{z}$=$4.41, finding a steeper attenuation curve in the NIR relative to the \citet{reddy2020} parameterization \citep[see other works on the dust attenuation in high-\emph{z} galaxies by][]{reddy2023,markov2025}.

\begin{figure}[t]
   \centering
   \includegraphics[width=0.50\textwidth]{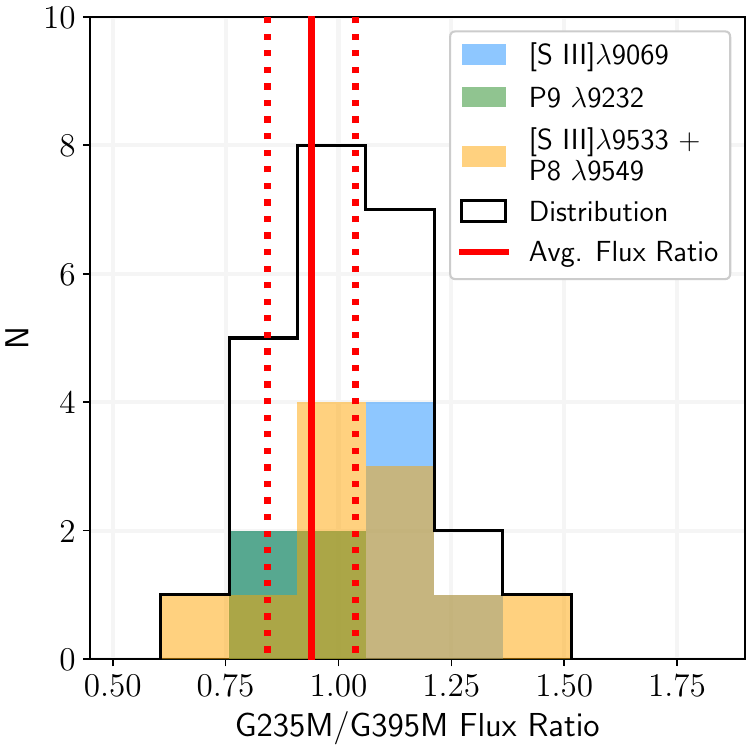}
   \caption{Comparison of the line fluxes measured simultaneously in G235M and G395M. The distribution of G235M flux to G395M flux ratios of \siii$\lambda$9071 (blue), P9 $\lambda$9232 (green), and \siii$\lambda$9533$+$P8$\lambda$9549 (yellow) are plotted along with the histogram distribution of the full comparison sample (solid black). The inverse variance weighted average and uncertainty of the G235M-to-G395M flux ratios for the full sample are plotted as solid and dotted red lines, respectively. While the average G235M-to-G395M flux ratio is consistent with unity, the uncertainty indicates that individual emission line flux measurements in each grating can differ by $\pm$9\%. We assign this percent uncertainty to any cross-grating line flux ratio.}
   \label{fig:ratios_appendix2}
\end{figure}

Another potential systematic is a relative flux calibration error between G235M and G395M. This is most evident when comparing the line fluxes measured in the overlap region between the G235M and G395M gratings. Depending on the galaxy's redshift, this can include the \siii\ nebular lines, P8, and P9. In Figure \ref{fig:ratios_appendix2}, we provide a histogram distribution of the flux ratios for these emission lines. Each ratio represents the flux measured for an emission line independently in G235M and G395M after global background subtraction, optimal extraction from the 2D spectra, and Gaussian emission line fitting. We note that \siii$\lambda$9533 and P8$\lambda$9549 are often blended in the G395M data, so we compare the G235M and G395M ratio of the total \siii$+$P8 flux. The F$_\lambda$(G235M)/F$_\lambda$(G395M) inverse variance weighted average is 0.95 with uncertainty of 0.09, indicating that the emission line fluxes measured from G235M and G395M generally agree but with non-negligible scatter.

These line flux variations can introduce systematic differences into the $E(B-V)$ derived from the G395M Paschen lines relative to the G235M H$\alpha$ flux. It is for this reason that we adopt the neighboring \ion{H}{1} line approach for a pseudo-reddening correction (see Section \ref{sec:tecalcs}), as this bypasses the uncertainties in the attenuation curve and relative flux calibration. Additionally, when reporting $E(B-V)$ we prioritize measurements from the Balmer decrement in NIRSpec and MOSFIRE since the \citet{reddy2020} attenuation curve is calibrated at these wavelengths. In instances when H$\alpha$ and H$\beta$ are not significantly detected, we use $E(B-V)$ constrained from the Paschen lines and H$\alpha$. Finally, we adopt a 9\% uncertainty on all flux ratios involving lines in G235M and G395M (e.g., Paschen-to-H$\alpha$ ratios) to account for the observed scatter in the individual line flux measurements. In the future, it may be possible to constrain the average shape of the NIR attenuation curve for galaxies at 2$\leq$\emph{z}$\leq$3 from the CECILIA spectra, but we do not attempt this calibration in the present work owing to the persistent fluxing issues.

\section{CECILIA Properties and Auroral Line Detections}\label{app:auroral}

\begin{figure*}[!t]
\epsscale{1.0}
   \centering
   \includegraphics[width=0.93\textwidth]{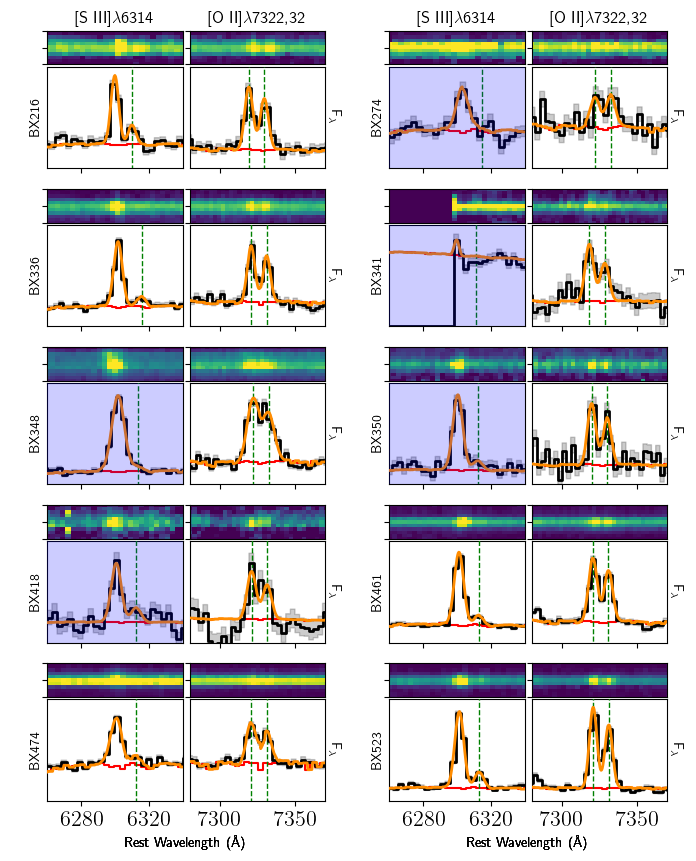}
   \caption{Auroral line detections in the CECILIA galaxies. Each set of four panels provides the NIRSpec G235M 2D spectrum (top row) and the 1D spectrum (bottom row) around the \siii\ (left) and \oii\ (right) auroral lines. The 1D spectra are plotted in units of erg/s/cm$^2$/\AA, and the vertical axes are adjusted for each spectrum to highlight the faint auroral lines. The model SED continuum (red) and Gaussian fit to the emission lines (orange) are plotted on top of the 1D spectrum (black with gray errors), and the position of each auroral line is highlighted as a vertical dashed green line. Galaxies with non-detections of the \siii\ auroral line are shaded in blue. Continued in Figure \ref{fig:auroral_2}.}
   \label{fig:auroral_1}
\end{figure*}

\begin{figure*}[!t]
\epsscale{1.0}
   \centering
   \includegraphics[width=0.95\textwidth]{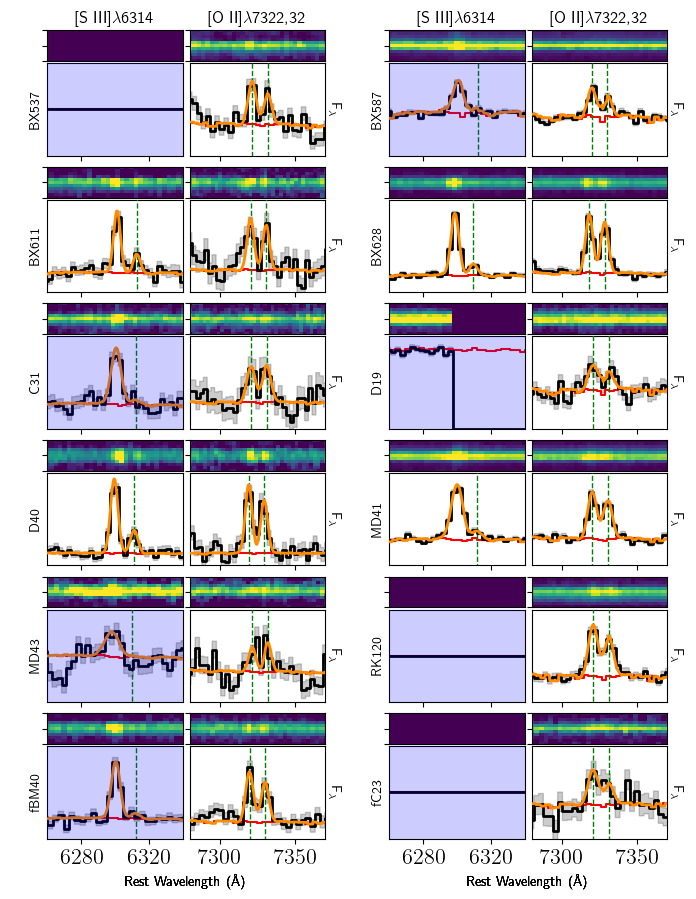}
   \caption{Auroral line detections in the CECILIA galaxies. Continued from Figure \ref{fig:auroral_1}, see above for details.}
   \label{fig:auroral_2}
\end{figure*}

In this Appendix, we provide tables with the physical conditions and abundances measured in the CECILIA galaxies with at least one direct \te\ (Table \ref{t:abun}). We also plot cutouts of the \siii\ and \oii\ auroral line detections in the CECILIA sample (Figures \ref{fig:auroral_1} and \ref{fig:auroral_2}).

\newpage

\startlongtable
\begin{deluxetable*}{lccccccc}  
\tablecaption{CECILIA Physical Conditions and Abundances \label{t:abun}}
\tablewidth{\columnwidth}
\tabletypesize{\footnotesize}
\tablehead{\multicolumn{8}{c}{CECILIA Physical Conditions}}
\startdata
Property  & Q2343-BX216 & Q2343-BX274 & Q2343-BX336 & Q2343-BX341 & Q2343-BX348 & Q2343-BX350 & Q2343-BX418 \\
\hline
\emph{z}   & 2.0875 & 2.1204 & 2.5455 & 2.5759 & 2.4494 & 2.5773 & 2.3059 \\
\vspace{-0.15cm} \\
$E(B-V)$   & 0.18$\pm$0.02 & 0.14$\pm$0.13 & 0.25$\pm$0.01 & 0.24$\pm$0.02 & 0.48$\pm$0.07 & 0.05$\pm$0.02 & 0.10$\pm$0.08 \\
\vspace{-0.15cm} \\
n$_e$[S\ii] (cm$^{-3}$)   &  85$_{- 35}^{+ 47}$ & 180$_{- 88}^{+116}$ & 278$_{- 32}^{+ 35}$ & 158$_{- 54}^{+ 66}$ & 347$_{- 32}^{+ 34}$ & 236$_{- 62}^{+ 74}$ &  78$_{- 40}^{+146}$ \\
n$_e$[O\ii] (cm$^{-3}$)   & \nodata & \nodata & 136$_{- 60}^{+ 60}$ & \nodata & 1077$_{-152}^{+168}$ & 831$_{-234}^{+271}$ & 393$_{-102}^{+109}$ \\
\vspace{-0.30cm} \\
n$_{e,Used}$ (cm$^{-3}$)   &  85$_{- 35}^{+ 47}$ & 180$_{- 88}^{+116}$ & 207$_{- 34}^{+ 35}$ & 158$_{- 54}^{+ 66}$ & 712$_{- 78}^{+ 86}$ & 533$_{-121}^{+141}$ & 235$_{- 55}^{+ 91}$ \\
\vspace{-0.15cm} \\
T$_e$[O\ii] (K)   & \nodata & 9000$\pm$1900 & 8200$\pm$ 300 & 11400$\pm$1000 & 8200$\pm$ 700 & 13100$\pm$1600 & 12000$\pm$2300 \\
T$_e$[N\ii] (K)   & \nodata & \nodata & $<$10600 & \nodata & $<$9700 & $<$21300 & \nodata \\
T$_e$[S\iii] (K)   & 11000$\pm$1100 & $<$22600 & 9100$\pm$ 800 & \nodata & $<$8400 & $<$11500 & $<$12700 \\
T$_e$[O\iii] (K)   & \nodata & \nodata & \nodata & \nodata & \nodata & \nodata & \nodata \\
\vspace{-0.40cm} \\
T$_{e,Low}$ (K)   & 10200$\pm$ 700 & 9000$\pm$1900 & 8200$\pm$ 300 & 11400$\pm$1000 & 8200$\pm$ 700 & 13100$\pm$1600 & 12000$\pm$2300 \\
T$_{e,Int}$ (K)   & 11000$\pm$1100 & 9100$\pm$2700 & 9100$\pm$ 800 & 12600$\pm$1400 & 7900$\pm$1100 & 15100$\pm$2300 & 13400$\pm$3400 \\
T$_{e,High}$ (K)   & 11100$\pm$1300 & 9700$\pm$2600 & 8900$\pm$1000 & 12900$\pm$1500 & 8700$\pm$1200 & 15100$\pm$2200 & 13600$\pm$3100 \\
\vspace{-0.15cm} \\
O$^{+}$/H$^+$ ($\times$10$^5$)   & 8.8$\pm$4.1 & 25.2$\pm$38.4 & 34.3$\pm$7.6 & 4.3$\pm$1.5 & 40.9$\pm$21.4 & 2.0$\pm$1.0 & 2.1$\pm$1.9 \\
O$^{2+}$/H$^+$ ($\times$10$^5$)   & 12.7$\pm$5.4 & 10.6$\pm$15.6 & 18.7$\pm$9.3 & 7.8$\pm$2.7 & 13.6$\pm$8.6 & 4.8$\pm$1.9 & 8.8$\pm$6.9 \\
12+log(O/H)   & 8.33$\pm$0.14 & 8.55$\pm$0.50 & 8.72$\pm$0.10 & 8.08$\pm$0.11 & 8.74$\pm$0.18 & 7.83$\pm$0.14 & 8.04$\pm$0.29 \\
\vspace{-0.15cm} \\
N$^{+}$/H$^+$ ($\times$10$^6$)   & 1.7$\pm$0.3 & 8.7$\pm$6.4 & 12.4$\pm$1.5 & 1.8$\pm$0.4 & 27.1$\pm$7.6 & 2.6$\pm$0.7 & 1.3$\pm$0.7 \\
12+log(N/H)   & 6.63$\pm$0.18 & 7.09$\pm$0.42 & 7.28$\pm$0.10 & 6.69$\pm$0.16 & 7.56$\pm$0.15 & 6.95$\pm$0.23 & 6.85$\pm$0.48 \\
log(N/O)   & $-$1.71$\pm$0.11 & $-$1.47$\pm$0.35 & $-$1.42$\pm$0.08 & $-$1.39$\pm$0.11 & $-$1.19$\pm$0.16 & $-$0.88$\pm$0.14 & $-$1.19$\pm$0.24 \\
\vspace{-0.15cm} \\
S$^{+}$/H$^+$ ($\times$10$^7$)   & 6.6$\pm$1.2 & 16.8$\pm$11.8 & 17.0$\pm$2.0 & 6.7$\pm$1.3 & 28.1$\pm$7.7 & 5.3$\pm$1.4 & 2.3$\pm$1.1 \\
S$^{2+}$/H$^+$ ($\times$10$^7$)   & 18.6$\pm$3.8 & 25.0$\pm$24.4 & 35.1$\pm$7.7 & 16.3$\pm$6.5 & 53.3$\pm$19.4 & 10.0$\pm$3.3 & 8.2$\pm$4.4 \\
S ICF   & 1.06$\pm$0.11 & 0.95$\pm$0.10 & 0.96$\pm$0.10 & 1.10$\pm$0.11 & 0.95$\pm$0.09 & 1.14$\pm$0.11 & 1.38$\pm$0.14 \\
12+log(S/H)   & 6.43$\pm$0.08 & 6.60$\pm$0.29 & 6.70$\pm$0.08 & 6.40$\pm$0.13 & 6.89$\pm$0.12 & 6.24$\pm$0.11 & 6.16$\pm$0.19 \\
log(S/O)   & $-$1.91$\pm$0.16 & $-$1.95$\pm$0.58 & $-$2.02$\pm$0.13 & $-$1.68$\pm$0.17 & $-$1.85$\pm$0.22 & $-$1.59$\pm$0.18 & $-$1.87$\pm$0.35 \\
\vspace{-0.15cm} \\
Ar$^{2+}$/H$^+$ ($\times$10$^7$)   & 3.7$\pm$0.8 & 5.2$\pm$6.4 & 6.2$\pm$1.5 & 2.4$\pm$0.6 & 9.6$\pm$4.1 & 2.5$\pm$0.8 & 1.9$\pm$1.1 \\
Ar ICF   & 1.08$\pm$0.11 & 1.19$\pm$0.12 & 1.16$\pm$0.12 & 1.07$\pm$0.11 & 1.21$\pm$0.12 & 1.08$\pm$0.11 & 1.10$\pm$0.11 \\
12+log(Ar/H)   & 5.60$\pm$0.11 & 5.79$\pm$0.53 & 5.86$\pm$0.11 & 5.42$\pm$0.12 & 6.06$\pm$0.19 & 5.44$\pm$0.14 & 5.32$\pm$0.26 \\
log(Ar/O)   & $-$2.73$\pm$0.17 & $-$2.76$\pm$0.73 & $-$2.86$\pm$0.15 & $-$2.67$\pm$0.16 & $-$2.67$\pm$0.27 & $-$2.39$\pm$0.20 & $-$2.72$\pm$0.39 \\
\tablebreak \\
Property  & Q2343-BX429 & Q2343-BX461 & Q2343-BX474 & Q2343-BX523 & Q2343-BX587 & Q2343-BX611 & Q2343-BX628 \\
\hline
\emph{z}   & 2.1753 & 2.5680 & 2.2273 & 2.2708 & 2.2438 & 2.7571 & 2.2798 \\
\vspace{-0.15cm} \\
$E(B-V)$   & 0.37$\pm$0.03 & 0.44$\pm$0.01 & 0.38$\pm$0.22 & 0.19$\pm$0.01 & 0.17$\pm$0.03 & 0.11$\pm$0.02 & 0.33$\pm$0.13 \\
\vspace{-0.15cm} \\
n$_e$[S\ii] (cm$^{-3}$)   & 144$_{- 40}^{+ 44}$ & 415$_{- 33}^{+ 34}$ & 394$_{- 45}^{+ 48}$ & 521$_{- 50}^{+ 43}$ & 345$_{- 51}^{+ 56}$ &  55$_{- 55}^{+110}$ & 137$_{- 26}^{+ 31}$ \\
n$_e$[O\ii] (cm$^{-3}$)   & 196$_{- 83}^{+ 88}$ &  67$_{- 67}^{+150}$ & 516$_{-249}^{+286}$ & 678$_{-236}^{+277}$ & 165$_{- 59}^{+ 70}$ & \nodata & \nodata \\
\vspace{-0.30cm} \\
n$_{e,Used}$ (cm$^{-3}$)   & 170$_{- 46}^{+ 49}$ & 241$_{- 37}^{+ 77}$ & 455$_{-127}^{+145}$ & 599$_{-121}^{+140}$ & 255$_{- 39}^{+ 45}$ &  55$_{- 55}^{+110}$ & 137$_{- 26}^{+ 31}$ \\
\vspace{-0.15cm} \\
T$_e$[O\ii] (K)   & \nodata & 8700$\pm$ 500 & 6600$\pm$1000 & 9200$\pm$1000 & 10700$\pm$1500 & \nodata & 15900$\pm$4100 \\
T$_e$[N\ii] (K)   & $<$24100 & $<$8800 & 13900$\pm$1400 & 13300$\pm$1100 & $<$9700 & \nodata & $<$12100 \\
T$_e$[S\iii] (K)   & 12100$\pm$2000 & 8800$\pm$ 600 & 10500$\pm$1000 & 10600$\pm$ 400 & $<$9500 & 13800$\pm$3000 & 9500$\pm$ 500 \\
T$_e$[O\iii] (K)   & \nodata & \nodata & \nodata & \nodata & \nodata & \nodata & \nodata \\
\vspace{-0.40cm} \\
T$_{e,Low}$ (K)   & 11000$\pm$1400 & 8700$\pm$ 500 & 13900$\pm$1400 & 13300$\pm$1100 & 10700$\pm$1500 & 12200$\pm$2100 & 15900$\pm$4100 \\
T$_{e,Int}$ (K)   & 12100$\pm$2000 & 8800$\pm$ 600 & 10500$\pm$1000 & 10600$\pm$ 400 & 11500$\pm$2200 & 13800$\pm$3000 & 9500$\pm$ 500 \\
T$_{e,High}$ (K)   & 12400$\pm$2500 & 8500$\pm$ 800 & 10600$\pm$1200 & 10700$\pm$ 700 & 11900$\pm$2100 & 14500$\pm$3700 & 9300$\pm$ 800 \\
\vspace{-0.15cm} \\
O$^{+}$/H$^+$ ($\times$10$^5$)   & 23.8$\pm$14.2 & 26.1$\pm$8.4 & 5.4$\pm$3.0 & 3.0$\pm$1.1 & 7.1$\pm$5.1 & 3.6$\pm$4.4 & 1.0$\pm$1.1 \\
O$^{2+}$/H$^+$ ($\times$10$^5$)   & 4.9$\pm$3.5 & 23.7$\pm$9.7 & 6.9$\pm$2.8 & 15.5$\pm$3.3 & 4.9$\pm$3.2 & 7.2$\pm$6.3 & 22.9$\pm$7.4 \\
12+log(O/H)   & 8.46$\pm$0.22 & 8.70$\pm$0.11 & 8.09$\pm$0.15 & 8.27$\pm$0.08 & 8.08$\pm$0.22 & 8.03$\pm$0.31 & 8.38$\pm$0.14 \\
\vspace{-0.15cm} \\
N$^{+}$/H$^+$ ($\times$10$^6$)   & 5.4$\pm$1.8 & 15.3$\pm$2.7 & 5.3$\pm$1.1 & 2.5$\pm$0.5 & 12.4$\pm$4.8 & 0.9$\pm$0.4 & 1.7$\pm$1.0 \\
12+log(N/H)   & 6.81$\pm$0.16 & 7.46$\pm$0.13 & 7.08$\pm$0.19 & 7.18$\pm$0.18 & 7.32$\pm$0.24 & 6.45$\pm$0.47 & 7.61$\pm$0.54 \\
log(N/O)   & $-$1.65$\pm$0.30 & $-$1.23$\pm$0.11 & $-$1.01$\pm$0.19 & $-$1.08$\pm$0.10 & $-$0.76$\pm$0.20 & $-$1.59$\pm$0.21 & $-$0.76$\pm$0.24 \\
\vspace{-0.15cm} \\
S$^{+}$/H$^+$ ($\times$10$^7$)   & 9.0$\pm$2.8 & 14.0$\pm$2.4 & 4.3$\pm$0.9 & 2.7$\pm$0.5 & 10.8$\pm$4.0 & 3.6$\pm$1.5 & 2.5$\pm$1.4 \\
S$^{2+}$/H$^+$ ($\times$10$^7$)   & 14.2$\pm$4.9 & 40.4$\pm$6.6 & 19.7$\pm$3.8 & 18.9$\pm$1.6 & 18.8$\pm$8.0 & 9.2$\pm$4.7 & 30.9$\pm$4.0 \\
S ICF   & 0.94$\pm$0.09 & 1.00$\pm$0.10 & 1.04$\pm$0.10 & 1.59$\pm$0.16 & 0.99$\pm$0.10 & 1.11$\pm$0.11 & 4.36$\pm$0.44 \\
12+log(S/H)   & 6.34$\pm$0.11 & 6.73$\pm$0.07 & 6.40$\pm$0.08 & 6.54$\pm$0.06 & 6.47$\pm$0.14 & 6.16$\pm$0.17 & 7.16$\pm$0.07 \\
log(S/O)   & $-$2.12$\pm$0.25 & $-$1.96$\pm$0.13 & $-$1.69$\pm$0.17 & $-$1.73$\pm$0.10 & $-$1.62$\pm$0.26 & $-$1.88$\pm$0.35 & $-$1.22$\pm$0.15 \\
\vspace{-0.15cm} \\
Ar$^{2+}$/H$^+$ ($\times$10$^7$)   & \nodata & 7.2$\pm$1.3 & 3.3$\pm$0.7 & 4.0$\pm$0.4 & 3.0$\pm$1.5 & 2.3$\pm$1.1 & 5.6$\pm$0.8 \\
Ar ICF   & \nodata & 1.11$\pm$0.11 & 1.08$\pm$0.11 & 1.11$\pm$0.11 & 1.13$\pm$0.11 & 1.07$\pm$0.11 & 1.81$\pm$0.18 \\
12+log(Ar/H)   & \nodata & 5.90$\pm$0.09 & 5.55$\pm$0.11 & 5.65$\pm$0.06 & 5.53$\pm$0.21 & 5.39$\pm$0.22 & 6.00$\pm$0.08 \\
log(Ar/O)   & \nodata & $-$2.79$\pm$0.14 & $-$2.54$\pm$0.18 & $-$2.62$\pm$0.10 & $-$2.55$\pm$0.30 & $-$2.65$\pm$0.38 & $-$2.37$\pm$0.16 \\
\tablebreak \\
Property  & Q2343-C31 & Q2343-D40 & Q2343-MD43 & Q2343-RK120 & Q2343-fBM40 & Q2343-fC23 \\
\hline
\emph{z}   & 3.0592 & 2.9628 & 2.5799 & 2.3522 & 2.1477 & 2.1734 \\
\vspace{-0.15cm} \\
$E(B-V)$   & 0.10$\pm$0.02 & 0.21$\pm$0.01 & 0.00$\pm$0.00 & 0.40$\pm$0.02 & 0.00$\pm$0.05 & 0.06$\pm$0.04 \\
\vspace{-0.15cm} \\
n$_e$[S\ii] (cm$^{-3}$)   & 113$_{- 67}^{+123}$ & 136$_{- 51}^{+ 56}$ & 633$_{-298}^{+541}$ & 275$_{- 26}^{+ 29}$ &  45$_{- 23}^{+ 71}$ &  26$_{- 26}^{+107}$ \\
n$_e$[O\ii] (cm$^{-3}$)   & \nodata & 335$_{-181}^{+536}$ & 224$_{-121}^{+156}$ & 381$_{-213}^{+317}$ & 270$_{-141}^{+162}$ & \nodata \\
\vspace{-0.30cm} \\
n$_{e,Used}$ (cm$^{-3}$)   & 113$_{- 67}^{+123}$ & 235$_{- 94}^{+269}$ & 428$_{-161}^{+281}$ & 328$_{-107}^{+159}$ & 157$_{- 72}^{+ 88}$ &  26$_{- 26}^{+107}$ \\
\vspace{-0.15cm} \\
T$_e$[O\ii] (K)   & 12700$\pm$1700 & 13200$\pm$2700 & 13200$\pm$2900 & 6700$\pm$ 500 & 14100$\pm$1700 & 8000$\pm$1100 \\
T$_e$[N\ii] (K)   & \nodata & $<$28400 & \nodata & $<$9200 & \nodata & \nodata \\
T$_e$[S\iii] (K)   & \nodata & 14700$\pm$1900 & \nodata & \nodata & $<$11600 & \nodata \\
T$_e$[O\iii] (K)   & \nodata & 14300$\pm$ 700 & \nodata & \nodata & \nodata & \nodata \\
\vspace{-0.40cm} \\
T$_{e,Low}$ (K)   & 12700$\pm$1700 & 13200$\pm$2700 & 13200$\pm$2900 & 6700$\pm$ 500 & 14100$\pm$1700 & 8000$\pm$1100 \\
T$_{e,Int}$ (K)   & 14500$\pm$2600 & 14700$\pm$1900 & 15200$\pm$4200 & 5700$\pm$ 700 & 16400$\pm$2400 & 7700$\pm$1500 \\
T$_{e,High}$ (K)   & 14500$\pm$2400 & 14300$\pm$ 700 & 15100$\pm$3800 & 6800$\pm$1000 & 16300$\pm$2300 & 8500$\pm$1600 \\
\vspace{-0.15cm} \\
O$^{+}$/H$^+$ ($\times$10$^5$)   & 3.2$\pm$1.8 & 1.7$\pm$1.6 & 2.0$\pm$2.1 & 173.7$\pm$83.6 & 1.8$\pm$0.8 & 46.5$\pm$39.2 \\
O$^{2+}$/H$^+$ ($\times$10$^5$)   & 6.0$\pm$2.9 & 7.3$\pm$0.9 & 4.2$\pm$3.5 & 34.4$\pm$29.1 & 3.6$\pm$1.4 & 17.9$\pm$17.3 \\
12+log(O/H)   & 7.96$\pm$0.16 & 7.96$\pm$0.09 & 7.80$\pm$0.28 & 9.32$\pm$0.18 & 7.74$\pm$0.12 & 8.81$\pm$0.29 \\
\vspace{-0.15cm} \\
N$^{+}$/H$^+$ ($\times$10$^6$)   & 1.2$\pm$0.4 & 1.0$\pm$0.5 & 0.8$\pm$0.5 & 44.8$\pm$11.5 & 1.8$\pm$0.5 & 10.6$\pm$4.8 \\
12+log(N/H)   & 6.53$\pm$0.25 & 6.73$\pm$0.39 & 6.38$\pm$0.46 & 7.73$\pm$0.13 & 6.74$\pm$0.20 & 7.17$\pm$0.25 \\
log(N/O)   & $-$1.40$\pm$0.16 & $-$1.22$\pm$0.23 & $-$1.41$\pm$0.24 & $-$1.61$\pm$0.16 & $-$1.00$\pm$0.13 & $-$1.65$\pm$0.25 \\
\vspace{-0.15cm} \\
S$^{+}$/H$^+$ ($\times$10$^7$)   & 4.8$\pm$1.5 & 3.2$\pm$1.5 & 4.7$\pm$2.5 & 47.1$\pm$11.7 & 4.7$\pm$1.1 & 20.9$\pm$9.0 \\
S$^{2+}$/H$^+$ ($\times$10$^7$)   & \nodata & 8.6$\pm$2.2 & \nodata & 104.0$\pm$43.4 & 9.2$\pm$2.4 & 66.6$\pm$42.5 \\
S ICF   & \nodata & 1.35$\pm$0.13 & \nodata & 0.94$\pm$0.09 & 1.10$\pm$0.11 & 0.95$\pm$0.10 \\
12+log(S/H)   & \nodata & 6.20$\pm$0.11 & \nodata & 7.15$\pm$0.14 & 6.18$\pm$0.09 & 6.92$\pm$0.22 \\
log(S/O)   & \nodata & $-$1.76$\pm$0.14 & \nodata & $-$2.17$\pm$0.23 & $-$1.55$\pm$0.16 & $-$1.89$\pm$0.36 \\
\vspace{-0.15cm} \\
Ar$^{2+}$/H$^+$ ($\times$10$^7$)   & 1.5$\pm$0.6 & 1.6$\pm$0.4 & 1.1$\pm$0.8 & 25.1$\pm$13.1 & 1.8$\pm$0.5 & 12.3$\pm$9.3 \\
Ar ICF   & 1.07$\pm$0.11 & 1.11$\pm$0.11 & 1.08$\pm$0.11 & 1.24$\pm$0.12 & 1.08$\pm$0.11 & 1.19$\pm$0.12 \\
12+log(Ar/H)   & 5.21$\pm$0.18 & 5.26$\pm$0.12 & 5.09$\pm$0.30 & 6.49$\pm$0.23 & 5.29$\pm$0.13 & 6.17$\pm$0.33 \\
log(Ar/O)   & $-$2.75$\pm$0.24 & $-$2.70$\pm$0.15 & $-$2.71$\pm$0.42 & $-$2.82$\pm$0.30 & $-$2.45$\pm$0.18 & $-$2.64$\pm$0.44 \\
\enddata
\tablecomments{Measured properties of the CECILIA galaxies. Rows from top to bottom are: Spectroscopic redshift (1); $E(B-V)$ measured from the available \ion{H}{1} lines (2); \den\ measurements from \sii\ and \oii\ (3-4); the adopted ISM \den\ (5); \te\ measurements from \oii, \nii, \siii, and \oiii\ (6-9); the adopted low-, intermediate-, and high-ionization zone \te\ (10-12); ionic and total abundances of O (13-15), N (16-18), S with ICF (19-23), and Ar with ICF (24-27). Each column corresponds to an individual galaxy in the Q2343+125 field observed as part of CECILIA.}
\end{deluxetable*}

\bibliography{strom_ref_library}
\bibliographystyle{aasjournal}

\newpage

\newpage

\end{CJK*}
\end{document}